\documentclass[acmsmall]{acmart}
\usepackage{graphicx}
\usepackage{todonotes}
\usepackage{xcolor}
\usepackage{subcaption}
\usepackage{wrapfig}
\usepackage{footmisc}
\usepackage{tabu}
\usepackage{algorithm}
\usepackage{algpseudocode}
\usepackage{multirow}
\usepackage{adjustbox}
\usepackage{amsmath,amsfonts}
\PassOptionsToPackage{normalem}{ulem}
\usepackage{ulem}
\algnewcommand\algorithmicforeach{\textbf{for each}}
\algdef{S}[FOR]{ForEach}[1]{\algorithmicforeach\ #1\ \algorithmicdo}

\definecolor{darkgreen}{RGB}{21,176,26}
\definecolor{orange}{RGB}{242,148,7}
\definecolor{revised}{RGB}{196, 37, 26}

\providecolor{newadded}{rgb}{0,1,0}
\providecolor{omidadded}{rgb}{0,0,1}%{0.75, 0, 0.2}
\providecolor{deleted}{rgb}{1,0,0}
\newcommand{\added}[1]{{\color{black}{}#1}}
\newcommand{\newadded}[1]{{\color{black}{}#1}}
\newcommand{\omidadded}[1]{{\color{black}{}#1}}

\newcommand*\fplus{{\fontfamily{pbk}\fontseries{db}\selectfont+}}

\graphicspath{{figures/}}

\setcopyright{acmcopyright}
\copyrightyear{2021}
\acmYear{2021}
\acmJournal{TAAS}
\begin{document}

\title{Deep Learning for Effective and Efficient  Reduction of Large Adaptation Spaces in Self-Adaptive Systems}

\author{Danny Weyns}
\affiliation{
	\institution{Katholieke Universiteit Leuven, Belgium}
	%\city{Leuven}
	\country{Linnaeus University Sweden}
}
\email{danny.weyns@kuleuven.be}

\author{Omid Gheibi}
\affiliation{
	\institution{Katholieke Universiteit Leuven}
	%\city{Leuven}
	\country{Belgium}
}
\email{omid.gheibi@kuleuven.be}

\author{Federico Quin}
\affiliation{
	\institution{Katholieke Universiteit Leuven}
	%\city{Leuven}
	\country{Belgium}
}
\email{federico.quin@kuleuven.be}

\author{Jeroen Van Der Donckt}
\affiliation{
    \institution{Ghent University (imec)}
    %\city{Zwijnaarde}
    \country{Belgium}
}
\email{jeroen.vanderdonckt@ugent.be}


\begin{abstract}
    Many software systems today face uncertain operating conditions, such as sudden changes in the availability of resources or unexpected user behavior. Without proper mitigation these uncertainties can jeopardize the system goals. Self-adaptation is a common approach to tackle such uncertainties. When the system goals may be compromised, the self-adaptive system has to select the best adaptation option to reconfigure by analyzing the possible adaptation options, i.e., the adaptation space. Yet, analyzing large adaptation spaces using rigorous methods can be resource- and time-consuming, or even be infeasible. One approach to tackle this problem is by using \added{online machine learning to reduce adaptation spaces.} However, existing approaches require domain expertise \added{to perform feature engineering to define the learner, and support online adaptation space reduction only for specific goals.} To tackle these limitations, we present ''Deep Learning for Adaptation Space Reduction Plus'' -- DLASeR\fplus{} in short. DLASeR\fplus{} offers an extendable learning framework for online adaptation space reduction that does not require feature engineering, while supporting three common types of adaptation goals: threshold, optimization, and set-point goals. We evaluate DLASeR\fplus{} on two instances of an Internet-of-Things application with increasing sizes of adaptation spaces for different combinations of adaptation goals. We compare DLASeR\fplus{} with a baseline that applies exhaustive analysis and two state-of-the-art approaches for adaptation space reduction that rely on learning. Results show that DLASeR\fplus{} is effective with a negligible effect on the realization of the adaptation goals compared to an exhaustive analysis \mbox{approach, and supports three common types of adaptation goals beyond the  state-of-the-art approaches.} 

\end{abstract}

% TODO: make sure that all right categories are covered




\maketitle

% "Any prior appearance should be noted on the title page and it is the obligation of the author to inform the Editors-in-Chief if there are any circumstances concerning the contribution that bear on this policy"
%\ctodo{CSS Concepts fixen}
%\ctodo{Add reference to SEAMS paper -> required by TAAS}

\section{Introduction}\label{sec:intro}

% DONE - Short intro to SaS
% DONE - Introduce the problem of analysis in large adaptation spaces
% DONE - The different types of adaptation goals
% DONE - how deep learning can be applied
%Illustrate the problem with DeltaIoT.

Many software systems today face changing and uncertain operating conditions. For such systems, employing a stationary approach that does not adapt to changes may jeopardize the quality goals of the system. Consider for instance a web-based system that runs a fixed number of servers based on an average load. This configuration will result in a waste of resources when the load is very low, but, the number of servers may be insufficient to handle the peak demand~\cite{klein2014brownout}.

Self-adaptation is one prominent approach to tackle such problems~\cite{cheng2009software, weyns2017software}. A self-adaptive system reasons about itself and its environment, based on observations, to determine whether adaptation is required. In the case that adaptation is required, the system adapts itself in order to meet its adaptation goals, or gracefully degrade if the goals may temporarily not be achievable. Self-adaptation has been applied in a wide range of application domains, ranging from service-based systems to cyber-physical systems, Internet-of-Things, the Cloud, and robotics~\cite{muccini2016self, castaneda2014self, athreya2013designing, edwards2009architecture, 7515437}.
In the example of the web-based system above, enhancing the system with self-adaptation enables it to increase and decrease the number of servers dynamically based on the monitored load. This results in higher user satisfaction as well as improved economical and ecological use of resources.

In this research, we apply architecture-based adaptation~\cite{Garlan:2004, Kramer2007, Weyns2012-1} that adds an external feedback loop to the system. The feedback loop uses up-to-date architectural models as first-class citizens to reason about changes. The feedback loop forms a managing system on top of managed software system and is structured according to the MAPE-K reference model, short for Monitor\,-\,Analyzer\,-\,Planner\,-\,Executor\,-\,Knowledge~\cite{kephart2003vision}. MAPE-K divides adaptation in four principle functions~\cite{kephart2003vision,empiricalMAPE}. The monitor monitors the system and its environment. The analyzer determines whether adaptation is required or not and if so it analyzes the \textit{adaptation options} for adapting the system. \added{An adaptation option is a configuration of the system that can be reached from the current configuration by changing elements of the system through \textit{adaptation actions}. Adaptation actions can range from adjusting a parameter of the system up to an architectural re-configuration of the system. We use the term \textit{adaptation space} as the set of all the possible adaptation options at some point in time, i.e., all the possible configurations that can be reached from the current configuration of the system by applying a set of adaptation actions to the system. The size of the adaptation space (i.e., the number of adaptation options) may be constant over time, or it may change dynamically.} The planner then selects the best adaptation option according to the adaptation goals and composes a plan for adapting the managed system. Finally, the executor applies the adaptation actions of the plan on the managed system. The four MAPE functions share common knowledge (K), e.g., models of the system, the adaptation goals, the set of adaptation options, an adaptation plan, among others. 
%
% Adaptation goals uitleggen
%In the MAPE-K feedback loop, the analyzer its decisions are guided by adaptation goals. Adaptation goals are quality constraints that should be met by the system. Thus, a violation of these goals will result in an adaptation action. One can depict three different types of adaptation goals; threshold goals, set-point goals, and optimization goals~\cite{shevtsov2019simca}.
%
%We will introduce formally the three different kinds of adaptation goals:
%\textbf{Threshold goal}: given a set of of values $V = {v_1, v_2, ..., v_n}$ and a threshold value $T$, $v_i$

% \subsection{The adaptation space problem}

This paper focuses on the analysis of the adaptation options of the adaptation space, which is a task of the analyzer, \added{and selecting the best option based on the analysis results and the adaptation goals,} which is a task of the planner. Both tasks are essential for the decision-making in self-adaptive systems. During the execution of these tasks the feedback loop estimates a set of quality properties for each adaptation option of the adaptation space; each quality property corresponding to an adaptation goal. We consider threshold goals that require a system parameter to stay above/below a given threshold, optimization goals that require to minimize or maximize a system parameter, and set-point goals that require a system parameter to stay as close as possible to a given value. 

Selecting an adaptation option from a large adaptation space is often computationally expensive~\cite{cheng2009software, delemos2017software, weyns2017perpetual}. A common technique used to find the best adaptation option is runtime verification of formal runtime models that represent the system and its environment for one or more quality properties. These quality models have parameters that can be instantiated for a particular adaptation option and the actual values of the uncertainties. \added{A classic approach to estimate the qualities of the adaptation option is runtime quantitative verification,} see for example~\cite{Calinescu2011, Moreno:2015, weyns2018applying}. 
%Consequently, these approaches are restricted to small-scale settings, which often limits their applicability in practice~\cite{weyns2018applying}. 
%
It is important to note that the adaptation space exhibits dynamic behavior that is difficult to predict upfront. On the one hand, the estimated quality properties of the adaptation options vary over time as the uncertainties the system is exposed to change over time. On the other hand, the system configuration itself dynamically changes as the system is adapted over time.  

%To determine the best adaptation option, many existing adaptation approaches apply exhaustive verification techniques at runtime. These approaches are restricted to small-scale settings, which often limits their applicability in practice~\cite{weyns2018applying}.
%
%Certain verification techniques provide an interesting trade-off between analysis time and the accuracy of the results obtained (e.g., by controlling the number of sample system executions generated)~\cite{zhang2006model}. This makes it appropriate for systems with strict time constraints in which accuracy is not a primary concern. 
%
%Furthermore, efficient analysis and planning is of utmost importance for self-adaptive systems to react quickly to changes at runtime.
%
%Finally, the adaptation space, i.e., the collection of adaptation options, has a dynamic behavior. This dynamic behavior is observed in terms of the qualities of the adaptation options. These qualities vary over time, as the uncertainties and the current system configuration changes. 
%The various factors presented above indicate that handling large adaptation spaces is challenging. It is this challenge that we call \textit{the adaptation space problem}.

Different techniques have been studied to find the best adaptation option in a large adaptation space. 
%One line of research tackles the problem from the point of view of improving verification, e.g.. For instance, in~\cite{filieri2011} the authors apply a symbolic approach where a set of expressions is generated before deployment with symbols that are assigned values at runtime enabling efficient verification. Another line of research tackles the problem by \added{reducing the adaptation space during operation}. 
\added{One particular approach to deal with the problem is adaptation space reduction that aims at retrieving a subset of relevant adaptation options from an adaptation space that %contains only relevant system configurations that
are then considered for analysis.} An adaptation option is labeled relevant when it is predicted to satisfy the adaptation goals. Techniques have been applied in this approach including search-based techniques~\cite{8469102} and feature-based techniques~\cite{metzger2019feature}. Recently, different machine learning techniques have been investigated to reduce the adaptation space \added{at runtime}, see for instance~\cite{elkhodary2010fusion, quin2019efficient, jamshidi2019machine}. %
Among the applied learning techniques are decision trees, classification, and regression~\cite{gheibi2021}. However, most of these techniques rely on domain expertise to perform feature engineering \added{to define the learner}, which may hamper the applicability in practice. Further, most existing approaches are limited to threshold goals and optimization goals. 
In this paper, we tackle the following research question:
\vspace{5pt}\\
%\begin{quote}
%\textit{Enable effective and efficient decision-making  based on the various types of goals in self-adaptive systems.}
\textit{\mbox{\ \ \ \ \ \ }How to reduce large adaptation spaces and rank adaptation options effectively and efficiently \\\mbox{\ \ \ \ \ } \added{at runtime} for self-adaptive systems with threshold, optimization, and set-point goals?}
%\textit{How to reduce large adaptation spaces of self-adaptive systems with multiple threshold goals and/or a set-point goal and/or an optimization goal effectively?}
%\end{quote}
\vspace{5pt}\\
With \textit{effectively}, we mean the solution should ensure that: (1) the reduced adaptation space is significantly smaller, (2) the relevant adaptation options should be covered well, \added{that is, the adaptation options that satisfy (most of) the adaptation goals should be included}, (3) the effect of the state space reduction on the realization of the adaptation goals is negligible. With \textit{efficiently} we mean the solution should ensure that: (4) the learning time is small compared to the time needed for analysis, (5) there is no notable effect on (1), (2), (3) and (4) for larger sizes of the adaptation space.

%Note that we explicitly use decision-making instead of adaptation space reduction. With decision-making we want to indicate that we focus not only on analysis, but reduce the adaptation space even further up to a decision, i.e., a ranking amongst the proposed adaptation options.

To answer the research question, we propose a novel approach for adaptation space reduction
called ``Deep Learning for Adaptation Space Reduction Plus'' -- DLASeR\fplus{} in short.  DLASeR\fplus{} leverages on deep learning that relies on deep artificial neural networks, i.e., neural networks with many layers. DLASeR\fplus{} offers an extendable framework that performs effective and efficient \added{online} adaptation space reduction for threshold, optimization, and set-point goals. %DLASeR\fplus{} exploits a deep learning neural network to cope with the dynamics of the system and its environment. 
While DLASeR\fplus{} can handle an arbitrary mix of goals, the concrete architecture we present in this paper is tailored to a set of threshold and set-point goals and one optimization goal. 
%This architecture reduces the adaptation spaces in two \added{runtime} stages. First, deep learning classification is applied to reduce the adaptation space for the threshold and set-point goals. Second, deep learning regression is applied to rank the options of the reduced adaptation space for the optimization goal. Depending on the types of adaptation goals at hand, one of the stages may be omitted. 
%DLASeR\fplus{}'s employs an offline training online learning to \added{continue training the deep learning network to deal with changing operating conditions.}
\newadded{DLASeR\fplus{}'s learning pipeline consists of an offline and online part. During the offline part, the learning model is selected and configured using training data. During the online part, the running system uses the model to reduce adaptation spaces, and exploits newly obtained data from analysis to continue the training and to update the learning model enabling the system to deal with changing operating conditions.} 
%form of retraining based on the last X data implemented or envisioned
%keep the deep learning network up to date with the changing operating conditions. 

We studied deep learning for four important reasons. First, classic learning techniques usually require some form of human input for feature engineering, whereas deep learning can handle raw data, without the need for feature engineering.  
Second, besides some exceptions, classic machine learning models are usually linear in nature, whereas deep learning can work with non-linear models. 
Third, learned features and even entire models can be reused across similar tasks. This type of transfer learning is a consequence of representation learning, which is the basic core concept that drives deep learning. We exploit representation learning in the  DLASeR\fplus{} neural network architecture. 
Fourth, given the success of deep learning in various other domains, e.g., computer vision~\cite{russakovsky2015imagenet} and natural language processing~\cite{radford2019language}, we were curious to explore how well deep learning could perform for an important runtime problem in self-adaptive systems.

%\subsection{DLASeR\fplus{}}
%We want to tackle this adaptation space problem by performing effective adaptation space reduction. Adaptation space reduction is the reduction of the complete adaptation space to the set of adaptation options that are relevant according to the adaptation goals.

%In this work we present ``Deep Learning for Adaptation Space Reduction Plus'' (DLASeR\fplus{}). DLASeR\fplus{} takes the DLASeR solution to the next level by supporting not only threshold and optimization goals (as DLASeR does), but also set-point goals. Instead of a separate DL model for each adaptation goal (as DLASeR does), DLASeR\fplus{} applies a single deep learning architecture to tackle the various adaptation goals. This architecture is also modifiable at runtime, i.e., we can add or remove adaptation goals at operation time.

In initial work, we explored the use of deep learning to reduce adaptation spaces for threshold and optimization goals~\cite{vanapplying}.\footnote{\added{This initial version was denoted DLASeR; the \fplus{} emphasizes that DLASeR\fplus{} significantly extends DLASeR.}} The goal of that initial work was to investigate the usefulness of deep learning for adaptation space reduction. Compared to that exploratory work, DLASeR\fplus{} supports besides threshold and optimization goals also set-point goals. Whereas the initial approach of~\cite{vanapplying} used a distinct model per goal, DLASeR\fplus{} works with an integrated learning architecture that uses a single model, where layers are shared and reused across different adaptation goals. Furthermore, DLASeR\fplus{} requires a single grid search process and a single prediction step in each adaptation cycle, whereas the initial approach required grid search and prediction for each goal in each cycle. 

We evaluate DLASeR\fplus{} on two instances of DeltaIoT, an  artifact for evaluating self-adaptive systems~\cite{iftikhar2017deltaiot}. The Internet-of-Things (IoT) is \added{a challenging} domain to apply self-adaptation, given its complexity and high degrees of uncertainties~\cite{weyns2018self}. The two instances of DeltaIoT differ in the size of their adaptation space, enabling us to evaluate the different aspects of effectiveness and efficiency. To that end, we define appropriate metrics to evaluate DLASeR\fplus{} and compare it with a baseline that applies exhaustive analysis, and two existing \added{learning-based} approaches for adaptation space reduction: ML4EAS~\cite{quin2019efficient} that uses classic learning techniques, and the initial DLASeR~\cite{vanapplying}. 

The contributions of this paper are: (1) DLASeR\fplus{}, a novel modular approach for adaptation space reduction in self-adaptive systems that is able to handle threshold, optimization, and set-point goals, and (2) a thorough evaluation of the effectiveness and efficiency of the approach in the domain of IoT, including a comparison with \added{a baseline} and two state of the art approaches. 

\added{Given the specific domain we use in the evaluation with relatively limited sizes of adaptation spaces, we want to emphasize that additional validation is required to generalize the findings.} 

%\subsection{Contribution and outline}
%We develop a framework to perform effective adaptation space reduction. In particular, we develop a single deep neural network architecture that is capable to tackle the three different kinds of adaptation goals.

The remainder of this paper is structured as follows. 
In Section~\ref{sec:background} we provide relevant background: we introduce DeltaIoT, \added{we present a high-level architecture of self-adaptation with adaptation space reduction,} we zoom in on the different types of adaptation goals, the adaptation space, and we introduce the essential concepts of deep learning.
Section~\ref{sec:methodology} gives a high-level  overview of the research methodology.
In Section~\ref{sec:metrics}, we introduce a set of metrics that we use to measure the effectiveness and efficiency of DLASeR\fplus{} and compare the approach with alternative approaches.
Sections~\ref{sec:dlaser+} and \ref{sec:learningpipeline} present the core technical contribution of this paper: the architecture and learning pipeline of DLASeR\fplus{} architecture respectively. 
In Section~\ref{sec:evaluation}, we use the metrics to evaluate DLASeR\fplus{} for two instances of DeltaIoT. Section~\ref{sec:relatedwork} positions DLASeR\fplus{} in the landscape of other related work. Finally, we draw conclusions and look at future work in Section~\ref{sec:conclusions}.

\section{Background}\label{sec:background}
This section introduces the necessary background for this paper. We start with introducing DeltaIoT. Then, \added{we explain 
the basic architecture of a self-adaptive system that integrates a verifier for runtime analysis and a learning module for online adaptation space reduction.} Next, we introduce the different types of adaptation goals that are supported by DLASeR\fplus{}. Then, we elaborate on the concept of adaptation space. Finally, we introduce the relevant concepts of deep learning.

\subsection{DeltaIoT}

DeltaIoT is a reference Internet-of-Things (IoT) application that has been deployed at the Campus Computer Science of KU Leuven~\cite{iftikhar2017deltaiot}. DeltaIoT has been developed to support research on self-adaptation, i.e, evaluate new self-adaptive approaches, e.g. to evaluate tradeoffs between non-functional requirements in self-adaptive systems~\cite{3194133.3194142} or perform cost-benefit-analysis in runtime decision-making for self-adaptation~\cite{vanderdonckt2018cost}. Next to the real physical setup deployed by VersaSense,\footnote{VersaSense website: \url{www.versasense.com}} DeltaIoT also offers a simulator for offline experimentation. We use DeltaIoT as evaluation case, but also as running example to illustrate the different parts that follow in this section. 

\begin{figure}[htbp]
    \centering
    \includegraphics[width=0.85\textwidth]{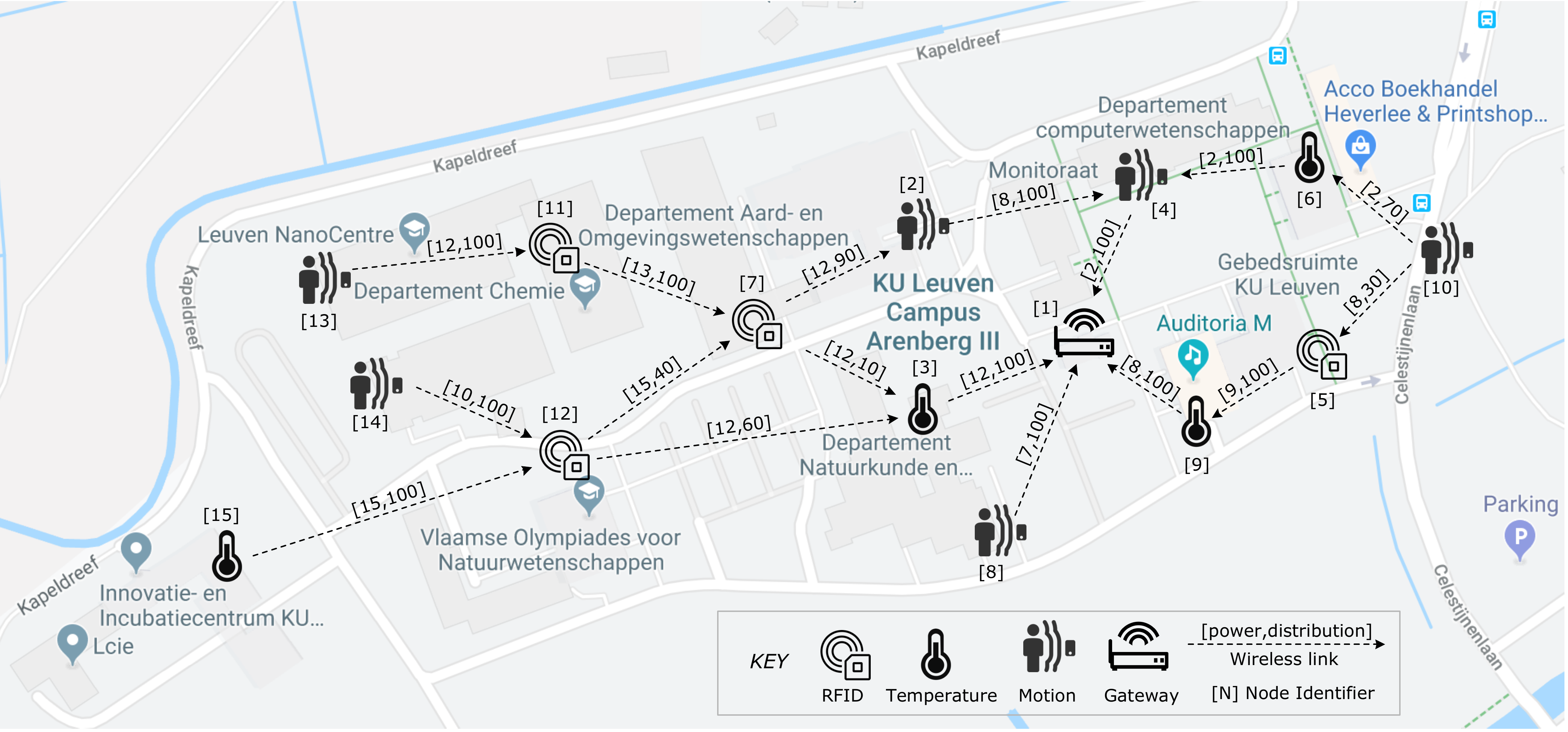}
    \caption{DeltaIoT deployment at the KU Leuven campus (borrowed from~\cite{vanderdonckt2018cost}). The gateway is marked by the blue icon in the center. The data collected by the sensors is sent over multiple wireless links to this gateway.}
    \label{fig:deltaiot}
\end{figure}

\autoref{fig:deltaiot} shows the physical setup of DeltaIoT. RFID sensors are used to provide access control to labs, passive infrared sensors monitor the occupancy of several buildings, and heat sensors are employed to sense the temperature. The data of these sensors is relayed to the gateway by means of wireless multi-hop communication. Each sensor is plugged into a battery-powered mote, i.e., a networked tiny embedded computer. The motes take care of the routing of sensor data to the gateway. The communication in DeltaIoT is time-synchronized and organized in cycles with a fixed number of slots. Neighboring motes are assigned such slots during which they can exchange packets. Motes collect data (locally generated or received from other motes) in a buffer. When a mote gets a turn to communicate with another mote, it forwards the packets to the other mote. Packets that cannot be sent remain in the buffer until the mote is assigned a next slot. 

DeltaIoT has three main quality requirements: packet loss, latency, and energy consumption. For these qualities we define corresponding adaptation goals, e.g, average latency of packages delivery should not exceed a predefined percentage and energy consumption of the network should be minimized. Ensuring such goals is challenging since the IoT network is subject to various types of uncertainties. Two main uncertainties are interference along network links caused by external phenomena, and changing load in the network, that is, motes only send packets when there is useful data, which may be difficult to predict.  
The IoT network can be adapted using two parameters: power setting and link distribution. The first parameter refers to the setting of the transmission power of each mote. The options are discretized as an integer in the range of 1 to 15. Increasing the transmission power will reduce packet loss, but increase energy consumption. The second parameter refers to the way data packets are relayed from the motes to the gateways. Observe in~\autoref{fig:deltaiot} that several motes have multiple links over which data can be sent. We refer to the distribution of packets sent by a mote to its parents as the link distribution. If a mote has only one parent, it is obvious that it  relays 100\% of its packets through that single parent. But when a mote has multiple parents, the distribution of packets over the different links to the parents can be selected and changed. Note that the sum of the distributions should remain 100\% (to optimize energy consumption). By changing the link distribution, paths with more interference can be avoided. However, this may cause delays at the buffers of the motes along these paths. 

%DeltaIoT comprises a real physical setup and a realistic simulator for offline experimentation. 
For practical reasons, we use the DeltaIoT simulator for the evaluation of DLASeR\fplus{}, since extensive experimentation on a physical IoT deployment is particularly time consuming. The DeltaIoT simulator offers a realistic alternative for the physical network where parameters of the simulator are based on field experiments. 
%also enables to easily alter the lay-out of the IoT network. 
We consider two instances of DeltaIoT. The first one, referred to as \textit{DeltaIoTv1}, consists of 15 motes (shown in Figure~\ref{fig:deltaiot}); the second instance, referred to as \textit{DeltaIoTv2}, consists of 37 motes. \added{The larger IoT network is more challenging in terms of the number of configurations that are available to adapt the system representing the adaptation space. In particular, the adaptation space of DeltaIoTv1 contains 216 possible adaptation options, while DeltaIoTv2 has 4096  adaptation options. These numbers are determined by the parameters of the IoT network that can be used for adapting the system: power setting and link distribution.\footnote{\added{Technically, we apply the following approach to determine the adaptation options for the IoT settings: first we determine the required power settings for each mote along the links to its children such that the signal-to-noise ratio is at least zero. These settings are determined based on the actual values of signal-to-noise along the links. The settings are then fixed for all adaptation options. The number of adaptation options is then determined by the combinations of all possible settings of link distributions in the network. This number is 216 for DeltaIoTv1 and 4096 for DeltaIoTv2.}} Hence, for both versions, the number of adaptation options is constant. However, as will explain in Section~\ref{subsec:adaptation_space}, the properties of the adaptation options change dynamically with changing conditions.} 

\added{
\subsection{Basic architecture self-adaptive system with adaptation space reduction}\label{subsec:basic-arch}

Figure~\ref{fig:basic_architecture} shows the basic architecture of a self-adaptive system that uses learning for adaptation space reduction. As explained in the introduction, in this research we apply architecture-based adaptation with a MAPE-K feedback loop.  
The figure highlights the main elements of architecture and high-level flow of interactions between the elements to realize adaptation space reduction. The \textit{managed system} takes input from the \textit{environment} and produces output to realize the \textit{user goals}. The \textit{managing system} manages the managed system to achieve a set of \textit{adaptation goals}. Central to the managed system are the \textit{MAPE} elements that share \textit{knowledge} and realize a feedback loop. The feedback loop \textit{senses} the managed system and \textit{adapts} it to achieve the adaptation goals. 

\begin{figure}[h!]
    \centering
    \includegraphics[width=0.65\textwidth]{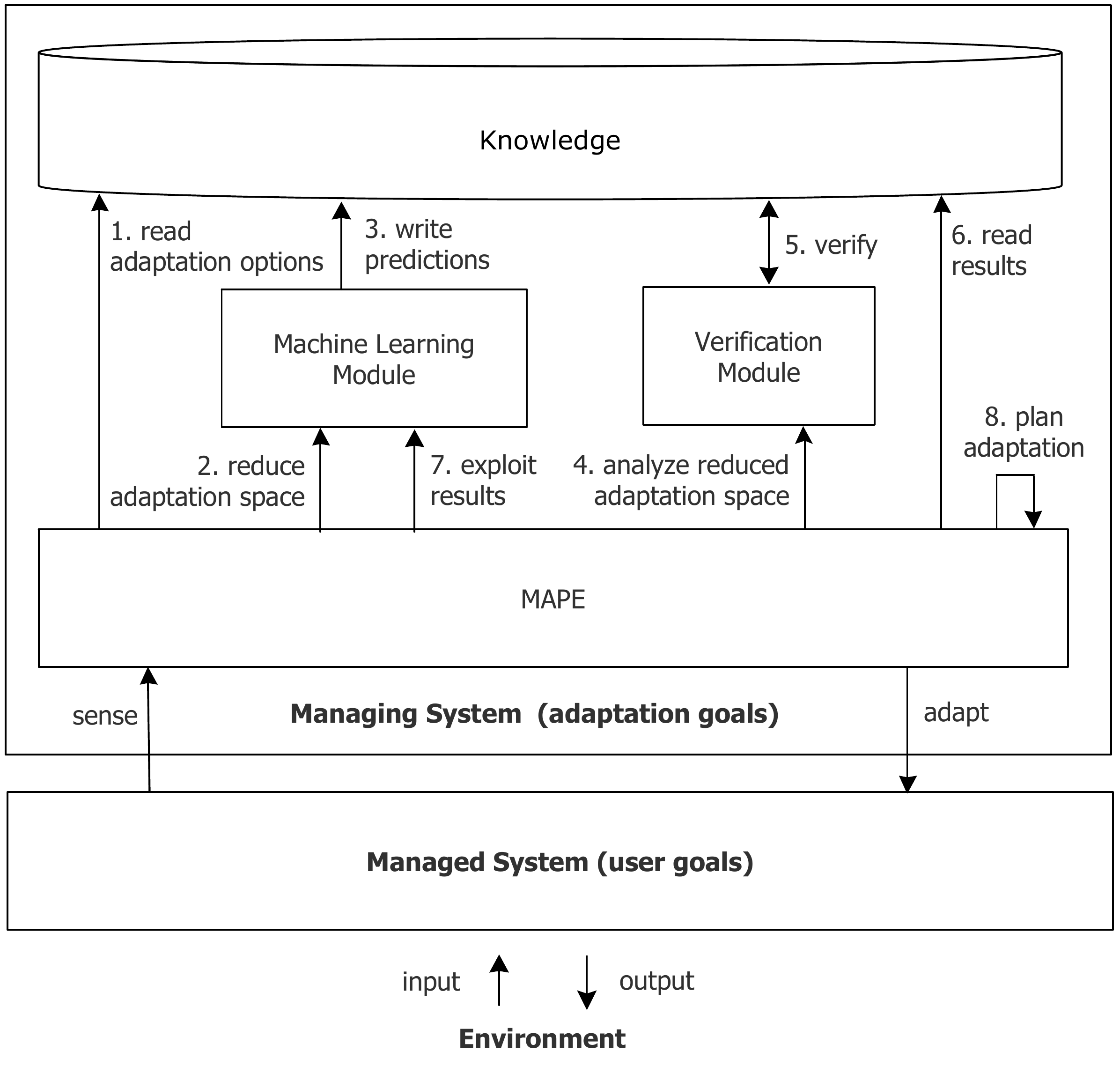}
    \caption{A self-adaptive system that uses learning for adaptation space reduction}
    \label{fig:basic_architecture}
\end{figure}

When the MAPE feedback loop detects that the adaptation goals are violated or may no longer be achievable, it reads the adaptation options from the knowledge (1). The feedback loop then instructs the \textit{machine learning module} to reduce the adaptation space (2). The machine learning module will use its learning model to make predictions about the expected quality properties for the different adaptation options (3). Based on these predictions, the adaptation space is reduced to the most relevant adaptation options. Next, the MAPE feedback loop instructs the \textit{verifier module} to analyze the adaptation options of the reduced adaptation space (4). When the verifier completes the verification (5), the MAPE feedback loop reads the results (6). It then forwards the results to the machine learning module that exploits the results to continue the training of its learning model (7). Finally, the MAPE feedback loop generates a plan (8) that is the used to adapt the managed system. 
\vspace{5pt}\\
\noindent \textbf{Analysis adaptation option.} We illustrate how an adaptation option is analyzed for DeltaIoT. In particular, we explain how packet loss is estimated by the analyzer of the MAPE loop using a statistical model checker  (we apply this approach in the evaluation in Section~\ref{sec:evaluation}). Figure~\ref{fig:runtime_quality_model_packet_loss} shows the quality model for packet loss that consists of two interacting automata: Topology and Network.

\begin{figure}[h!tb]
	\centering
	\includegraphics[width=0.75\textwidth]{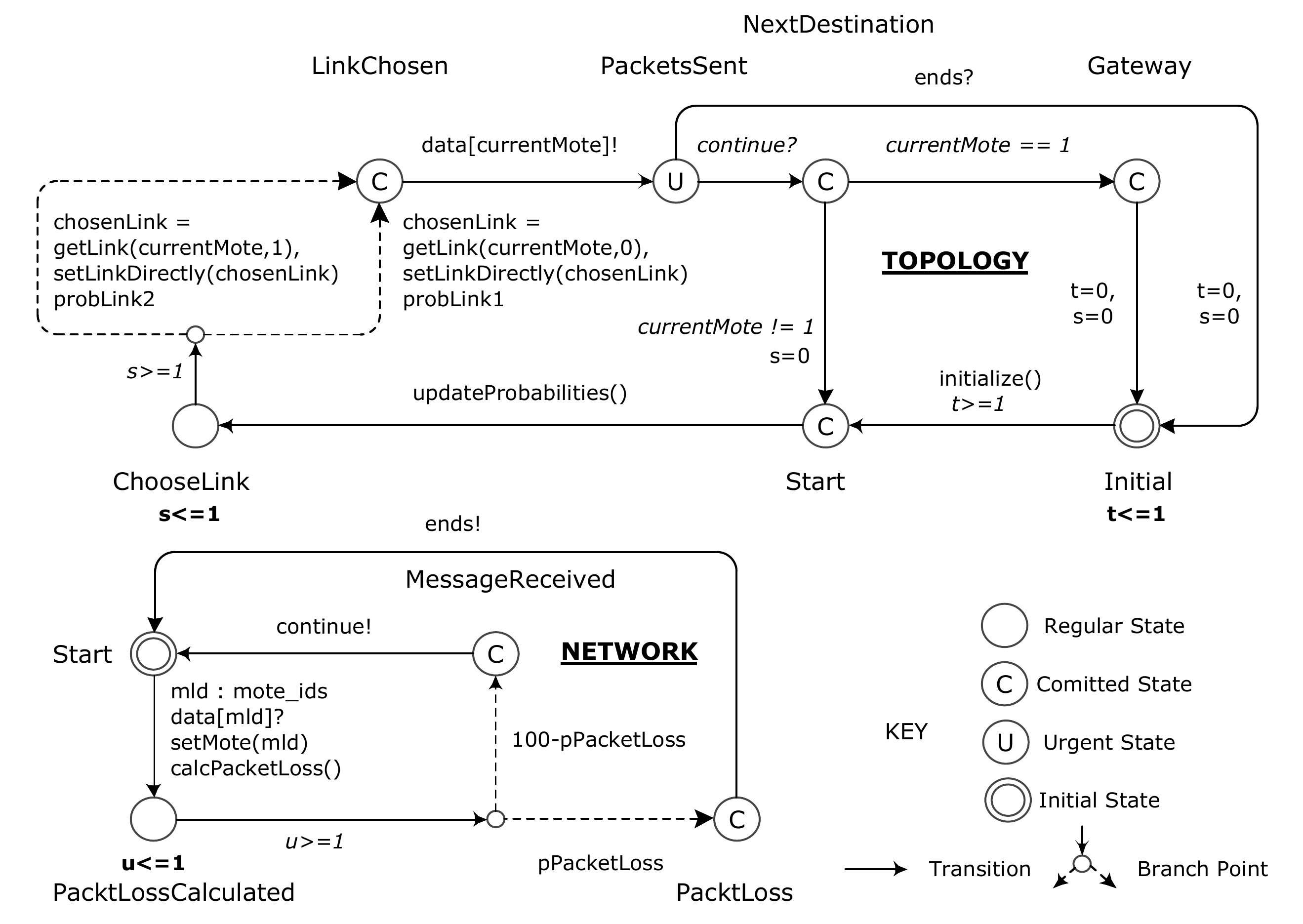}
	\caption{Runtime quality model for packet loss}
	\label{fig:runtime_quality_model_packet_loss}
\end{figure}

The automata have two sets of parameters: (i) parameters to configure the model for an adaptation option, and \added{(ii)} parameters of the uncertainties that need to be set based on the current conditions. 

To configure an adaptation option, the power settings of the motes per link need to be set (with values between 0 to 15) and the distributions factors for links of motes with two parents need to be set (each with a value of 0, 20, 40, 60, 80 or 100\%). The power of the links is set based on the current levels of interference along the links that are available in the knowledge repository. The power settings are applied to all adaptation options. The values for the distribution factors are different for each adaptation option; these values are assigned to the variables $probLink1$ and $probLink2$ of the topology model. 
Furthermore, the values of the uncertainties, network interference (SNR) and traffic load, need to be set. These values are available in the knowledge repository and are based on the recent observations. The uncertainties apply to all adaptation options. 

After initializing the model for a particular adaptation option, the \textit{Topology} automaton simulates the communication of data along a path selected for verification, i.e., a sequence of links from one mote via other motes to the gateway (see also Figure~\ref{fig:deltaiot}). 
The current link to send data is selected probabilistically based on the distribution factors ($probLink1$ and $probLink2$). The model then signals the \textit{Network} automaton. Next, the probability for packet loss is calculated (based on the SNR). Depending on the result either the packet is lost or the message is received. In the latter case, the network automaton returns to the start location, continuing with the next hop of the communication along the path that is currently checked, until the gateway is reached. If the packet is lost, the verification of the communication along that path ends. The quality model allows determining the packet loss of the adaptation options using the following query: 
\vspace{5pt}
\begin{center}
$Pr [$<=$1](<>Network.PacketLoss)$
\end{center}
\vspace{5pt}
This query determines the probability that the state \textit{Network.PacketLoss} is reached for the different paths of an adaptation option (see the Network automaton in  Figure~\ref{fig:runtime_quality_model_packet_loss}). To that end, the verifier performs a series of simulations and returns the expected packet loss with a required accuracy and confidence. These estimates together with the estimates of other quality properties for the different adaptation options are then used to select an adaptation option using the adaptation goals. 

This example illustrates that rigorous runtime analysis can be a resource- and time-consuming activity, implying the need for adaptation space reduction if the number of adaptation options, i.e., the size of the adaptation space, is too big to be completely verifiable within the available time period to make an adaptation decision. We elaborate on the adaptation space below in Section~\ref{subsec:adaptation_space}.
}

\subsection{Adaptation goals}
An adaptation goal refers to a quality requirement of the managed system that the managing system should realize. A violation (or an expected violation) of one or more of the adaptation goals triggers an adaptation, aiming to satisfy all the adaptation goals again.

%Shevstov S. et al. state, in the context of control-theoretic adaptation, that one can depict three different types of adaptation goals~\cite{shevtsov2019simca}; 
In this research, we consider three types of adaptation goals: (1) threshold goals, (2) set-point goals, and (3) optimization goals. 
Intuitively, a threshold goal states that the value of some quality property of the system should be below (or above) a certain threshold value. A set-point goal states that some value of a quality property of the system should be kept at a certain value, i.e., the set-point, with a margin of at most $\epsilon$. Finally, an optimization goal states that some value of a quality property of the system should be minimized (or maximized). Formally, we define the satisfaction of goals by the adaptation options as follows.\footnote{For the explanation, we consider only threshold goals below a value and optimization goals that minimize a value; the other variants are defined similarly.} Consider $C$ %= \{c_0, \ldots, c_n\}$ 
the set of possible configurations, each configuration representing an adaptation option of the adaptation space. We refer to a particular quality property $q_x$ of an adaptation option $c_i \in C$ as $c_i[q_x]$, with $x \in \{t,s,o\}$ referring to quality properties related to threshold, set-point, and optimization goals respectively. Further, consider a threshold goal $g_t$, a set-point goal $g_s$, and an optimization goal $g_o$. The set of adaptation options $T$ that satisfy $g_t$, the set $S$ that satisfy $g_s$, and the set $O$ that satisfy $g_o$ are then defined as follows:\footnote{\added{While a set-point goal may conceptually be modeled as two threshold goals, there are good arguments to differentiate them as a distinct type of goal. In particular, using a set-point to express the corresponding goal is more straightforward and natural for stakeholders compared to using two thresholds. Further, it makes maintenance easier, e.g., when updating the set-point value of the goal. Lastly, from a learning perspective, if we use two threshold goals instead of one set-point goal, we require two times the processing resources to train and infer.}}  
%
%\begin{quote}
 %   \textbf{Threshold goal}: $\forall s_i \in S : s_i[q] < Threshold$ \\
%\textbf{Setpoint goal}: $\forall s_i \in S : Setpoint - \epsilon < s_i[q] < Setpoint + \epsilon$ \\
%\textbf{Optimization goal}: $s_i \in S : s_i[q] < s_j[q]$, $\forall s_j \in S/s_i$
%\end{quote}
%
%\begin{quote}
%\newline%\vspace{5pt}
%\noindent\vspace{5pt}
%\hspace{30pt}$T = \{ c_i \in C \mid c_i[q_t] < g_t\}$
%
\begin{align}
    T & = \{ c_i \in C \mid c_i[q_t] < g_t\} \\
    S & = \{ c_i \in C \mid g_s - \epsilon\ \leq c_i[q_s] \leq g_s + \epsilon\} \\
    O & = \{ c_i \in C \mid c_i[q_o] <  c_j[q_o],~~\forall c_j \in C \setminus \{c_i\}\}
\end{align}

%\end{quote}

While DLASeR\fplus{} can handle an arbitrary mix of adaptation goals, in this paper, we focus on systems that can have multiple threshold and set-point goals, but only one optimization goal. In particular, the DLASeR\fplus{} architecture we present in this paper maps each adaptation goal to a rule. Decisions are made by first applying the rules of the threshold and set-point goals, and then the rule of the optimization goal. Handling multiple optimization goals with different decision-making techniques, e.g., based on Pareto optimality~\cite{hwang2012multiple}, are outside the scope of this paper.

\begin{figure}[htbp]
  \centering
  \begin{subfigure}[b]{0.305\textwidth}
    \includegraphics[width=\textwidth]{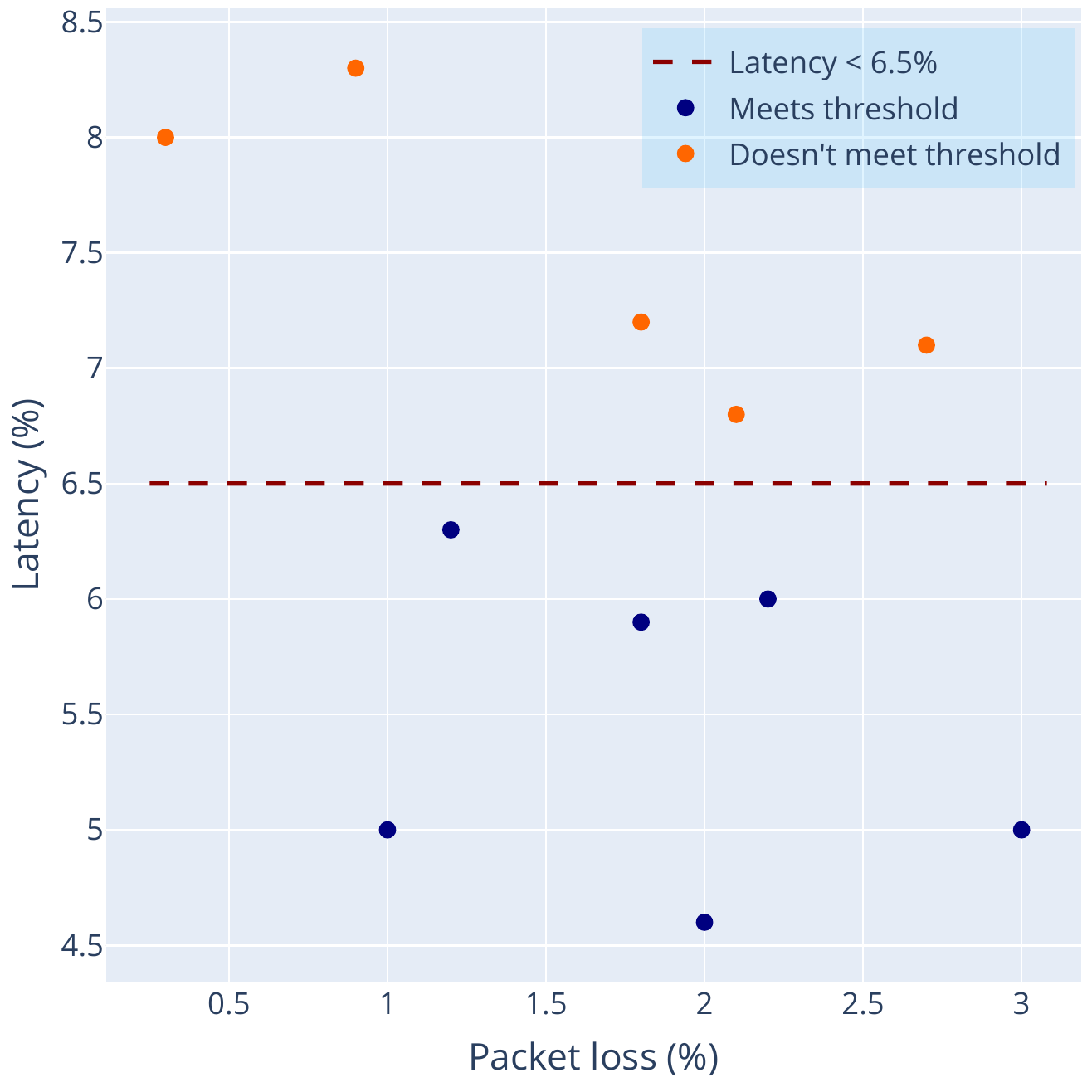}
    %\caption{Threshold goal: Quality 1 < 6.}
    \caption{Threshold goal.}
    \label{fig:adapt-goals-threshold}
  \end{subfigure}
  \hfill
  \begin{subfigure}[b]{0.305\textwidth}
    \includegraphics[width=\textwidth]{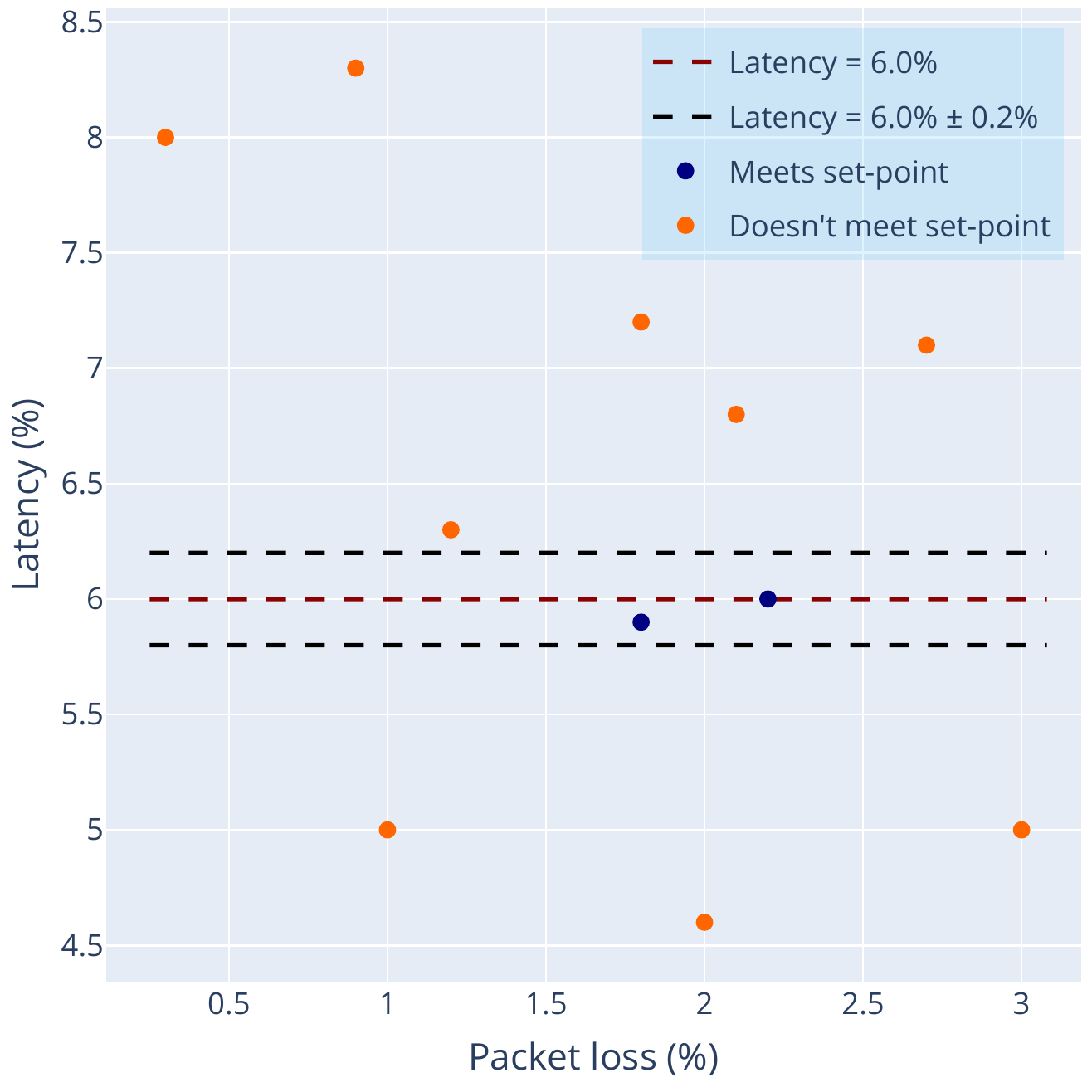}
    %\caption{Set-point goal: Quality 1 = 6.}
    \caption{Set-point goal.}
    \label{fig:adapt-goals-setpoint}
  \end{subfigure}
  \begin{subfigure}[b]{0.365\textwidth}
    \includegraphics[width=\textwidth,height=4.25cm]{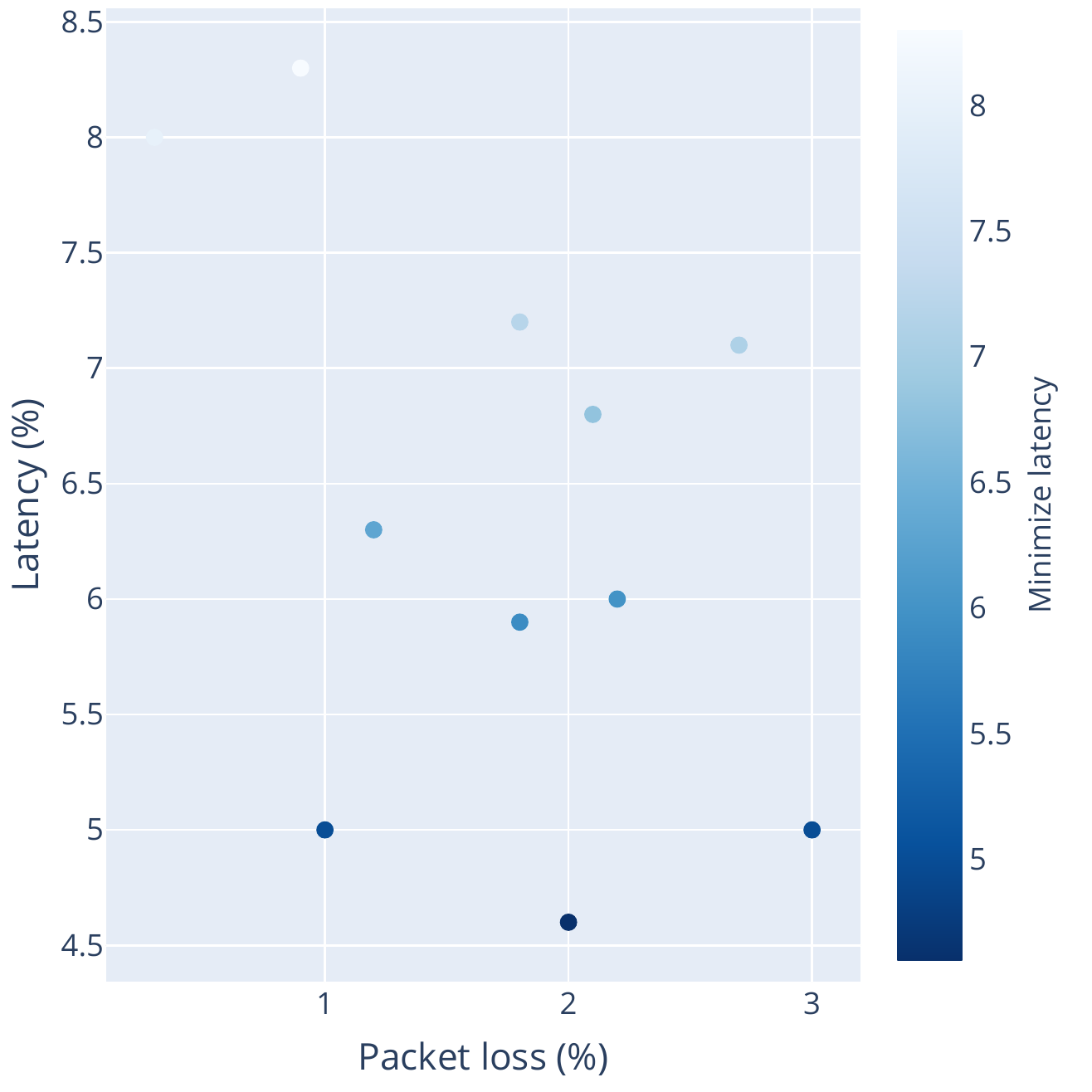}
    %\caption{Optimization goal: minimize Quality 2.}
    \caption{Optimization goal.}
    \label{fig:adapt-goals-optimization}
  \end{subfigure}
  \caption{Illustration of the three types of adaptation goals.}\vspace{-10pt}
  %\textbf{Top left:} a threshold goal for Quality 1. \textbf{Top right:} a set-point goal for Quality 1. \textbf{Bottom:} an optimization goal for Quality 2.
  \label{fig:3-types-adapt-goals}
\end{figure}

As an illustration, \autoref{fig:3-types-adapt-goals} shows a latency goal for DeltaIoT specified in three formats corresponding with the three types of goals (the diagrams are simplified for didactic reasons).\footnote{In DeltaIoT, latency is defined as a relative part of the cycle time of the time-synchronized communication in the network. E.g., a latency of 5\% means that the average time packets stay in the network is 5\% more as the cycle time.} Each dot in a diagram represents an adaptation option based on the values of latency and packet loss (we are here mainly interested in the values of latency). 
\autoref{fig:adapt-goals-threshold} shows the latency as a threshold goal \verb|Latency|\,<\,\verb|6.5%|. All the dots below the red dotted line represent adaptation options that satisfy the threshold goal. The adaptation options above the red dotted line do not meet this goal. 
\autoref{fig:adapt-goals-setpoint} shows the latency as a set-point goal \verb|Latency|\,=\,\verb|6%|$~\pm~$\verb|0.2%|. The two blue dots are the only adaptation options that meet this set-point goal, since both are in the range defined by the set-point value and the margin $\epsilon$. The other adaptation options have values out of the range $[5.8\%,6.2\%]$ and do not satisfy the goal. 
Finally,~\autoref{fig:adapt-goals-optimization} shows the latency as an optimization goal. The darker the dot, the lower the latency, hence, the dot at the bottom has the lowest value for latency and this adaptation option optimizes the latency.

%As a concrete example for adaptation goals on our running example (DeltaIoT) we present the goals used by Quin F. et al.~\cite{quin2019efficient}:
%\vspace{5pt}\\
%\textbf{T1}: The average packet loss over 12 hours should not exceed 10\,\%.\\
%\textbf{T2}: The average latency over 12 hours should not exceed 5\,\%.\\
%\textbf{O1}: The energy consumption should be minimized.
%\vspace{5pt}

For the evaluation with DeltaIoT in Section~\ref{sec:evaluation}, we consider three combinations of adaptation goals: (1) two threshold goals: packet loss and latency, with energy consumption as optimization goal, (2) the same threshold goals, with energy consumption as set-point goal, and (3) latency as threshold goal, energy consumption as set-point goal, and packet loss as optimization goal. 

\subsection{Adaptation space}\label{subsec:adaptation_space}
An adaptation space comprises the set of adaptation options at some point in time. The adaptation space is determined by the effectors  (also called actuators or ``knobs'') that are available to adapt the managed system. The  actuators allow to set system variables. These variables are usually characterized by a discrete domain (e.g., start/stop a number of servers, launch a number of new virtual machines, select a power setting, send a number of messages over a link, etc.). For variables with a continuous domain, we assume in this work that the domain can be discretized. 

If adaptation is required, it is the task of the analyzer and the planner of the MAPE-K feedback loop to find the best adaptation option. This task involves estimating the expected quality properties related to the adaptation goals for all or at least a relevant subset of the adaptation options. Different techniques can be used to make these predictions. \added{One approach is to represent the quality of the system as a network of stochastic timed automata~\cite{legay2010statistical, younes2006statistical,weyns2016model} as we illustrated in Sub-section~\ref{subsec:basic-arch}. In this approach, the model is configured for the effector settings of a particular adaptation option and parameters that represent uncertainties are assigned up-to-date values. By utilizing statistical model checking one can determine the expected quality of an adaptation option. Another approach is to represent the quality of the system as a parameterized Markov model~\cite{Calinescu2011,Moreno:2015,camara2020quantitative} and apply probabilistic model checking~\cite{legay2010statistical, younes2006statistical} to estimate the quality property of interest.} Statistical model checking is more efficient than probabilistic model checking, but offers results bounded to a given accuracy and confidence, depending on the number of simulation runs. 

\begin{figure}[htbp]
    \centering
    \includegraphics[width=0.85\textwidth]{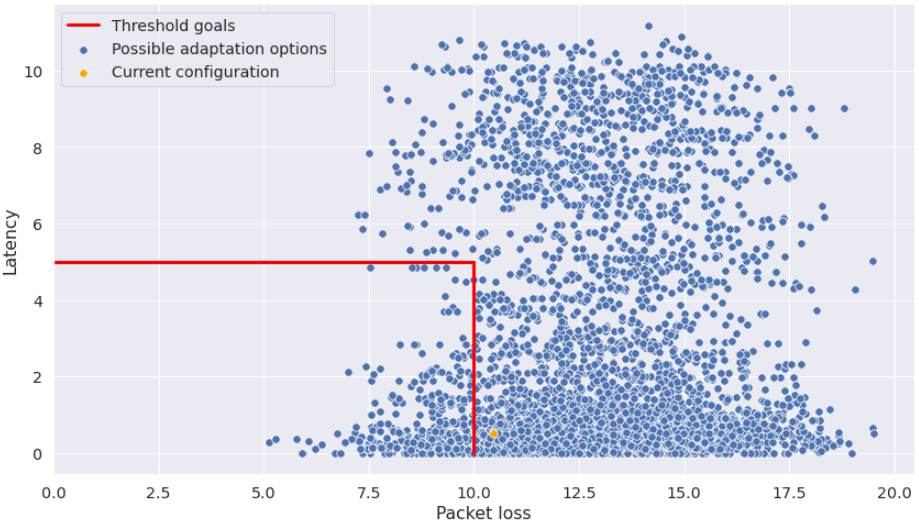}
    \caption{Adaptation space of DeltaIoTv2 at some point in time.}
    \label{fig:adaptation-space-example}
\end{figure}

In the case of DeltaIoT, we are interested in the failure rate of the network, its latency and the energy consumed to communicate the data to the gateway. Figure~\ref{fig:adaptation-space-example} shows a representation of the adaptation space for DeltaIoTv2 at some point in time. 

The red lines denote two threshold goals for this particular scenario (latency and packet loss). Each blue dot in this graph represents one adaptation option. The dot in \newadded{orange} represents the current configuration. For this particular instance, the number of adaptation options is over 4000. Analyzing all these options within the available time slot may not be feasible. Hence, the analysis should focus on the relevant options for adaptation, i.e., those that are compliant with the goals, represented by the dots in the box bottom left as determined by the two threshold goals. This paper is concerned with analyzing such a large adaptation space in an effective and efficient manner. 

\begin{figure}[htbp]
    \centering
    \includegraphics[width=0.95\textwidth]{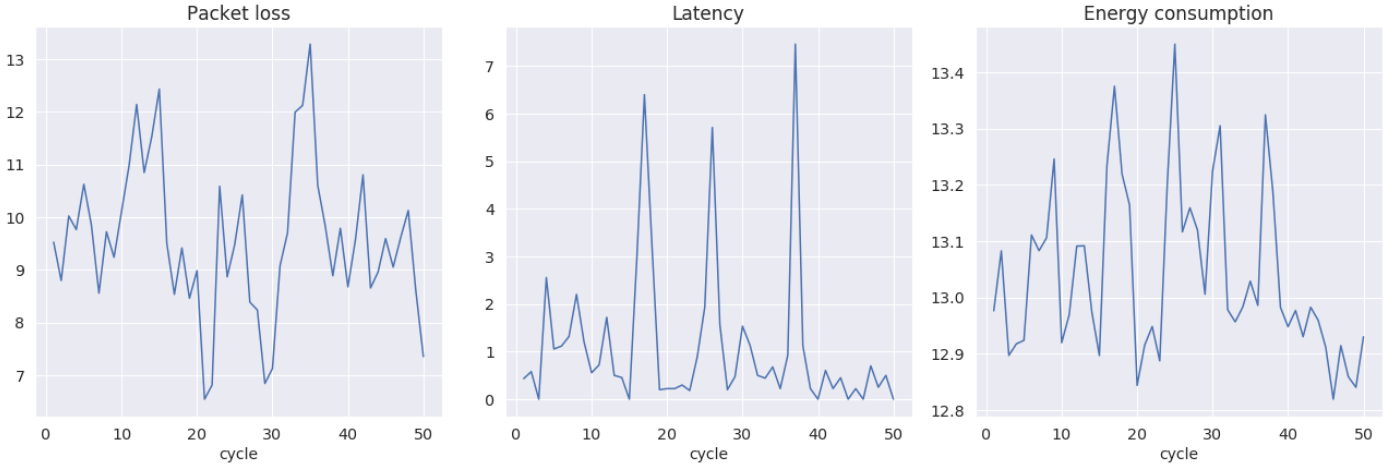}
    \caption{Dynamic behavior of an adaptation option over 50 cycles reflected in changes of its quality properties.}
    \label{fig:dynamic-behavior}
\end{figure}

It is important to notice that the adaptation space is dynamic, i.e., the qualities of each adaptation option may change over time. In Figure~\ref{fig:adaptation-space-example}, this means that the position of the adaptation options on the diagram move over time. The main cause for this dynamic behavior are the uncertainties the system is subjected to. It is due to this dynamic behavior that the analysis and verification needs to be performed at runtime, when the actual values of the uncertainties are available. \autoref{fig:dynamic-behavior} illustrates the dynamic behavior for an instance of DeltaIoT. The figure illustrates how three qualities of one of the adaptation options change over a series of cycles. These graphs are based on field tests. 

\subsection{Deep learning in a nutshell}
Deep learning (DL) refers to a subset of machine learning mechanisms in which the learning methods are based on \textit{deep artificial neural networks} (ANNs) with \textit{representation learning}~\cite{goodfellow2016deep}. 
%The two key terms in this definition are artificial neural networks and representation learning. We will elaborate on both of these concepts. 
%First we will outline the principal differences between machine learning and deep learning. Then, we introduce the fundamental concepts of deep learning.

%\ttodo{Verschil deep learning en classical machine learning outlinen?}

\subsubsection{Artificial neural networks}
%As stated in the definition at the start of this section, deep learning is driven by two main concepts; (1) artificial neural networks, and (2) representation learning. We will elaborate on both in the following paragraphs. 
%
ANNs are the core of deep learning. These networks are commonly described in terms of their input and output, which are the only externally observable parts of an ANN. Both the input and the output are represented as a vector of numbers. A neural network applies a chain of mathematical operations aiming to transform the input to the desired output. 

The basic building block of a neural network is a neuron, also called a perceptron. In the case of a fully connected neural network, a neuron is connected to all the inputs and produces a single output.
To obtain its output, the neuron applies two  operations:
\begin{equation}\label{eq:weighted-sum}
    z = \vec{w} \cdot \vec{x} + b
\end{equation}
\begin{equation}\label{eq:activation-func}
    y = f(z)
\end{equation}
with $\vec{x}$ the current input vector, $\vec{w}$ the weights associated with the inputs, $b$ a constant, $z$ an intermediate computed value, $y$ the output, and $f(\cdot)$ an activation function. First, the weighted sum $z$ of all the inputs is calculated (Equation~\ref{eq:weighted-sum}). The weights associated with the inputs are learnable. Intuitively, these weights represent the relative importance of the inputs. When the weighted sum is computed, the neuron applies an activation function $f$  (Equation~\ref{eq:activation-func}). This function allows introducing non-linearity in the neural network. Since, a weighted sum is a linear function, and a linear combination of linear functions is still a linear combination, a neural network without a non-linear activation function will only be able to learn linear combinations of input values. Non-linearity in learning is very important to learn more complex concepts. Examples of common activation functions are hyperbolic tangent (tanh), sigmoid, and rectified linear unit (ReLU).

A deep neural network is structured in two dimensions; width and depth. The depth of a network corresponds to the number of layers, whereas each layer is defined by its width, i.e., the number of neurons present in that layer.

\subsubsection{Representation learning} A key concept of ANNs is representation learning~\cite{goodfellow2016deep}. Representation learning learns representations of input data, typically through transformations, which makes it easier to perform certain tasks~\cite{bengio2013representation}. An important advantage of representation learning is that it enables transfer learning, i.e., reuse of the learned representations, complemented with fine-tuning for  specific learning tasks~\cite{pan2009survey}. In deep learning, representations are learned through multiple non-linear transformations on the input data. These learned representations provide abstract and useful representations for the learning task. A simple linear layer can then be stacked on top of the complex block with many non-linear layers to tailor the learning network to the specific learning task at hand, e.g., classification, regression, or prediction~\cite{devlin2018bert}. 

%In deep learning, representations are learned through multiple non-linear transformations on the input data. These learned representations provide abstract and useful representations for the learning task at hand, e.g., classification, regression, or prediction. 

%Representation learning is important because of various reasons. Fist of all, the performance of any ML model is critically dependent on the representations it learns to output. Next to this, representation learning explains why in many solutions a simple linear layer is stacked on top of a complex block with many non-linear layers, e.g., state-of-the-art BERT for NLP-tasks~\cite{devlin2018bert}. A simple linear layer is thus sufficient to perform the (hard) task described by the output on the highly rich learned representations of the input.  Finally, the most important property of representation learning, which is a direct consequence of the previous characteristic, is that it enables transfer learning, i.e., reuse the learned representation and fine-tune for a specific task~\cite{pan2009survey}.

\subsubsection{Training a deep neural network}
We distinguish three steps in the training of a ANN: (1) forward propagation, (2) loss calculation, and (3) back propagation.

In the \textit{forward propagation} step, the input data is passed through the layers of the network, from input to output. The neurons apply their transformations to the data they receive from all the connected neurons of the previous layer and pass the result to the connected neurons of the next layer. In the final layer, the neurons apply their transformations to obtain the (predicted) output.

In the second step, a loss function estimates the loss (or error). The loss captures how good or bad the predicted result is (i.e., the output predicted by the network) compared to the correct result (i.e., the expected output). To obtain a loss as close as possible to zero, the weights of the connections between the neurons (that determine the weighted sums) are gradually adjusted in the next step.

% As final step, the loss value is propagated backwards through the network, hence \textit{back propagation}. 
% You can think of this step as recursively applying the chain rule to compute the gradients all the way to the input of the network. These gradients tell you in which direction the loss is increasing and indicate the influence of the neurons on the loss. Hence, in the final step we will update our weights (also called parameters) using some optimizer that exploits these gradients. Examples of commonly used optimizers are Adam, RMSprop, and Nadam~\cite{ruder2016overview}.
%Each neuron receives only a fraction of the total loss, based on the relative contribution of that neuron on the predicted output. Then, for each neuron, the gradient for the retrieved loss will be calculated. You can think of this process as recursively applying the chain rule to compute the gradients all the way to the input of the network. These gradients tell you in which direction the loss is increasing. Based on this information, the neurons will update their weights (also called parameters) using some optimizer that exploits these gradients. Examples of commonly used optimizers are Adam, RMSprop, and Nadam~\cite{ruder2016overview}.

As final step, the loss value is propagated backwards through the network, hence \textit{back propagation}. 
You can think of this step as recursively applying the chain rule to compute the gradients all the way from the output to the input of the network. These gradients tell in which direction the loss is increasing and indicate the influence of the computations of the neurons on the loss. An optimizer exploits these gradients to update the weights of the neuron connections aiming to minimize the loss. Examples of commonly used optimizers are Adam, RMSprop, and Nadam~\cite{ruder2016overview}.

\subsubsection{Classification and regression}
Classification and regression are two important learning tasks that require different predictions. %tasks of machine learning. 
%The difference between both lies in what should be predicted. 
Suppose that we have some machine learning model that is described by the following equation:
\begin{equation}
    \vec{y} = M(\vec{x})
\end{equation}
with $\vec{x}$ the input and $\vec{y}$ the output of the learner, and $M$ a function that maps input to output.

In the case of classification, $M$ needs to map the input to a set of classes that are represented by different labels with different encodings (e.g., rainy = 0 and sunny = 1). A classification with only two labels is called binary classification; 
while multi-class classification has more than two labels. 
%Next to this, a distinction is often made in terms of the number of possible labels for an input instance. If only one label is possible for each input, we use the term single-label classification. In the other case, when multiple labels are possible for each instance, the task is called multi-label classification.  \todo{Is dit wel allemaal interessant. Ik kan dit mogelijk weglaten}
%
%In the case of classification, model $M$ is only allowed to map the input to multiple categorical classes, i.e., discrete values. The output data is categorized under different labels, e.g., rainy or sunny. In this example, there are only two labels (rainy and sunny), such a task is called binary classification (rainy = 0 and sunny = 1). When there are more than two labels, we call the task multi-class classification. 
%
Regression on the other hand, maps the input data to continuous real values instead of classes or discrete values. Whereas for classification the predicted output is not ordered, for regression the output is ordered. 
In sum, the main difference between both tasks is that the output variable $\vec{y}$ for classification is categorical (or discrete), while for regression it is numerical (or continuous).

\section{Overview of the Research methodology}\label{sec:methodology}

\begin{figure}[htbp]
    \centering
    \includegraphics[width=\textwidth]{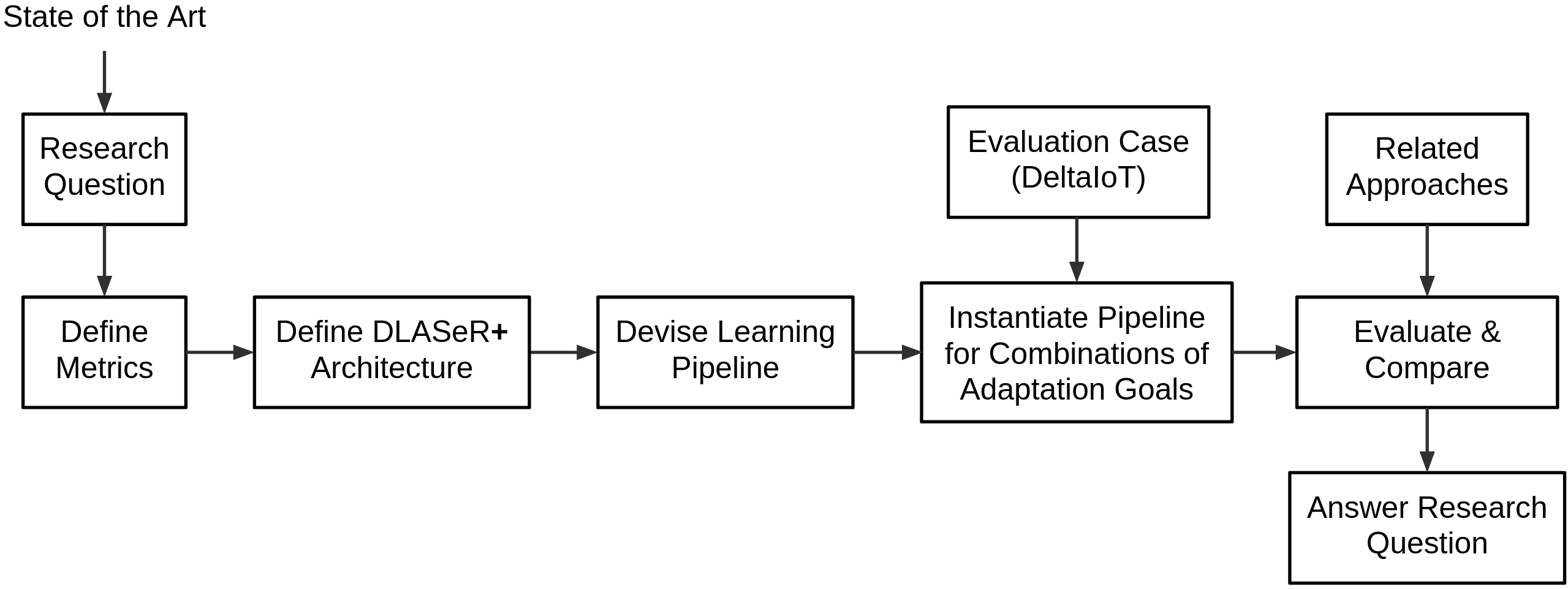}
    \caption{Overview of research methodology.}
    \label{fig:methdodology}
\end{figure}

To tackle the research question, we followed a systematic methodology as shown in~\autoref{fig:methdodology}. 

Based on a study of the state-of-the-art and our own experiences with applying machine learning in self-adaptive systems, we defined the research question (see Section~\ref{sec:intro}). 
%Earlier work applied machine learning techniques to reduce large adaptation spaces based on only threshold goals~\cite{quin2019efficient}. 
%In this work, we focus on threshold, set-point and optimization goals. Inspired by recent progress in the area of deep learning, we decided to research whether we can use deep learning to effectively reduce large adaptation spaces, without compromising the system goals. 
% 
Once we determined the research question, we specified the metrics that enabled us to measure the effectiveness and efficiency of DLASeR\fplus{} and compared it with other approaches. Then, we defined the DLASeR\fplus{} architecture that is able to deal with threshold goals, set-point goals, and an optimization goal. 
%Next, we decided how to combine the three types of adaptation goals into the framework. This defines how we will structure the DLASeR\fplus{}'s neural network architecture.
Next, we devised DLASeR\fplus{}'s learning pipeline that applies deep learning for the three types of adaptation goals. We  instantiated this pipeline for various setups for the two evaluation cases of DeltaIoT applying different combinations of adaptation goals. We then evaluated DLASeR\fplus{} and compared it with representative related approaches. Finally, we answered the research question, and reflected and discussed the extent to which we met our objectives.

In the next sections, we zoom in on the different steps of the methodology. We start with explaining the metrics we used to design and evaluate DLASeR\fplus{} (Section~\ref{sec:metrics}). Then we zoom in on the DLASeR\fplus{} architecture (Section~\ref{sec:dlaser+}) and we explain how to engineer a solution with DLASeR\fplus{} (Section~\ref{sec:learningpipeline}). Next, we present the evaluation of DLASeR\fplus{}, answering the research question (Section~\ref{sec:evaluation}). Finally, we discuss related work (Section~\ref{sec:relatedwork}) and conclude the paper (Section~\ref{sec:conclusions}). 
\section{Metrics}\label{sec:metrics}
To answer the research question, we determined metrics that allow us to evaluate the different aspects of effectiveness and efficiency of the proposed solution. %DLASeR\fplus{}. 
Table~\ref{tab:metrics-research-question} summarizes the metrics.% for each aspect.
\small
\begin{table}[htbp]
\rule{\textwidth}{0.3mm}
\begin{tabular}{@{}ll@{}}
\multicolumn{2}{c}{\textbf{Effectiveness}} \\ \bottomrule
\multicolumn{1}{l|}{Coverage relevant adaptation options} & \begin{tabular}[c]{@{}l@{}}F1-score (for threshold and set-point goals)\\ Spearman's rho (for optimization goal)\end{tabular} \\ \hline
\multicolumn{1}{l|}{Reduction adaptation space} &  
\begin{tabular}[c]{@{}l@{}}Average adaptation space reduction (AASR) \\ \added{Average analysis effort reduction (AAER)}\end{tabular} \\ \hline
\multicolumn{1}{l|}{Effect on realization adaptation goals} & \begin{tabular}[c]{@{}l@{}}Differences in mean values of quality properties of \\ the goals with and without learning\end{tabular} \\ \toprule
\multicolumn{2}{c}{\textbf{Efficiency}} \\ \bottomrule
\multicolumn{1}{l|}{Learning time} & \begin{tabular}[c]{@{}l@{}}Time used for learning \added{(training + prediction, expressed in s)} \\ Time reduction for analysis with learning (\added{expressed in \%})\end{tabular} \\ \hline
\multicolumn{1}{l|}{Scalability} & Metrics above for increasing size of adaptation space
\end{tabular}
\rule{\textwidth}{0.3mm}\vspace{10pt}
\caption{Left: different aspects of research question to be evaluated; right: metrics for each aspect.}
\label{tab:metrics-research-question}
\end{table}
\normalsize

We use different metrics to capture the coverage of relevant adaptation options for classification and regression. In particular, we use the F1-score for classification and Spearman's rank correlation coefficient (Spearman's rho) for regression.\footnote{A common metric to evaluate regression models is mean squared error (MSE). However, since we are more interested in the ranking of the regressed output, we use Spearman'rho to capture this ranking, which is not covered by MSE.} We define average adaptation space reduction (AASR) and \added{average analysis effort reduction (AAER)} to capture and compare the reduction of adaptation spaces. To capture the effect on the realization of the adaptation goals, we use the differences in mean values over time for the corresponding quality properties with and without learning. We measure the time used for learning\footnote{\added{The learning time refers to the time the system uses for training and making predictions of the quality properties of interest for the set of adaptation options that are considered for the respective quality properties.}} and compare this with the time that would be necessary to analyze the complete adaptation space.  Finally, for scalability, we apply the different metrics for scenarios with an increasing size of adaptation spaces. We further elaborate on F1-score, Spearman's rho, AASR, and AAER. The other basic metrics are further explained in the evaluation section. 

\subsection{F1-score: precision and recall}
The quality of a classification algorithm is often expressed in terms of precision and recall. Applied to the classification of adaptation options (relevant or not relevant according to the adaptation goals), precision is the fraction of selected adaptation options that are relevant, while recall is the fraction of the relevant options that are selected from the total number of relevant options. Hence, precision and recall are a measure of relevance.
The F1-score combines both metrics with equal importance into a single number. Concretely, the F1-score is defined as the harmonic mean of precision and recall~\cite{geron2019hands}; 
\begin{equation}
    F1 = 2*\frac{precision * recall}{precision + recall}
\end{equation}

\subsection{Spearman correlation}
Spearman correlation, or Spearman's rho, is a non-parametric metric that can be used to measure rank correlation. We use this as a metric to capture the ranking of the predicted values of quality properties of regression models. The Spearman correlation essentially converts the predicted values to numeric ranks and then assesses how well a monotonic function describes the relationship between the predicted ranks and the true ranks. Spearman's rho is defined as~\cite{kokoska2000crc}:
\begin{equation} 
    \rho_{xy} = \frac{n \sum x_i y_i - \sum x_i \sum y_i}{\sqrt{n \sum x_i^2 - (\sum x_i)^2} \sqrt{n \sum y_i^2 - (\sum y_i)^2}}
\end{equation}
This formula computes the Spearman correlation between $x$ (the predictions of the selected adaptation options) and $y$ (the true values for the selected adaptation options), with $n$ the number of observations (the number of selected adaptation options). The result is a value between 0 and 1 (for an increasing monotonic trend, or -1 for a decreasing monotonic trend). Large errors are penalized harder. For example, a swapping of the first and third rank in the  prediction results is worse (lower Spearman's rho) compared to a swapping of the first and second rank.

%Spearman correlation, also called Spearman's rho, is a useful non-parametric measure of rank correlation. We will use this metric to capture the ranking quality of regression ML models.  Spearman's rho simply converts the values of the predicted (regressed) quality to numeric ranks. Then, the linear relationship is calculated with the true ranks. This assesses how well a monotonic function describes the relationship between the predicted and the true ranks.  Spearman's rho is expressed as follows~\cite{kokoska2000crc}:
%\begin{equation} 
 %   \rho_{xy} = \frac{n \sum x_i y_i - \sum x_i \sum y_i}{\sqrt{n \sum x_i^2 - (\sum x_i)^2} \sqrt{n \sum y_i^2 - (\sum y_i)^2}}
%\end{equation}
%The above formula calculates the Spearman correlation between $x$ and $y$, with $n$ the number of observations. Large errors are penalized harder. For example, swapping the first and third rank in your prediction results in a worse association than if the first and second rank were swapped

\subsection{Average Adaptation Space Reduction}
To capture how well the adaptation space is reduced, we define a new metric called average adaptation space reduction ($AASR$).
%$AASR$ captures the percentage of adaptation options that are not considered relevant. 
$AASR$ is defined as: 
\begin{equation} 
    AASR = (1 - \frac{selected}{total}) \times 100   
\end{equation}
with $selected$ the number of adaptation options selected by learning (over multiple adaptation cycles) and $total$ the total number of adaptation options (of multiple adaptation cycles). 
%of the adaptation space. 
For instance, an average adaptation space reduction of 70\% means that after learning only 30\% of the original adaptation space is considered for analysis. $AASR$ is a particularly suitable metric for stakeholders as it covers the high-end goal of adaptation space reduction and allows comparing solutions.

%It is important to note that the interpretation of the relative adaptation space reduction depends on the types of adaptation goals considered. For threshold and set-point goals, the relative adaptation space reduction corresponds to the percentage of adaptation options that are predicted to conform with the given threshold and/or set-point goals. However, for an optimization goal, the relative adaptation space reduction corresponds to 100\% minus the percentage of adaptation options that have to be analyzed until an adaptation option is found that meets all the threshold and set-point goals. We explain this in detail in the next section. 

Remark that the average adaptation space reduction is determined by the system's adaptation goals. In particular, for the three types of goals considered in this work, the $AASR$ is determined by the threshold and set-point goals, corresponding to the percentage of adaptation options that are predicted to be conform with these goals. For systems with only optimization goals, the $AASR$ is zero (since the selected and total number of adaptation options are the same). 

%to 100\% minus the percentage of adaptation options that have to be analyzed until an adaptation option is found that meets all the threshold and set-point goals. We explain this in detail in the next section. 
%
It is also interesting to note that the $AASR$ depends on the restrictiveness of both threshold and set-point goals. Suppose that the threshold and set-point goals are not very restrictive, thus many adaptation options will comply with the adaptation goals. In this case, the reduction will be rather small. In the other case, for very restrictive threshold and set-point goals the opposite is true. A larger reduction for these goals can be expected. 
%, and a smaller reduction for only an optimization goal.

\subsection{\added{Average Analysis Effort Reduction}}
To capture the effect of the adaptation space reduction on the effort required for analysis, we define a new metric called \added{average analysis effort reduction ($AAER$). 
$AAER$} is defined as: 
\begin{equation} 
    \added{AAER} = (1 - \frac{analyzed}{selected}) \times 100  
\end{equation}
with $analyzed$ the number of adaptation options that have been analyzed (over multiple adaptation cycles) and $selected$ the number of adaptation options selected by learning (over multiple adaptation cycles). For instance, an \added{average analysis effort reduction} of 90\% means that only 10\% of the adaptation options selected by learning were analyzed to find an option to adapt the system.
Similarly to $AASR$, \added{$AAER$} also covers a high-end goal of adaptation space reduction.

Note that the \added{$AAER$} depends on the analysis approach that is used to find an adaptation option from the reduced adaptation space (and not on the constraints imposed by the adaptation goals as for $AASR$).   
In particular, the \textit{analysis reduction} corresponds to 100\% minus the percentage of selected adaptation options that have to be analyzed until an adaptation option is found that meets all the threshold and set-point goals.
For systems that include threshold and set-point goals and an optimization goal, the selected adaptation options are analyzed in the order predicted for the optimization goal. For systems with only threshold and set-point goals, the most basic method to analyze the selected adaptation options is random order. For the evaluation of this specific case, we randomly shuffle the selected options per cycle, resulting in a more representative \added{$AAER$} score.
Remark that for systems with a large adaptation space and only an optimization goal, $\added{AAER} \approx 100$\% (since $analyzed$ is 1 and $selected$ is a large number equal to the size of the adaptation space).

\section{DLASeR\fplus{} Architecture}\label{sec:dlaser+}
%\todo{Jeroen tot hier geraakt}
We introduce now the novel adaptation space reduction approach ``Deep Learning for Adaptation Space Reduction Plus'' -- DLASeR\fplus{}. We start with outlining how this approach deals with different types of adaptation goals, and how learning for these adaptation goals is combined into a unified architecture. Finally, we zoom in on the neural network architecture of DLASeR\fplus{}. 

%First we will focus on how each type of goal is tackled individually. Then, we present how the adaptation space is reduced if there is a combination of multiple types of adaptation goals. Next, we present the deep learning architecture of DLASeR\fplus{}. Finally, we introduce the learning pipeline, defining the DLASeR\fplus{} approach, in which all of the previous aspects are combined.

\subsection{Adaptation space reduction for different types of adaptation goals}
DLASeR\fplus{} can reduce the adaptation space for adaptation goals based on thresholds, set-points, and optimization goals. 
%that we discuss in We will explain the specific learning technique that is applied to handle each type of goal. We will focus on two techniques; classification and regression.
%Note that each goal is defined for a certain quality (that is dependent on the adaptation option). 

\subsubsection{Threshold goals}
To deal with threshold goals, DLASeR\fplus{} relies on classification deep learning. Concretely we apply binary classification using class $1$ (true) if an adaptation option meets the threshold goal and class $0$ (false) otherwise. We say that an adaptation option meets a threshold goal when the associated quality property of that adaptation option is below (or above) the given threshold value $g_t$ of that goal. Hence, DLASeR\fplus{} reduces the adaptation space to the adaptation options that are valid, i.e., that are classified as $1$. 
In case of multiple threshold goals, the adaptation space is reduced to the intersection of the subsets of adaptation options with predicted values that are classified as $1$ for each of the different threshold goals.

\subsubsection{Set-point goals}
For set-points goals, DLASeR\fplus{} also relies on classification deep learning. When a user defines a set-point goal, he or she has to specify a set-point value $g_s$ and a bound $\epsilon$. This bound defines the range $[g_s-\epsilon, g_s+\epsilon]$ in which adaptation options are considered valid.\footnote{From a control theoretic perspective, this bound corresponds to the steady state error.} 
Hence, DLASeR\fplus{} again applies binary classification; class 1 for adaptation options that are predicted within the interval, and class 0 otherwise.
In case the self-adaptive system has multiple set-point goals, the reduced adaptation space is the intersection of the subsets of adaptation options with predicted values that are classified as 1 for each of the different set-point goals.

\subsubsection{Optimization goals}
DLASeR\fplus{} handles optimization goals using regression deep learning. Based on the regressed (predicted) values of a quality property, the adaptation options are ranked. From this ranking the adaptation option that maximizes or minimizes the adaptation goal can be derived. 
%\footref{foot:note-adapt-goal}
%
%The adaptation space reduction occurs by selecting the adaptation that is predicted to optimize the goal. Thus, in the case of a single optimization goal, the adaptation space is reduced to only one adaptation option.
%
%We limit ourselves to only one optimization goal. Handling multiple optimization goals requires to make trade-offs in order to find the Pareto optimal solution~\cite{hwang2012multiple}. Such problems are outside the scope of this work.
%
The adaptation space reduction is then determined by the number of adaptation options that need to be analyzed to make an adaptation decision. In the case of a single optimization goal, as we consider in this work, the adaptation space reduction is determined by the number of ranked adaptation options that need to be analyzed before one is found that complies with the other goals.

\subsection{Unified DLASeR\fplus{} architecture}\label{sec:dlaser+_unified_architecture}
\begin{figure}[htbp]
    \centering
    \includegraphics[width=0.9\textwidth]{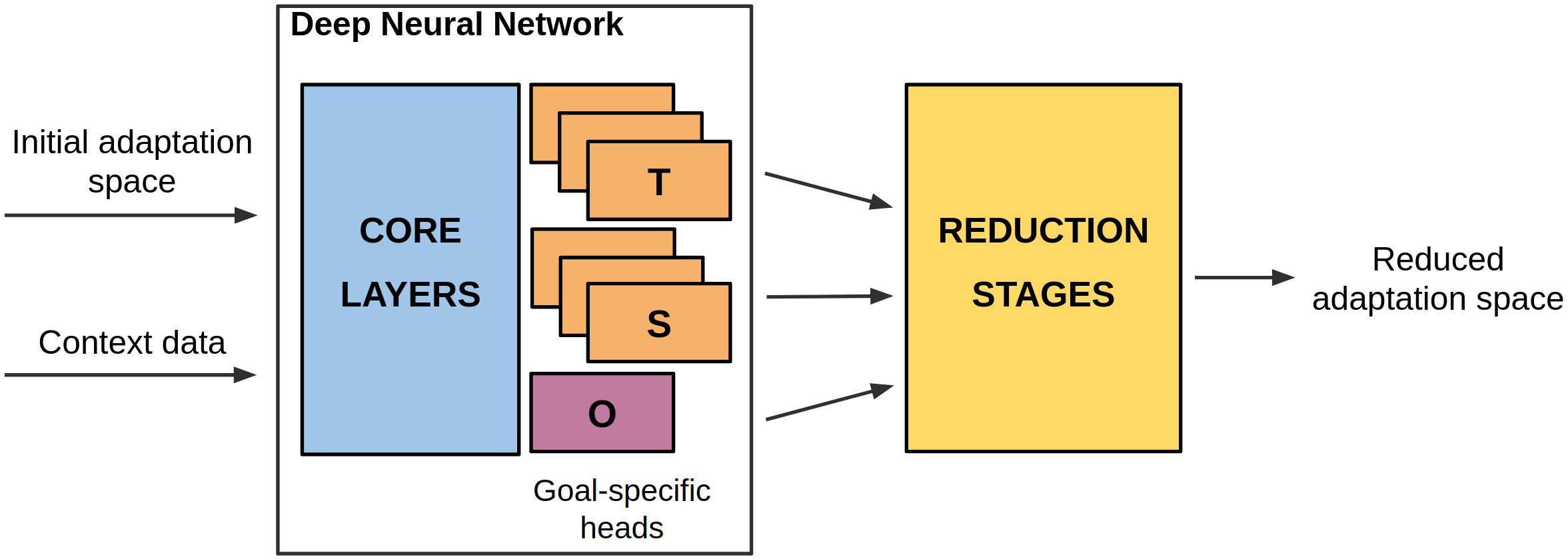}
    \caption{Unified architecture of DLASeR+ for a mix of threshold and set-point goals and one  optimization goal.}
    \label{fig:unified-architecture}
\end{figure}

We explained how DLASeR\fplus{} handles single types of adaptation goals, however, practical systems usually combine different types of adaptation goals. Over the years, different techniques have been developed to combine adaptation goals for the decision-making of self-adaptation. Classic examples are goals, utility functions, and rules~\cite{8-3-642-04425-0_36,1078411,3380965}. DLASeR\fplus{} offers a unified architecture for adaptation space reduction for a class of systems that combine multiple threshold and set-point goals with a single optimization goal. Our focus is on rule-based goals that are representative for a large number of practical self-adaptive systems. 
Figure~\ref{fig:unified-architecture} shows the unified DLASeR\fplus{} architecture. 

The deep neural network takes as input the initial adaptation space that consists of all the adaptation options, together with context data, such as the actual values of uncertainties in the environment and the current values of the relevant qualities. 
Internally, the DLASeR\fplus{} architecture is centered on a single deep neural network that covers all the adaptation goals. The deep neural network consists of a number of core layers, complemented with goal specific heads. The core layers are shared among the different goals. Each head deals with a specific goal. 
%according to its type. 
The output produced by the deep neural network is then combined to produce the reduced adaptation space. 
\added{The layered structure of DLASeR+ with shared core layers and goal specific heads adds to the modularity of the learning architecture supporting modifiability and extensibility. }

\begin{figure}[htbp]
  \centering
  \begin{subfigure}[b]{\textwidth}
    \centering
    \includegraphics[width=0.82\textwidth]{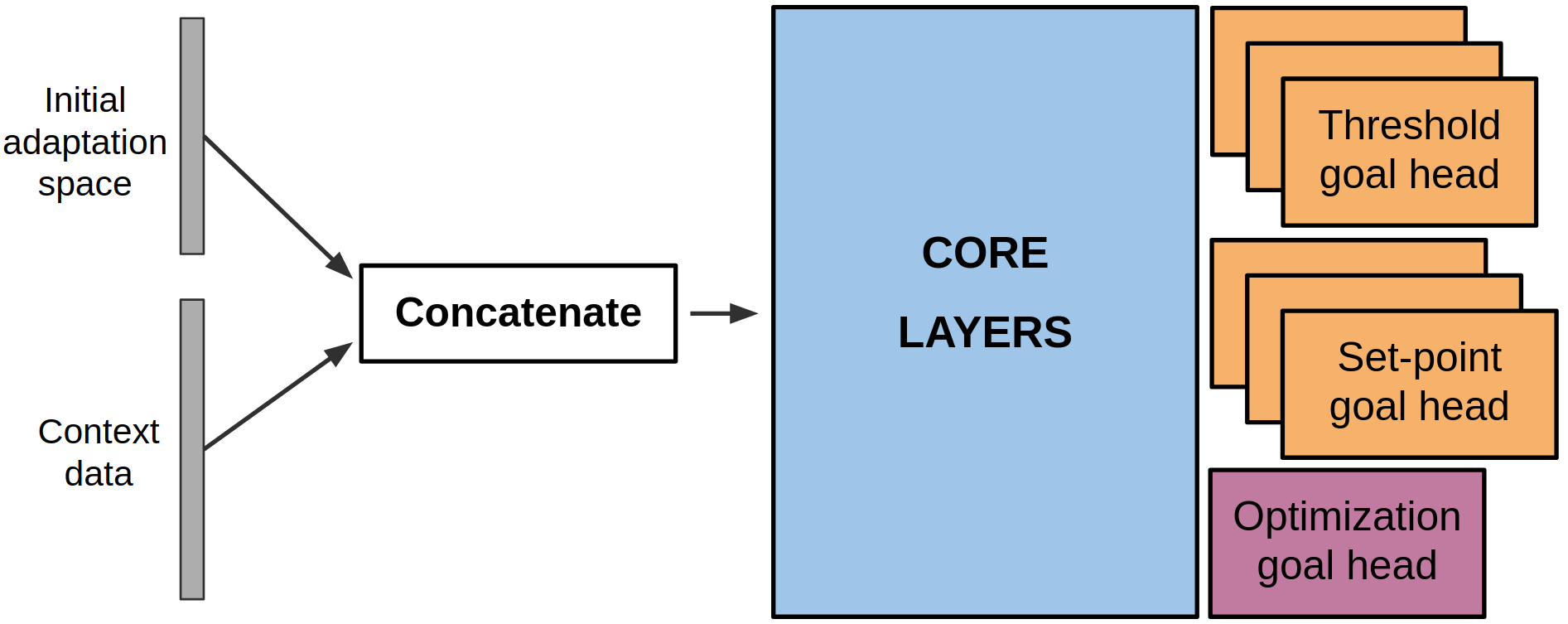}
    \caption{Architecture deep neural network.}\vspace{8pt}
    \label{fig:dlaser-architecture}
  \end{subfigure}
  \hfill
  \begin{subfigure}[b]{\textwidth}
    \centering
    \includegraphics[width=\textwidth]{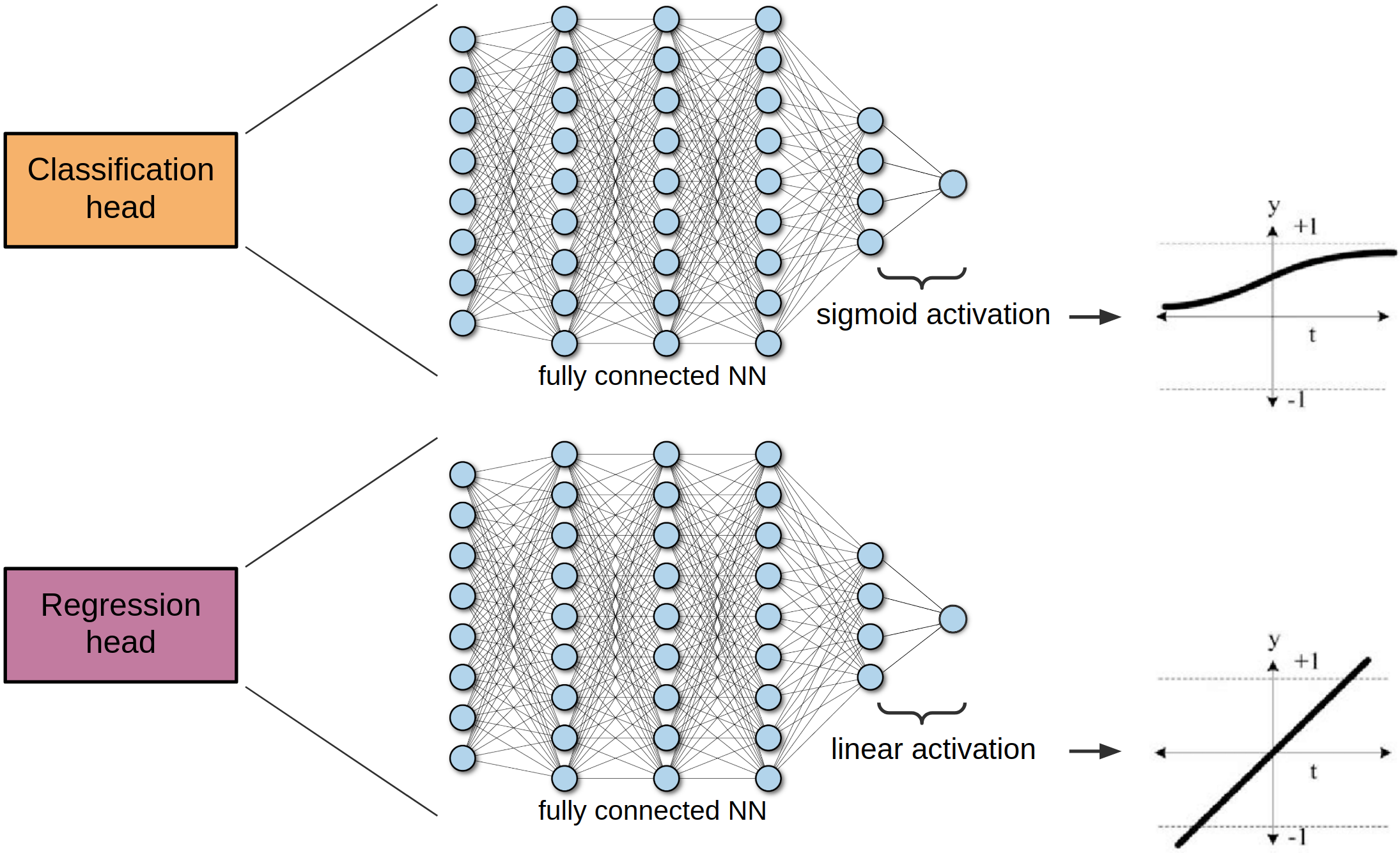}
    \caption{Internal structure of the classification and regression heads in the DLASeR+ architecture.}
    \label{fig:dlaser-legend}
  \end{subfigure}
  \caption{Overview of the internal deep neural network (NN) architecture of DLASeR\fplus{}.} 
  %The adaptation goal specific heads correspond to the ones displayed in the legend at the bottom.}
  %\textbf{Bottom:} legend of the two types of output heads. Both neural network heads have an output of dimensionality 1, with sigmoid and linear activation for respectively classification and regression.}
  \label{fig:dlaser-architecture-overview}
\end{figure}

\subsubsection{Internal neural network architecture}

The core of the DLASeR\fplus{} approach is a deep neural network that is kept updated at runtime. 
%In contrast with previous work \cite{vanapplying, quin2019efficient}, we utilize a single (deep learning) architecture for all the goals. 
%
%In the neural network architecture we can distinguish two parts. The first layers of the model are so-called shared layers, we refer to this part as the \textit{core layers}. On top of these core layers, we add for each of the goals a separate head. These \textit{goal-specific heads} are either classification heads for threshold and set-point goals, and a regression head for an optimization goal.
%
\autoref{fig:dlaser-architecture-overview} shows the internal architecture of the neural network. The figure at the top shows the structure of the network and the flow of the data through the network. The network starts with concatenating the input into \added{input} vectors, one vector per adaptation option. \added{The \added{input} vectors include high-level data relevant to adaptation in two parts: data of the adaptation options, such as the settings of the system, and data of the context, such as the current configuration, the recent load of the system and recent values of uncertainty parameters.}  The data of the \added{input} vectors are then fed to the core layers of the network that are modeled as a fully connected network of neurons. The output of the last layer of the core layers encodes the input of goal-specific heads. 
DLASeR\fplus{} supports two types of goal-specific heads that can be added on top of the core layers: classification heads for threshold and set-point goals, and a regression head for an optimization goal. \autoref{fig:dlaser-architecture-overview} at the bottom illustrates the difference between the two types. Both types of heads are fully connected neural networks that produce a single output, i.e., the output layer has a dimensionality of 1. However, the heads differ in the output they produce. Classification heads that use a sigmoid activation function produce values between 0 and 1. These values are classified based on predefined thresholds; for instance, all values below 0.5 are classified as class 0 and all values above (and including) 0.5 as class 1. The regression head that uses a linear activation function produces values that predict the quality property that needs to be optimized.  %regression head, we apply a linear activation since we want to regress a quality.

\subsubsection{Adaptation space reduction stages}\label{subsec:reduction_stages}
By using a single deep neural network, the input data passes only once through the neural network to produce the output for the different adaptation goals. This output is then combined to reduce the adaptation in two stages. The first stage uses classification deep learning to reduce the adaptation space to the adaptation options that are classified as being valid according to the threshold and set-point goals. We refer to the first stage as the \textit{classification stage}. The second stage that deals with the optimization goal further reduces the adaptation space obtained from stage 1 by ranking the regressed values of the adaptations options. The adaptation options are ranked from low to high in the case of a minimization goal, and vice versa for a maximization goal. The ranked adaptation options can then be analyzed one by one until an option is found that satisfies (e.g., determined by verification) all the other goals. We refer to the second stage as the \textit{regression stage}. 

It is important to note that either of these stages can be omitted when the corresponding type(s) of goals are not present for a problem at hand. For instance, for a system with only threshold goals (for which DLASeR\fplus{} uses classification), the second stage can be omitted. 

\begin{figure}[htbp]
    \centering
    \includegraphics[width=\textwidth]{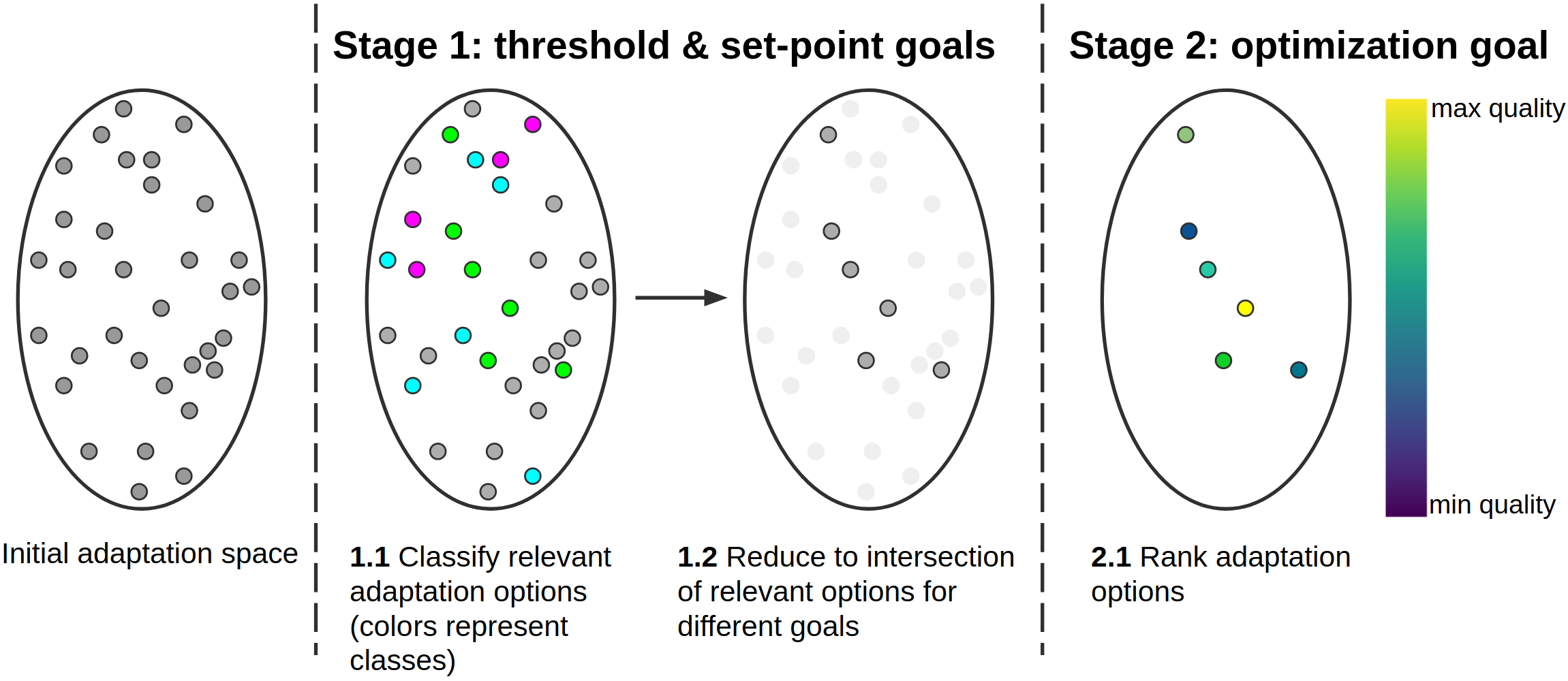}
    \caption{Illustration of the two stages of DLASeR\fplus{}. Each dot represents an adaptation option. In the first stage the initial adaptation options are classified for threshold and set-point goals (\textbf{1.1}): blue dots satisfy the threshold goals, pink dots satisfy the set-point goals, and green dots satisfy both types of goals. Then the intersection of adaptation options that satisfy all threshold and set-point goals are kept, i.e., the green dots, (\textbf{1.2)}. In the second stage the relevant subset of adaptation options obtained in stage 1 are ranked (\textbf{2.1}).}
    \label{fig:dlaser+-_stages_example}
\end{figure}

%We explain now the two stages and illustrate the combination with a simple example.
%\subsubsection{Illustration of the adaptation space reduction stages}
\vspace{6pt}\noindent
Figure~\ref{fig:dlaser+-_stages_example} schematically illustrates the two reduction stages, showing how the predictions of the neural network for different adaptation goals are combined to reduce an adaptation space. 
%, see . 
%Suppose that the DLASeR\fplus{} neural network predicted (once) for all the adaptation options. The illustration presents how the predictions are used to perform the adaptation space reduction.

%with a simple example shown in
%\autoref{fig:dlaser+-_stages_example} illustrates the two stages of adaptation space reduction for a system with a mix of the three types of goals. 

%Suppose that the DLASeR\fplus{} neural network predicted (once) for all the adaptation options.\footref{note:dlaser+_predict_once} This illustration presents how the output predicted by the NN is processed in order to enhance decision-making. 
%
% Stage 1
%\textbf{Stage 1}: When threshold and/or set-point goals are present, we can perform adaptation space reduction by means of the two steps of stage 1. First, for all the threshold and set-point goals we use to predicted output to classify them as either relevant or not relevant. Then, as second step, we aggregate the results and reduce the adaptation space to the adaptation options that were classified relevant for all of the aforementioned goals. Note that in the illustration we considered only one threshold goal and one set-point goal (for the purpose of visibility).
%
In stage 1, all the adaptation options for all the threshold and set-point goals are classified relevant or not relevant. Then, the results are aggregated, reducing the initial adaptation space to the intersection of the adaptation options that were classified relevant for all the aforementioned goals. 
%Note that in the illustration we considered only one threshold goal and one set-point goal (for the purpose of visibility).
%
%Stage 2
%\textbf{Stage 2}: Then, if there is one optimization goal present, we perform the second stage. Here we select from the predicted output the regressed quality (for which the optimization goal was defined) for the subset of adaptation options obtained in stage 1. We rank the adaptation options based on the regressed values. It is in this ranked order that the adaptation options are verified until one is found that satisfies all the adaptation goals. 
%
In stage 2, DLASeR\fplus{} ranks the subset of adaptation options obtained from stage 1. This ranking is based on the regressed quality for which the optimization goal was defined and depends on the optimization goal, i.e., ascending and descending order for respectively minimization and maximization. The adaptation options of the reduced adaptation space can then be analyzed in the order of the ranking until one is found that satisfies all the adaptation goals.

\section{Engineering with DLAS{e}R\fplus{}}\label{sec:learningpipeline}

We explain now how to engineer a solution for adaptation space reduction with DLASeR\fplus{}. Central to this is the learning pipeline that consists of an offline and online part. During the offline part, an engineer selects a model for the deep neural network model and a scaler. During the online part, the running system uses the model and scaler to reduce adaptation spaces, and exploits newly obtained data from analysis to update the model and the scaler. 

\subsection{Offline part of the DLASeR\fplus{} learning pipeline}\label{subsec:offline_learning_pipeline}
\begin{figure}[htbp]
    \centering
    \includegraphics[width=0.9\textwidth]{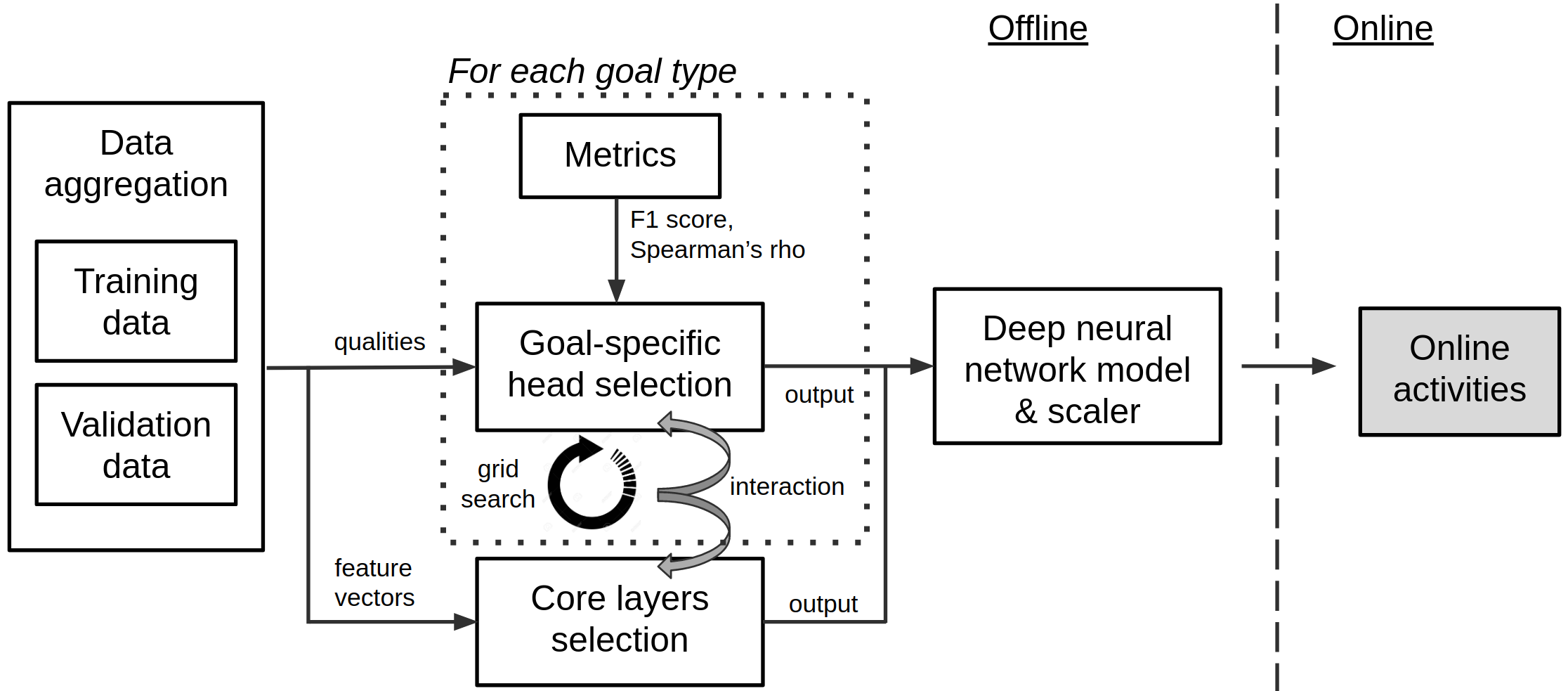}
    \caption{The offline part of the DLASeR\fplus{} learning pipeline. During model selection the optimal hyper-parameters are determined for the shared layers (\textit{Core layers selection}) and for the goal-specific heads (\textit{Goal-specific head selection}). The output of the offline part of the pipeline is a completely configured deep neural network model (of core layers complemented with the various goal-specific heads) and a scaler.}
    \label{fig:dlaser+_offline_pipeline}
\end{figure}

During the offline part of the pipeline, shown in Figure~\ref{fig:dlaser+_offline_pipeline}, the engineer collects data of a series of adaptation cycles (via observation or simulation of the system). This data comprises the adaptation options, context info, and also the associated qualities of the adaptation options. The collected data is then  concatenated in \textit{\added{input} vectors}, one per adaptation option. \added{Each item of \added{an input} vector refers to a} measurable property or characteristic of the system that affects the effectiveness of the learning algorithms.
\added{As an example, \added{an input} vector for DeltaIoT contains: 
\begin{itemize}
    \item Data of an adaptation option, i.e., the settings of the transmission power for each link and the distribution factors per link that determine the percentage of messages sent over the links;
    \item Context data, i.e., the traffic load generated per mote and the signal-to-noise ratio over the links (uncertainties), the current system configuration.   
\end{itemize}}
The qualities associated with the adaptation options are used to validate the output of the generated goal-specific heads. \added{As an example, the current qualities (one per adaptation goal) for DeltaIoT are:  
\begin{itemize}
    \item The packet loss along the links;
    \item The energy consumed by the motes.   
\end{itemize}}
\noindent To successfully train a deep neural network, it is important that all relevant data is collected.\footnote{\added{It is important to note that deep learning can work with ``raw data'' without the need for transforming and aggregating features, etc., as with classic machine learning that requires a substantial effort of engineers. Using the raw data, the deep neural network will learn complex relations automatically.}}

The aggregated data is then used to perform so called back-testing on the application to evaluate the performance of candidate learning solutions. To that end, the aggregated data is split in two sets: training data and validation data. \added{It is crucial that both sets do not overlap. An overlap of the two sets would introduce data-leakage, where knowledge of the test set leaks into the dataset used to train the model. This would jeopardize the validity of the results.~\cite{Brownley2020}} 

The main activity of the offline stage of the pipeline is selecting a deep neural network model based on finding optimal hyper-parameter settings. Hyper-parameters are non-learnable parameters that control the learning process. The main hyper-parameters are the number of layers of the deep neural network and the number of neurons per layer. Other hyper-parameters that apply for deep learning architectures are the scaler algorithm (normalizes the data, improving the learning process), the batch size (defines how much of the data samples are considered before updating the model affecting the learning speed), the learning rate LR (determines the impact of on update, affecting the learning speed), and the optimizer (influences the updates of model parameters utilizing the learning rate and the local gradient at the neurons). 

In DLASeR\fplus{}, we distinguish between hyper-parameters for the shared core layers and for goal-specific heads. \textit{Core layers selection} deals with the hyper-parameters of the shared core layers, while \textit{goal-specific head selection} deals with the hyper-parameters of the goal-specific heads.  
To determine the optimal values of the hyper-parameters for the DLASeR\fplus{} neural network model we applied grid search~\cite{geron2019hands}. With grid search, the neural network model is trained and then evaluated for every combination of hyper-parameters.\footnote{\added{Technically, we consider the layout of the core layers and the goal-specific heads \newadded{(number of layers and neurons per layer)} as distinct hyper-parameters. However, these hyper-parameters are optimized together in a single grid search process, since the loss function from the goal-specific heads guides the learning process, including the learning of core layers.}}
%\footnote{\added{Technically, the gradients flow backwards, from the heads to the core layers. Thus, first the gradients for each head are calculated, then they are added together, then the result of the addition is forwarded down the stack of core layers.}} 
We used different metrics during evaluation: (i) F1-score that combines precision and recall of predicted classes of adaptation options, and \added{(ii)} Spearman correlation that measures the ranking of predicted values of quality properties of regression models, see the explanation in Section~\ref{sec:metrics}. 
Once the models are trained (on the training data set), they are evaluated on the validation data set, i.e., the predictions are evaluated using the validation data. 
%The effectiveness of the models is measured by the F1-score for the classification heads and Spearman's rho for the regression head. 
The best model is then selected based on the validation loss that is determined by a loss function \mbox{that sums the losses of the heads, capturing the overall quality of the neural network model.}

When the core layers and goal-specific heads are fine-tuned and a proper scaler is found the integrated solution can be deployed, which brings us to the second stage of the learning pipeline. 

\subsection{Online part of the DLASeR\fplus{} learning pipeline}
\begin{figure}[htbp]
    \centering
    \includegraphics[width=0.9\textwidth]{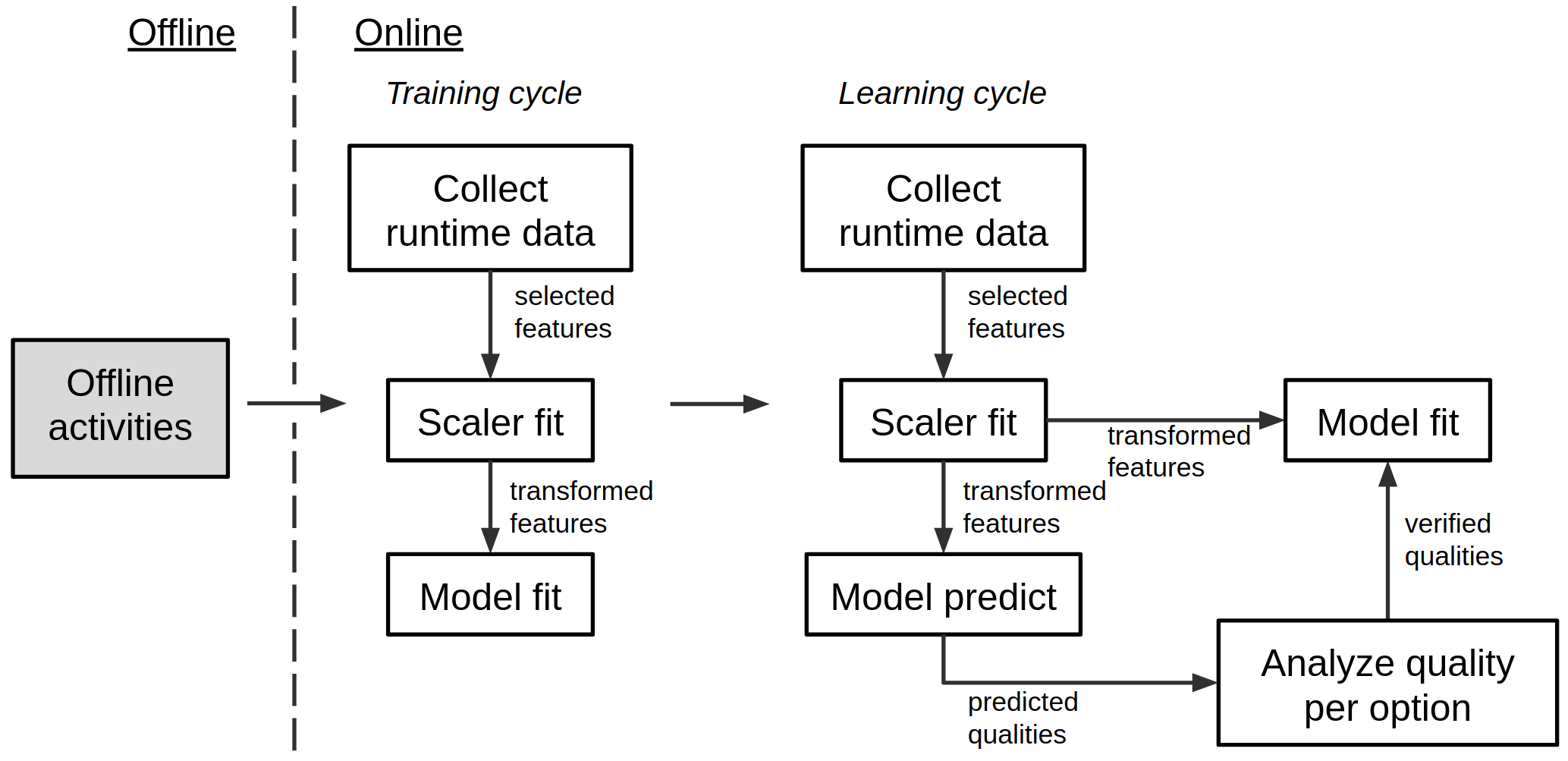}
    \caption{The online part of the DLASeR\fplus{} learning pipeline. In the training phase (training cycles), the deep neural network model is initialized for the problem at hand. In the learning phase (learning cycles), adaptation space reduction is applied and new data is used to update the deep neural network model (online learning).}
    \label{fig:dlaser+_online_pipeline}
\end{figure}

During the online part of the learning pipeline, the deep neural network supports the decision-making process of a self-adaptive system with reducing large adaptation spaces. The online part, shown in~\autoref{fig:dlaser+_online_pipeline}, consists of two consecutive phases. \added{In the training phase, which consists of a series of training cycles, the deep learning model is initialized based on the current state of the system exploiting all available runtime data. In the learning phase, which consists of learning cycles, the deep learning model performs adaptation space reduction and is updated using online learning. }
%It is in this stage that the deep learning model is deployed to aid decision-making in the self-adaptive system.

\subsubsection{Training phase}
The goal of the training phase is to initialize the learnable parameters of the model for the problem at hand. During the training cycles, relevant runtime data is collected to construct \added{input} vectors. This data is used to update the scaler (e.g., updating min/max values of parameters, the means, etc.) and the \added{input} vectors are adjusted accordingly. The deep neural network model is then trained using the transformed data; i.e., the parameters of the model are initialized, e.g., the weights of the contributions of neurons. The training phase ends when the validation accuracy stagnates, i.e., when the difference between the predicted values with learning and the actual verification results are getting small.\footnote{\added{For the DeltaIoT, the training phase ended after 45 training cycles (one per adaptation cycle), see Section~\ref{sec:evaluation}.}} During the training cycles, \added{the complete time slot available for adaptation is used for analyzing adaptation options to generate data} that is used to initialize the deep neural network model; hence there is no reduction of the adaptation space yet. 

\subsubsection{Learning phase}
In the learning phase, \added{the deep neural network model is actively used to make predictions about the adaptation options for the various adaptation goals aiming to reduce the adaptation space before the analysis of the reduced adaptation space.} Furthermore, the results of the analysis of the reduced adaptation space are then used to incrementally update the learning model.
Concretely, \added{in each learning cycle, an input} vector is composed for each adaptation option that is scaled using the scaler. 
Scaling normalizes the range of independent variables of data, ensuring that each element contributes approximately proportionately to the result. 
The deep neural network model then makes predictions for the \added{input} vectors. Based on these predictions, the adaptation space is reduced and the adaptation options of the reduced adaptation space are then analyzed. In the evaluation of DLASeR\fplus{} (Section~\ref{sec:evaluation}), we use runtime statistical model checking~\cite{david2015uppaal,Agha:2018,7573167,weyns2019activforms,weyns2022AF} to analyze the adaptation options, however other analysis techniques can be applied. 

The analysis of the reduced adaptation space depends on the types of adaptation goals at hand. For a self-adaptive system with a mix of threshold and set-point goals and an optimization goal, analysis consists of two stages: the classification stage and the regression stage, see Section~\ref{subsec:reduction_stages}. For systems without optimization goal, only the classification stage applies. Similarly, for systems with only an optimization goal, only the regression stage applies.

%\vspace{6pt}\noindent
In the \textit{classification stage} the adaptation space is reduced to the relevant subset of adaptation options that comply with the threshold and set-point goals. Algorithm~\ref{algo:dlaser+_classification} shows how the adaptation space reduction is applied  and how analysis is performed in the classification stage. 

%If there is no optimization goal (thus no optimization stage), than these selected adaptation options are used for verification. In addition, we add a fixed percentage of the other (unselected) adaptation options, i.e., the \textit{exploration rate}. This leads to a more representative distribution of the various adaptation options. As a final step, the verification results are used to further improve the learning model.

%In the other case, i.e., there is an optimization goal (thus an optimization stage), than we perform no verification in this stage and pass the reduced subset to the optimization stage.  Note that if there are no threshold nor set-point goals, we pass the complete set of adaptation options, i.e., the adaptation space, to the optimization stage.

\begin{algorithm}[htbp]
    \caption{Stage 1: classification stage}
    \label{algo:dlaser+_classification}
    \begin{algorithmic}[1]
        %Prediction relevant subspace
        \State $pred\_subspace \gets K.adaptation\_options$ \label{line:algo-dlaser+_classification-start}
        \State $DNN\_output \gets K.DNN\_model$.predict($input\_vectors$) \label{line:dlaser+_classification-predict}
        \If{$K.threshold\_goals$ == None \& $K.set\_point\_goals$ == None} \label{line:check-classification-goals}
            \State Proceed to the regression stage: Algorithm~\ref{algo:dlaser+_optimization}
        \EndIf \label{line:end-check-classification-goals}
        \ForEach{$class\_goal \gets K.classification\_goals$} \label{line:algo-dlaser+_classification-start-reduction}
            \State $pred\_subspace \gets pred\_subspace \cap DNN\_output.class\_goal$
        \EndFor \label{line:algo-dlaser+_classification-end-reduction}
        %Verifying adaptation options
        \If{$K.optimization\_goal \neq$ None} \label{line:dlaser+check-optimization-goal}
            \State Proceed to the regression stage: Algorithm~\ref{algo:dlaser+_optimization} \label{line:algo-dlaser+_classification-to-optimization-stage}
        \EndIf %\label{line:algo-dlaser+_classification-to-not-optimization-stage}
        \State $pred\_subspace$.shuffle()~\label{line:shuffle}
        % \added{
        % \If{length$(K.set\_point\_goals) == 1$}
        %     \State $set\_point\_goal \gets K.set\_point\_goals[0]$ 
        %     \State $ranking \gets |DNN\_output.set\_point\_goal - SET\_POINT\_VALUE|$
        %     \State $ranked\_subspace \gets pred\_subspace$.sort($ranking$)
        %     \State $pred\_subspace \gets ranked\_subspace$ 
        % \Else
        %     \State $pred\_subspace$.shuffle() \label{line:algo-dlaser+_classification-start-verify}
        % \EndIf
        % }
        \State $valid\_found \gets$ False, $idx \gets$ 0~\label{iterate-start} 
        \State $verified\_subspace \gets \emptyset$~\label{iterate-end} 
        \While{not $valid\_found$ \& idx $< pred\_subspace.size$} \label{line:classification_start_while}
             \State $adapt\_opt \gets pred\_subspace$[$idx$]
             \State $Analyzer$.analyzeAdaptationOptions($adapt\_opt$) \Comment{Knowledge}
             \State $\_, qualities \gets K.verification\_results$[$idx$]
             \If{$qualities$ meet all threshold and set-point goals}
                 \State $valid\_found \gets$ True
             \EndIf
             \State $verified\_subspace$.add($adapt\_opt$)
             \State $idx \gets idx$ + 1
        \EndWhile \label{line:classification_end_while}
        \State Select valid adaptation option $adapt\_opt$, if none valid use fall back option \label{line:algo-dlaser+_classification-select}
        \State $unselected \gets K.adaptation\_options \setminus pred\_subspace$  \label{line:algo-dlaser+_classification-start-verify2}
        \State $explore \gets unselected$.randomSelect($exploration\_rate$)\label{line:algo-dlaser+_explore}
        \State $Analyzer$.analyzeAdaptationOptions($explore$) \Comment{Knowledge} \label{line:algo-dlaser+_classification-end-verify}
        % Update the model
        \State $input\_vectors$, $qualities \gets K.analysis\_results$ \label{line:algo-dlaser+_classification-start-update}
        \State $K.DNN\_model$.update($input\_vectors$, $qualities$) \label{line:algo-dlaser+_classification-end-update}
    \end{algorithmic}
\end{algorithm}

%\vspace{-5pt}
In lines~\ref{line:algo-dlaser+_classification-start}~and~\ref{line:dlaser+_classification-predict}, we initialize the variable $pred\_subspace$ to the complete adaptation space and store the predicted output of the deep neural network in the $DNN\_output$ variable. Both these variables are used in  Algorithm~\ref{algo:dlaser+_classification} and Algorithm~\ref{algo:dlaser+_optimization}. $K.DNN\_model$ refers to the deep neural network model that is stored in the Knowledge module ($K$). % that is shared among the monitor, analyzer, planner and executor of the feedback loop.  
Since DLASeR\fplus{} uses a single deep learning model for all the goals, prediction can be done in a single step. Note that the \textit{predict()} function in line~\ref{line:dlaser+_classification-predict} scales the elements of the \added{input} vectors before making predictions. Lines~\ref{line:check-classification-goals}~to~\ref{line:end-check-classification-goals} check whether there are any threshold or set-point goals. If there are no threshold nor set-point goals, we proceed to the regression stage, i.e., Algorithm~\ref{algo:dlaser+_optimization}. In the other case, i.e., there are threshold and/or \added{set-points} goals, the adaptation space ($pred\_subspace$) is reduced in lines~\ref{line:algo-dlaser+_classification-start-reduction}~to~\ref{line:algo-dlaser+_classification-end-reduction}. Concretely, we reduce the adaptation space to the intersection of the predicted subspaces that are relevant for each of these two types of classification goals. This step represents the actual adaptation space reduction for all threshold and set-point goals. The adaptation option is analyzed to check whether it complies with the threshold and set-point goals. If this is the case this option is selected for adaptation. If not, analysis is continued until an adaptation option is found that satisfies the threshold and set-point goals. 
%Verifying adaptation options
Line~\ref{line:dlaser+check-optimization-goal} checks whether there is an optimization goal. If there is such a goal, the classification stage ends and the system continues the regression stage using the reduced subset, i.e., $pred\_subspace$ (see line~\ref{line:algo-dlaser+_classification-to-optimization-stage}). If there is no optimization goal, the adaptation options for analysis are selected.
%in lines~\ref{line:algo-dlaser+_classification-start-verify}~to~\ref{line:classification_end_while}. 
We first shuffle the relevant subspace \added{to avoid bias in the way the adaptation options are determined} (see line~\ref{line:shuffle}), prepare the analysis (see lines~\ref{iterate-start} and~\ref{iterate-end}), and then iterate over the adaptation options top-down (see lines~\ref{line:classification_start_while}~to~\ref{line:classification_end_while}).
In this iteration, an adaptation option is analyzed to check whether it complies with the threshold and set-point goals. If this is the case this option is selected for adaptation and the iteration halts. If not, analysis is continued until an adaptation option is found that satisfies the threshold and set-point goals. 
% Select the adaptation option
After iterating over the subspace of adaptation options, line~\ref{line:algo-dlaser+_classification-select} selects a valid option to adapt the system, i.e., an adaptation option that satisfies all the threshold and set-point goals. If no valid adaptation option is found according to the goals, a fall-back option is used that implements a graceful degradation strategy.
Then, lines~\ref{line:algo-dlaser+_classification-start-verify2}~and~\ref{line:algo-dlaser+_explore} use the \textit{exploration\_rate} to select a random sample of adaptation options from the options that were not analyzed. Adding this random sample aims at anticipating potential concept drifts that might occur in dynamic environments after a large number of adaptation cycles.\footnote{Intuitively, one may argue to select this sample nearer to the boundaries set by the thresholds rather than random, yet, this may reduce the intended effect on potential concept drifts. Further study is required to clarify this issue.} These additional options are then also analyzed (see line~\ref{line:algo-dlaser+_classification-end-verify}) and the analysis results are stored in the knowledge. 
%Updating threshold model
Finally, the analysis results are exploited to update the deep neural network model (see  lines~\ref{line:algo-dlaser+_classification-start-update}~and~\ref{line:algo-dlaser+_classification-end-update}). To that end, the \added{input} vectors (configurations etc.) and the analysis results (i.e., the qualities per goal obtained by verification) of the analyzed adaptation options are retrieved from the knowledge. Based on this data, the neural network model is updated using the same learning mechanism as used in the training cycle.

%\vspace{6pt}\noindent
The \textit{regression stage} starts either from the adaptation options selected in the classification stage or from the complete adaptation space in case there is only an optimization goal. 
%In this stage a task of the planner is performed; select the adaptation option (see~\autoref{sec:dlaser+_optimization_stage}). Therefore this stage leans more towards decision-making than adaptation space reduction. 
The set of adaptation options are then ranked according to the predicted quality of the optimization goal. Algorithm~\ref{algo:dlaser+_optimization} shows how the adaptation options are ranked and how one of the options is selected based on the result of the analysis and its compliance with the threshold and set-point goals.

\begin{algorithm}
    \caption{Stage 2: regression stage}
    \label{algo:dlaser+_optimization}
    \begin{algorithmic}[1]
        %Prediction relevant subspace
        \State $pred\_subspace$ retrieved from  Algorithm~\ref{algo:dlaser+_classification} \label{line:algo-dlaser+_optimization-start}
        %Ranking the adaptation subspace
        \State $opt\_goal \gets K.optimization\_goal$ \label{line:algo-dlaser+_optimization-start2}
        \State $ranking \gets DNN\_output.opt\_goal$ \Comment{Use prediction of Algorithm~\ref{algo:dlaser+_classification}, line~\ref{line:dlaser+_classification-predict}}
        \State $ranked\_subspace \gets pred\_subspace$.sort($ranking$) \label{line:algo-dlaser+_optimization-end-ranking}
        %Verifying adaptation options
        \State $valid\_found \gets$ False, $idx \gets$ 0 \label{line:algo-dlaser+_optimization-start-verify}
        \State $verified\_subspace \gets \emptyset$
        \While{not $valid\_found$ \& idx $< ranked\_subspace.size$}
             \State $adapt\_opt \gets ranked\_subspace$[$idx$]
             \State $Analyzer$.analyzeAdaptationOptions($adapt\_opt$) \Comment{Knowledge}
             \State $\_, qualities \gets K.verification\_results$[$idx$]
             \If{$qualities$ meet all thresholds and set-point goals}
             
                 \State $valid\_found \gets$ True
             \EndIf
             \State $verified\_subspace$.add($adapt\_opt$)
             \State $idx \gets idx$ + 1
        \EndWhile \label{line:end_while}
        \State Select valid adaptation option $adapt\_opt$, if none valid use fall back option \label{line:algo-dlaser+_optimization-select}
        \State $unselected \gets K.adaptation\_options \setminus verified\_subspace$ \label{line:unselected}
        \State $explore \gets unselected$.randomSelect($exploration\_rate$)\label{line:explore}
        \State $Analyzer$.analyzeAdaptationOptions($explore$) \Comment{Knowledge} \label{line:algo-dlaser+_optimization-end-verify}
        %Updating the model
        \State $input\_vectors$, $qualities \gets K.verification\_results$ \label{line:algo-dlaser+_optimization-start-update}
        \State $K.DNN\_model$.update($input\_vectors$, $qualities$) \label{line:algo-dlaser+_optimization-end-update}    
    \end{algorithmic}
\end{algorithm}

%This algorithm continues on the reduced adaptation space of algorithm~\ref{algo:dlaser+_classification}. 
%Ranking the adaptation options (by predicting)
In lines~\ref{line:algo-dlaser+_optimization-start2}~to~\ref{line:algo-dlaser+_optimization-end-ranking}, the predictions of the previous stage are reused to obtain a ranking of the relevant adaptation options. Since we have a single deep neural network model, we require only a single prediction 
(see algorithm~\ref{algo:dlaser+_classification}, line~\ref{line:dlaser+_classification-predict}).
%Verifying adaptation options
Lines~\ref{line:algo-dlaser+_optimization-start-verify}~to~\ref{line:end_while} iterate over the ranked adaptation options in descending order of the predicted value for the quality of the maximization goal (the opposite order is used for a minimization goal). The adaptation option is analyzed to check whether it complies with the threshold and set-point goals. If this is the case this option is selected for adaptation. If not, analysis is continued until an adaptation option is found that satisfies the threshold and set-point goals. 
%Select the adaptation option
After iterating through the ranked adaptation (sub)space, a valid option is selected to adapt the system, i.e., an adaptation option that satisfies all the threshold and set-point goals  (see line~\ref{line:algo-dlaser+_optimization-select}). If no valid adaptation option is found according to the goals, a fall-back option is used ensuring graceful degradation of the system. 
Then, lines~\ref{line:unselected}~and~\ref{line:explore} use the \textit{exploration\_rate} to select a random sample of adaptation options from the options that were not analyzed. These options are then also analyzed in line~\ref{line:algo-dlaser+_optimization-end-verify} and the analysis results are stored in the knowledge. 
%Updating threshold model
Finally, lines~\ref{line:algo-dlaser+_optimization-start-update}~and~\ref{line:algo-dlaser+_optimization-end-update} exploit the analysis results to update the deep neural network model. This enables the model to cope with the dynamic behavior of the adaptation space.

%\vspace{6pt}\noindent
%Note that in the case where we only consider threshold and/or set-point goals, the adaptation space reduction occurs prior to analysis. In case there is also an optimization goal, the adaptation space reduction is conducted by cleverly and efficiently analyzing adaptation options, i.e., in the order of their predicted ranking with respect to the optimization goal until an option is found that satisfies all the threshold and/or set-point goals.

\subsubsection{Runtime integration of DLASER\fplus{} with MAPE-K} 
\begin{figure}[h!]
    \centering
    \includegraphics[width=\textwidth]{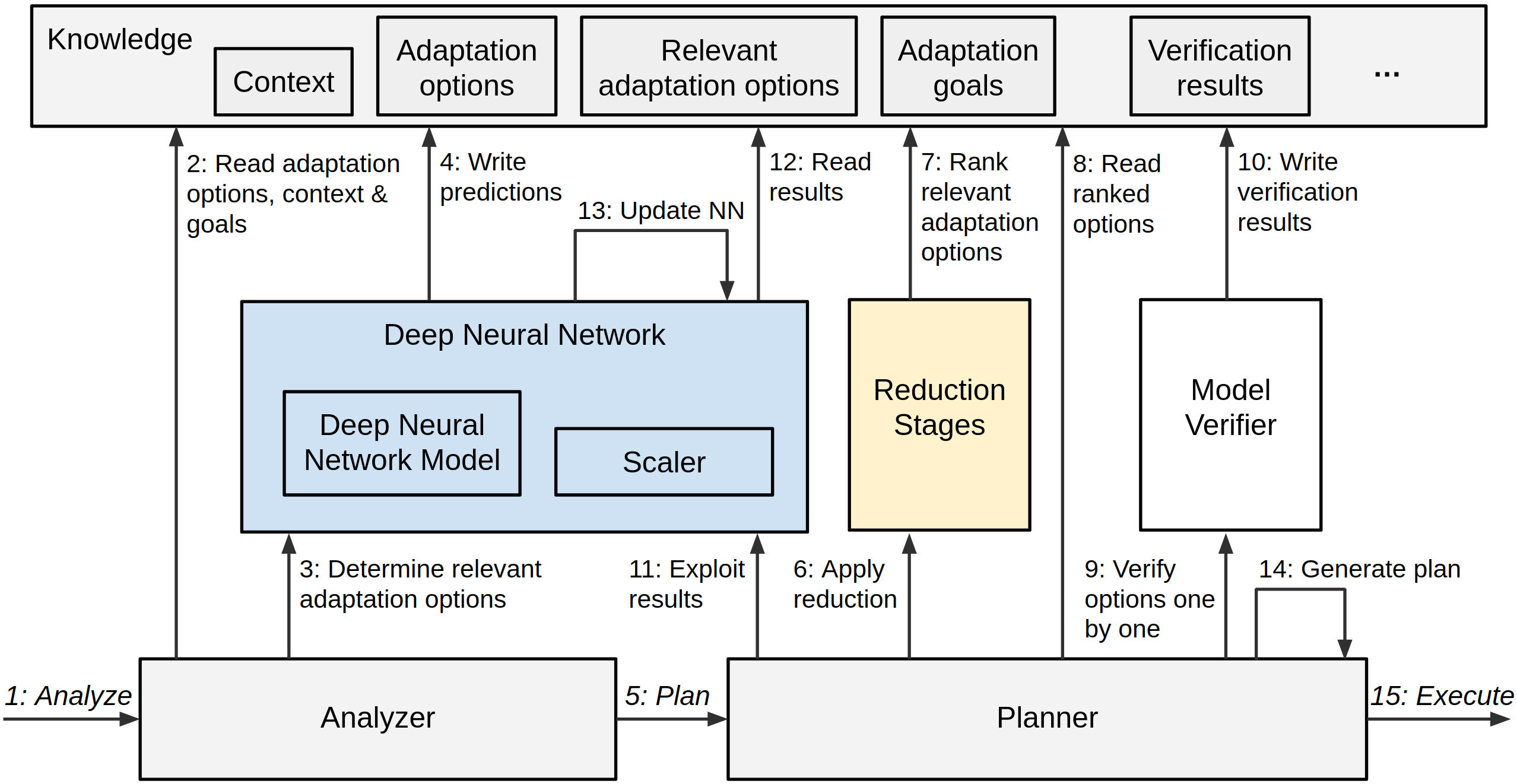}
    \caption{Runtime integration of DLASER\fplus{} with MAPE}
    \label{fig:runtime_MAPE}
\end{figure}

Figure~\ref{fig:runtime_MAPE} shows how DLASER\fplus{} is integrated with a MAPE-K feedback loop. 

We follow here the MAPE model and the tasks associated with the different MAPE elements as specified in~\cite{WeynsBook2020}. When the monitor finishes an update of the knowledge with new runtime data, it triggers the analyzer. The analyzer reads the data that is necessary for the deep neural network to determine the relevant adaptation options. The deep neural network then produces predictions that are written to the knowledge (steps 1 to 4). Next, the analyzer triggers the planner that initiates adaptation space reduction; the ranked adaptation options are then written to the knowledge (steps 5 to 7). Next, the planner reads the ranked options and invokes the model verifier to analyze the options one by one and writes the results to the knowledge (steps 8 to 10). When a suitable adaptation is found, the planner invokes the deep neural network to read the verification results and update the learning model accordingly (steps 11 to 13).  
%
%\added{During these steps, the deep neural network exploits the verification results generated by the model verifier to continue online training improving the learner over time to deal with new conditions that the system encounters.} 
%
\newadded{During these steps, the deep neural network trains the learning model using the most recent verification results that are generated by the model verifier. This online learning in each adaptation cycle helps improving the learner over time to deal with new conditions that the system encounters. } 
Finally, the planner generates a plan for the adaptation option that is selected to adapt the system and triggers the executor to enact the actions of the plan on the managed system (steps 14 and 15). 
%\section{MAPE-K architecture with DLASeR}

\section{Evaluation}\label{sec:evaluation}

We start with describing the evaluation setup and specify the various combination of adaptation goals that are evaluated. Then we explain the offline results we obtained. Finally, we zoom in on the online results for effectiveness and efficiency of DLASeR\fplus{}. \added{All material used for the evaluation, incl. all configurations and  settings, and all evalution results are available at the DLASeR website.}\footnote{\added{  \url{https://people.cs.kuleuven.be/danny.weyns/software/DLASeR/index.html}}} 

\subsection{Evaluation setup}
\added{We evaluated DLASeR\fplus{} on two instances of DeltaIoT, as described in Section~\ref{sec:background}. For both instances we used the same type of network settings. The stochastic uncertainty profiles for the traffic load generated by the motes ranged from 0 to 10 messages per mote per cycle, while the network interference along the links varied between -40~dB and +15~dB. The configurations of these profiles are based on field tests. The MAPE-K feedback loop was designed using a network of timed automata. These runtime models were executed by using the ActivFORMS execution engine~\cite{iftikhar2014activforms, weyns2022AF}. The quality models were specified as stochastic timed automata. These models are used to determine the quality estimates for the adaptation options. Section~\ref{sec:background} explains an example model for packet loss. We applied runtime statistical model checking using Uppaal-SMC for the verification of adaptation options~\cite{david2015uppaal}. The exploration rate was set to 5\%. 
For both instances of DeltaIoT, we considered 275 online adaptation cycles, corresponding with a wall clock time of 77 hours. 
We used 45 cycles to train the network parameters. The remaining 230 cycles are evaluated as learning cycles.

To evaluate the effect on the realization of the adaptation goals using DLASeR\fplus{}, we compare the results with a reference approach that analyzes the whole adaptation space without learning. For the evaluation of coverage of relevant adaptation options and the reduction of the adaptation space, we could only compare the results for settings with threshold goals and an optimization goal with the basic DLASeR approach~\cite{vanapplying} and ML4EAS (short for ``Machine Learning for Efficient Adaptation Space Reduction'') proposed by Quin et al.~\cite{quin2019efficient}. The latter approach applies classic machine learning techniques to reduce adaptation spaces for threshold goals.

\autoref{tab:evaluation-settings} summarizes the combinations of adaptation goals we evaluated for both instances of DeltaIoT: \textit{TTO} (2 Threshold goals and 1 Optimization goal), \textit{TTS} (2 Threshold goals and 1 Set-point goal), and \textit{TSO} (1 Threshold, 1 Set-point, and 1 Optimization goal).\footnote{For TTO setting, ML4EAS applies classifiers to reduce the adaptation space for the two threshold goals and then searches within the reduced adaptation space to find the best adaptation option for the optimization goal.} } 

\begin{table}[htbp]
\centering
\begin{tabular}{l|l|l}
\toprule
Reference & Goals DeltaIoTv1 & Goals DeltaIoTv2 \\ \midrule
\multirow{3}{*}{\textbf{TTO}} & \textbf{T1}: $PL < 10$\% & \textbf{T1}: $PL < 10$\% \\
 & \textbf{T2}: $LA < 5$\% & \textbf{T2}: $LA < 5$\% \\
 & \textbf{O1}: minimize $EC$ & \textbf{O1}: minimize $EC$ \\ \midrule
\multirow{3}{*}{\textbf{TTS}} & \textbf{T1}: 
%\deleted{$PL < 10$}\added{$PL < 15$} 
$PL < 15$\% & \textbf{T1}: %\deleted{$PL < 10$}\added{$PL < 15$} 
$PL < 15$\% \\
 & \textbf{T2}: 
 %\deleted{$LA < 5$}\added{$LA < 10$}
 $LA < 10$\% & \textbf{T2}: %\deleted{$LA < 5$}\added{$LA < 10$}
 $LA < 10$\% \\
 & \textbf{S1}: $EC \in [12.9 \pm 0.1]$ & \textbf{S1}: $EC \in [67 \pm 0.3]$ \\ \midrule
\multirow{3}{*}{\textbf{TSO}} & \textbf{T1}: 
%\deleted{$LA < 5$}\added{$LA < 10$} 
$LA < 10$\% & \textbf{T1}: 
%\deleted{$LA < 5$}\added{$LA < 10$} 
$LA < 10$\% \\
 & \textbf{S1}: $EC \in [12.9 \pm 0.1]$ & \textbf{S1}: $EC \in [67 \pm 0.3]$ \\
 & \textbf{O1}: minimize $PL$ & \textbf{O1}: minimize $PL$ \\
 \bottomrule
\end{tabular}\vspace{5pt}
\caption{Overview of combinations of adaptation goals we evaluated for both instances of DeltaIoT. The adaptation goals are defined for packet loss (PL), latency (LA), and energy consumption (EC).}
\label{tab:evaluation-settings}
\end{table}

\added{
For the implementation of DLASeR\fplus{} we used the scalers from \textit{scikit-learn}~\cite{pedregosa2011scikit} and the neural networks from \textit{Keras} and \textit{Tensorflow}~\cite{abadi2016tensorflow}.
The simulated IoT networks are executed on an i7-3770 \added{CPU} @ 3.40GHz with 12GB RAM; the deep learning models are trained and maintained on 
an i7-3770 \added{CPU} @ 3.40GHz with 16GB RAM.

We start with presenting the results of the offline stage of the DLASeR\fplus{} learning pipeline. Then we present the results for effectiveness and efficiency of the online stage of the learning pipeline for each combination of goals. 

\subsection{Results offline settings}
As explained in Section~\ref{subsec:offline_learning_pipeline}, DLASeR\fplus{} uses grid search for configuring and tuning the deep neural network. We performed grid search on 30 sequential adaptation cycles.\footnote{\added{Concretely, we used the data of all adaptation options with their verification results over 30 adaptation cycles, i.e., 216 adaptation options with verification results per cycle for DeltaIoTv1 and 4096 adaptation options with verification results per cycle for DeltaIoTv2. Figure~\ref{fig:adaptation-space-example} and Figure~\ref{fig:dynamic-behavior} illustrate the performance of different adaptation options (for DeltaIoTv2).}}
\autoref{tab:dlaser+_grid_search}~shows the best parameters of the network for each of the three combinations of adaptation goals. Each row in the table corresponds to one grid search process. In total, 6 grid search processes were completed; one for each of the three combinations of adaptation goals and this for the two instances of DeltaIoT. 
}
\begin{table}[b!]
\centering
\begin{adjustbox}{width=1\textwidth}
\begin{tabular}{l|l|lllllll}
\toprule
\multirow{2}{*}{Problem} & \multirow{2}{*}{Goals} & \multicolumn{7}{c}{Hyper parameters} \\ \cline{3-9} 
 &  & Scaler & Batch size & LR & Optimizer & Core layers & Class. layers & Regr. layers \\ \midrule
DeltaIoTv1 & \textbf{TTO} & Standard & 64 & 5e-3 & Adam & {[}50, 25, 15{]} & {[}20,10,5{]} & {[}40, 20, 10, 5{]} \\
DeltaIoTv2 & \textbf{TTO} & Standard & 512 & 5e-3 & Adam & 
{[}150, 120, 100, 50, 25{]}%{[}200, 100, 50, 25{]} 
& {[}40, 20, 10, 5{]} & 
{[}30, 40, 50, 40, 15, 5{]} %{[}30, 40, 15, 5{]} 
\\ \midrule
DeltaIoTv1 & \textbf{TTS} & MaxAbsScaler & 64 & 5e-3 & Adam & {[}50, 25, 15{]} & {[}40, 20,10,5{]} & / \\
DeltaIoTv2 & \textbf{TTS} & MaxAbsScaler & 512 & 5e-3 & Adam & 
{[}200, 100, 50, 25{]}%{[}200, 100, 50, 25{]} 
& {[}40, 20, 10, 5{]} & / %{[}30, 40, 15, 5{]} 
\\ \midrule
DeltaIoTv1 & \textbf{TSO} & StandardScaler & 16 & 5e-3 & Adam & {[}50, 80, 35, 15{]} & {[}20,10,15{]} & {[}40,20,10,5{]} \\
DeltaIoTv2 & \textbf{TSO} & StandardScaler & 512 & 2e-3 & RMSprop & {[}150, 120, 100, 50, 25{]} & {[}40, 20, 10, 5{]} & {[}30, 40, 50, 40, 15, 5{]} \\
\bottomrule
\end{tabular}
\end{adjustbox}\vspace{8pt}
\caption{Best grid search results for DLASeR\fplus{} on \textit{TTO}, \textit{TTS}, and \textit{TSO} goal combinations (see~\autoref{tab:evaluation-settings}). LR refers to learning rate. The values between brackets for the different layers represent the number of neurons.}
\label{tab:dlaser+_grid_search}
\end{table}
\added{
Given that DLASeR\fplus{} comprises a single integrated neural network architecture with multiple classification and regression heads, we use the validation loss to select the best hyper-parameter configuration. The validation loss captures how good the predictions are of the neural network model compared to the true data (of the validation set). Here, the validation loss corresponds to the sum of the losses for each head. 
In total, grid search evaluated 4120 configurations for DeltaIoTv1 (1728 for TTO and TSO goals and 768 for TTS goals) and 3456 configurations for DeltaIoTv2 (1296 for TTO and TSO goals and 864 for TTS goals). 
\added{The average time that was required for \newadded{the offline training of} the deep neural network for a configuration was on 25s for DeltaIoTv1 and 90s for DeltaIoTv2.} 
We observe that the loss for the configuration with TTO goals is significantly lower for DeltaIoTv1. Overall, the loss for DeltaIotv2 is somewhat lower. Yet, as we will show in the following subsection, the differences do not lead to significantly inferior results.

We used the configurations with the best results for the different instances of DLASeR\fplus{} shown in Table~\ref{tab:dlaser+_grid_search} to perform adaptation space reduction for the three configurations with different adaptation goals  (see~\autoref{tab:evaluation-settings}) for both versions of DeltaIoT. 

\subsection{Results online setting}
We present the results for the different settings shown in~\autoref{tab:evaluation-settings}, starting with the metrics for effectiveness, followed by the metrics for efficiency. The results allow us to to answer the different aspects of the research question in a structured manner.

\subsubsection{Effectiveness - Coverage of relevant adaptation options}
To assess the first aspect of the effectiveness of DLASeR\fplus{}, we look at the F1-score for the threshold and set-point goals, and Spearman's rho for the optimization goal. \autoref{tab:effectiveness-metrics}~presents the results.

Note that the result of  DLASeR\fplus{} for the classification task is a continuous value that needs to be rounded to an integer class number. Hence, the classification error can be computed in two ways. The first method  measures the error before rounding, e.g., if the real class of an input data is ``1'' and the predicted class is ``0.49'', the error will be ``$1 - 0.49 = 0.51$''. The second method measures the error after rounding, e.g., if the real class of an input data is ``1'' and the predicted class is ``0.49'', the error will be ``$1 - 0 = 1$'' as ``0.49'' has been fixed to ``0''. Depending on which method is selected for computing the classification error, the F1-score can have different values. 
We use the second method for calculating the classification error, enabling a comparison of the results of DLASeR\fplus{} with ML4EAS, where the output is an integer class number instead of a continuous value. 
}

\begin{table}[]
\begin{tabular}{l|l|l|lll}
\toprule
\textbf{Problem} & \textbf{Setting} & \textbf{Method} & \textbf{F1 threshold} & \textbf{F1 set-point} & \begin{tabular}[c]{@{}l@{}}\textbf{Spearman's rho} \\ \textbf{optimization}\end{tabular} \\ \midrule

\multirow{5}{*}{DeltaIoTv1} &  & DLASeR\fplus{} & 86.92\% & /         & 83.43\% \\
                            & TTO & DLASeR         & 83.92\% & /         & 43.34\% \\
                            &  & ML4EAS         & 75.02\% & /         & /         \\\cmidrule(r){2-6}
                            & TTS & DLASeR\fplus{} & 57.12\% & 33.77\% & /         \\\cmidrule(r){2-6}
                            & TSO & DLASeR\fplus{} & 75.60\% & 43.75\% & 96.49\% \\\midrule

\multirow{5}{*}{DeltaIoTv2} &  & DLASeR\fplus{} & 63.16\% & /         & 21.53\% \\
                            & TTO & DLASeR         & 62.35\% & /         & 5.81\%  \\
                            &  & ML4EAS         & 89.87\% & /         & /         \\\cmidrule(r){2-6}
                            & TTS & DLASeR\fplus{} & 60.48\% & 35.14\% & /         \\\cmidrule(r){2-6}
                            & TSO & DLASeR\fplus{} & 64.19\% & 35.20\% & 95.99\% \\
\bottomrule
\end{tabular}\vspace{5pt}
\caption{Results for the coverage of relevant adaptation options for the different evaluation settings.}
\label{tab:effectiveness-metrics}
\end{table}

\added{
For DeltaIoTv1, we notice an F1-score for the threshold goals of 86.92\% for the setting TTO, 57.12\% for TTS, and 75.60\% for TSO (with 100\% being perfect precision and recall).
The F1-scores for the set-point goal for TTS and TSO are respectively 33.77\% and 43.75\%. 
The differences can be explained by the constraints imposed by the types of goals combined in the different settings, with setpoint goals being most constraining, followed by threshold goals and then optimization goals. Consequently, the F1-score is highest for TTO with two threshold goals, followed by TSO with one threshold goal and one setpoint goal, and finally TTS that combines two threshold goals with a setpoint goal. 
For Spearman's rho, we observe a difference between the settings with an optimization goal, with 83.43\% and 96.49\% for TTO and TSO respectively. This shows that regression is more difficult for energy consumption (TTO) compared to packet loss (TSO). 

For DeltaIoTv2, we observe an F1-score for the threshold goals of 63.16\% for the TTO setting, 60.48\% for TTS, and 64.19\% for TSO. The F1-score for the two set-point goals, are around 35\%. These results for F1-score are slightly lower compared to the results for DeltaIoTv1, indicating that the setting is more challenging for the learners. For Spearman's rho, we observe values of 21.53\% and 95.99\% for TTO and \newadded{TSO} respectively. 
%
%\added{The weak score for the setting with TTO might point to a negative effect caused by the two threshold goals during the training of the core layers on the result of the optimization goal. This may indicate that the knowledge shared in the core layers for the threshold goals and the optimization goal is limited.} 
%
The weak score for the setting with TTO may point to a negative effect on the optimization goal caused by the training of the core layers for the two threshold goals. In particular, this may indicate that the knowledge shared in the core layers for the threshold goals and the optimization goal is limited.}
While this may seem problematic, the evaluation will show that lower Spearman's rho values do not necessarily imply inferior results. 

Overall, we obtained acceptable to excellent results for the F1-score and Spearman's rho (with one exception). Compared to the results obtained for the initial version of DLASeR~\cite{vanapplying} and ML4EAS~\cite{quin2019efficient},
we observe similar results for the TTO setting \newadded{(the results with ML4EAS were somewhat better for DeltaIoTv2, but somewhat worse for DeltaIoTv1)}. However, it is important to emphasize that these approaches do not consider the other combinations of adaptation goals. 

\subsubsection{Effectiveness - Reduction adaptation space}
\autoref{tab:dlaser+-AASR} presents the results for adaptation space reduction obtained with the different approaches for the different configurations. 
We observe that the highest average adaptation space reduction (AASR) with DLASeR\fplus{} is achieved for the TTO setting and this for both IoT instances. The results are slightly better compared to the initial DLASeR but slightly worse compared to ML4EAS. The AASR is lower for the other combinations of goals, which indicates that reducing the adaptation space with DLASeR\fplus{} for settings with a setpoint goal is more challenging. 
On the other hand, for the average analysis effort reduction (AAER), we notice that only a limited number of adaptation options need to be analyzed from the selected subspace before a valid option is found that complies with the classification goals. This is particularly the case for TTS and TSO settings that include a setpoint goal. This means that the adaptation space reductions captured by AASR is of high quality. 

Besides AASR and AAER, we also measured the 
fraction of the adaptation space that was analyzed of the total adaptation space, which combines $AASR$ and $AAER$,\footnote{\textit{The definition of total reduction can be rewritten as: Total Reduction} $ = 100 - (100 - AASR) \times \left(1 - \frac{AAER}{100}\right)$} defined as: 

\begin{equation} 
%\added{
\textit{Total Reduction}\ = (1 - \frac{analyzed}{total}) \times 100 
%\ = 100 - (100 - AASR) \times \left(1 - \frac{AAER}{100}\right)
%}
\end{equation} \vspace{2pt}

For DeltaIoTv1, we measured a total reduction of 95.72\%, \omidadded{99.16\%}, and 96.41\% for TTO, TTS and TSO respectively, while the total reduction for DeltaIoTv2 was 99.52\%, \omidadded{99.94\%}, and 99.98\%. These excellent results show that deep learning with DLASeR\fplus{} is particularly effective in reducing adaptation spaces, i.e., the total reduction is near to the optimum of what can be achieved.

\begin{table}[htbp]
\centering
\begin{tabular}{l|l|l|lll}
\toprule
\textbf{Problem} & \textbf{Setting} & \textbf{Method} & \textbf{AASR} & \textbf{AAER} & \textbf{Total Reduction} \\ \midrule
\multirow{5}{*}{DeltaIoTv1} &  & DLASeR\fplus{} & 56.77\% & 90.11\% & 95.72\% \\
                            & TTO & DLASeR         & 54.84\% & 88.88\% & 94.98\% \\
                            &  & ML4EAS         & 62.61\% & 0.00\%  & 62.61\% \\\cmidrule(r){2-6}
                            & TTS & DLASeR\fplus{} &
                            \omidadded{42.88\%} & \omidadded{98.54\%} & \omidadded{99.16\%} \\\cmidrule(r){2-6}
                            & TSO & DLASeR\fplus{} & 38.27\% & 94.18\% & 96.41\% \\\midrule

\multirow{5}{*}{DeltaIoTv2} &  & DLASeR\fplus{} & 89.03\% & 95.57\% & 99.52\% \\
                            & TTO & DLASeR         & 84.55\% & 72.41\% & 95.73\% \\
                            &  & ML4EAS         & 92.86\% & 0.00\%  & 92.86\% \\\cmidrule(r){2-6}
                            & TTS & DLASeR\fplus{} & \omidadded{41.79\%} & {\omidadded{99.90\%}} & {\omidadded{99.94\%}} \\\cmidrule(r){2-6}
                            & TSO & DLASeR\fplus{} & 51.00\% & 99.95\% & 99.98\% \\\midrule
\bottomrule
\end{tabular}\vspace{5pt}
\caption{Adaptation space reductions %for DLASeR\fplus{} and other approaches 
on the three configurations of Table~\ref{tab:evaluation-settings}.}
\label{tab:dlaser+-AASR}
\end{table}

\begin{figure}[htbp]
\centering
\begin{subfigure}{\textwidth}
    \centering
    \includegraphics[width=0.98\textwidth]{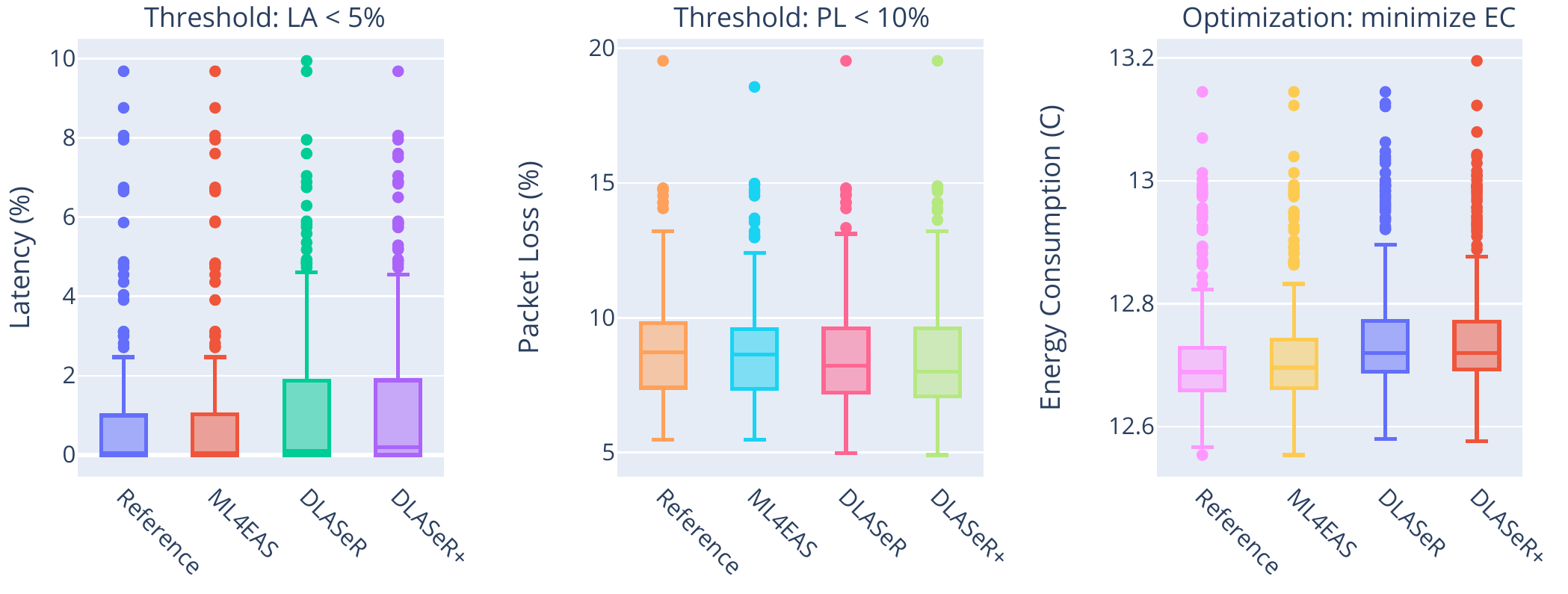}
    \caption{DeltaIoTv1}
\end{subfigure}
%%%%% DeltaIoTv1 TTO %%%%%
%                         min   q1     median      q3    upper_fence      max    
% Latency|  Reference     0.0   0.0     0.0       1.0         2.5         9.7
%           ML4EAS        0.0   0.0     0.0       1.0         2.5         9.7
%           DLASeR        0.0   0.0     0.1       1.9         4.6         9.9
%           DLASeR+       0.0   0.0     0.2       1.9         4.5         9.7

%                         min   q1     median      q3    upper_fence      max    
% Packet|   Reference     5.5   7.4     8.7       9.8         13.2        19.5
% Loss  |   ML4EAS        5.5   7.4     8.6       9.6         12.4        18.6
%           DLASeR        5.0   7.2     8.2       9.6         13.1        19.5
%           DLASeR+       4.9   7.1     8.0       1.9         13.2        19.5

%                         min   q1     median      q3    upper_fence      max    
% Energy|   Reference     12.6  12.7    12.7      12.7        12.8        13.1
% Consum|   ML4EAS        12.6  12.7    12.7      12.7        12.8        13.1
% ption |   DLASeR        12.6  12.7    12.7      12.8        12.9        13.1
%           DLASeR+       12.6  12.7    12.7      12.8        12.9        13.2

%**** In most cases, the lower_fence is equal to the minimum. 
%**** In all cases, the difference between the lower_fence and the minimum is less than 0.2.
%%%%%%%%%%%%%%%%
\begin{subfigure}{\textwidth}
    \centering
    \includegraphics[width=0.98\textwidth]{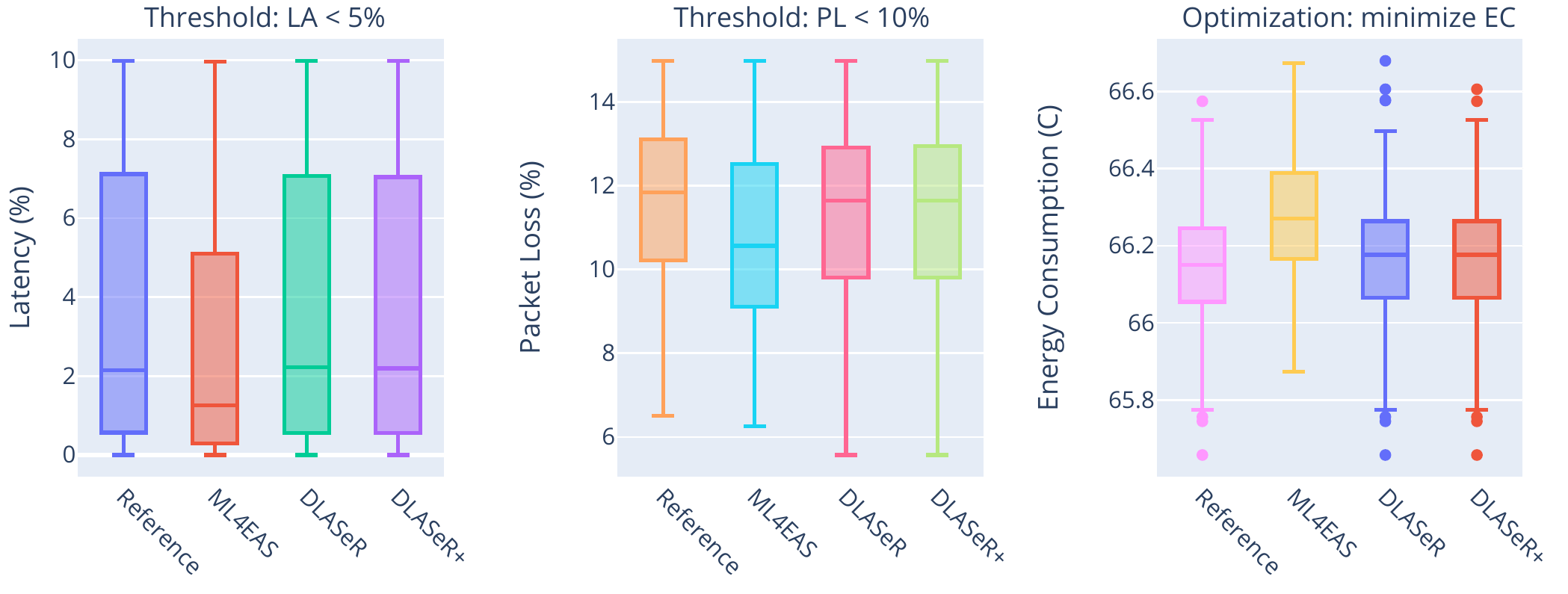}
    \caption{DeltaIoTv2}
\end{subfigure}
%%%%% DeltaIoTv2 TTO %%%%%
%                         min   q1     median      q3    upper_fence      max    
% Latency|  Reference     0.0   0.6     2.2       7.1         10.0        10.0
%           ML4EAS        0.0   0.3     1.3       5.1         10.0        10.0
%           DLASeR        0.0   0.6     2.2       7.1         10.0        10.0
%           DLASeR+       0.0   0.6     2.2       7.0         10.0        10.0

%                         min   q1     median      q3    upper_fence      max    
% Packet|   Reference     6.5   10.2    11.8      13.1        15.0        15.0
% Loss  |   ML4EAS        6.3   9.1     10.6      12.5        15.0        15.0
%           DLASeR        5.7   9.8     11.6      12.9        15.0        15.0
%           DLASeR+       5.6   9.8     11.6      13.0        15.0        15.0

%                         min   q1     median      q3    upper_fence      max    
% Energy|   Reference     65.7  66.1    66.2      66.2        66.5        66.6
% Consum|   ML4EAS        65.9  66.2    66.3      66.4        66.7        66.7
% ption |   DLASeR        65.7  66.1    66.2      66.3        66.5        66.7
%           DLASeR+       65.7  66.1    66.2      66.3        66.5        66.6

%**** In most cases, the lower_fence is equal to the minimum. 
%**** In all cases, the difference between the lower_fence and the minimum is less than 0.2.
%%%%%%%%%%%%%%%%
\caption{Effect on the realization of the adaptation goals for the TTO setting.}
\label{fig:dlaser+-TTO}
\end{figure}

\begin{figure}[htbp]
\centering
\begin{subfigure}{\textwidth}    
    \centering
    \includegraphics[width=\textwidth]{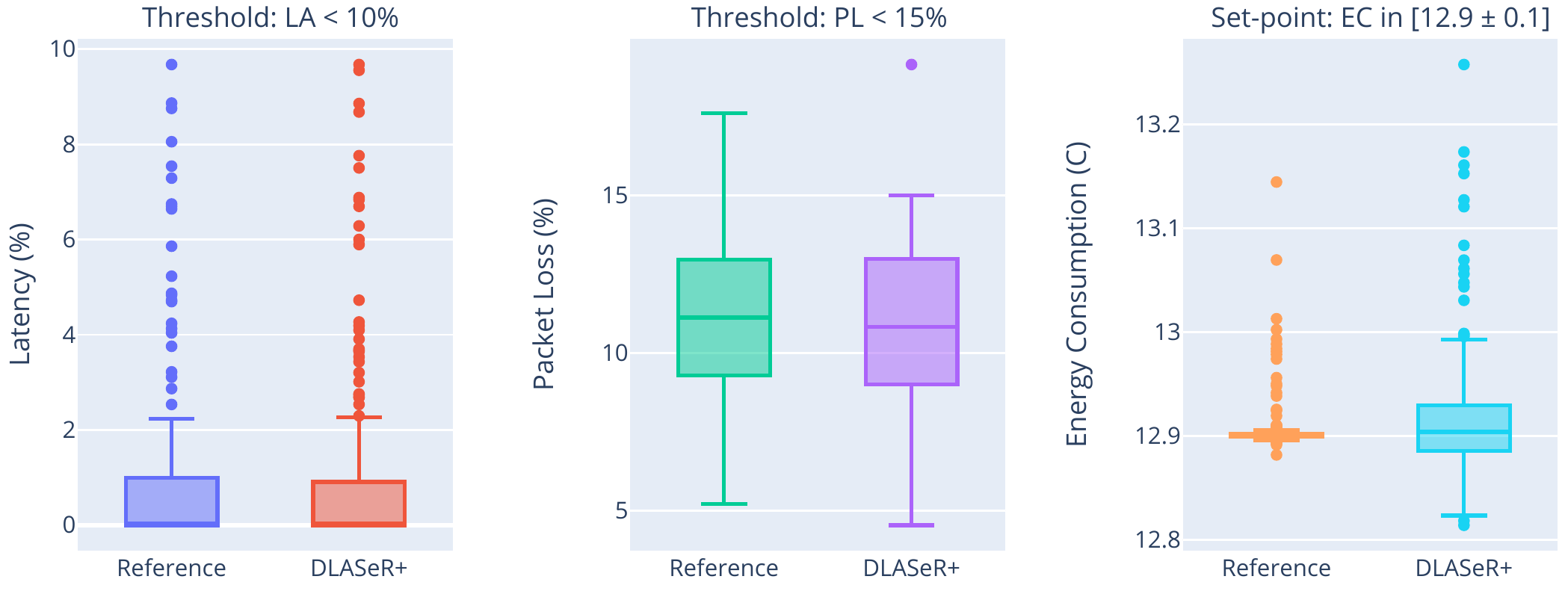}
    \caption{DeltaIoTv1}
\end{subfigure}
%%%%% DeltaIoTv1 TTS %%%%%
%                         min   q1     median      q3    upper_fence      max    
% Latency|  Reference     0.0   0.0     0.0       1.0         2.2         9.7
%           DLASeR+       0.0   0.0     0.0       0.9         2.3         9.7

%                         min   q1     median      q3    upper_fence      max    
% Packet|   Reference     5.2   9.3     11.1      13.0        17.6        17.6
% Loss  |   DLASeR+       4.5   9.0     10.8      13.0        15.0        19.1

%                         min   q1     median      q3    upper_fence      max    
% Energy|   Reference     12.9  12.9    12.9      12.9        12.9        13.1
% Consum|   DLASeR+       12.8  12.9    12.9      12.9        13.0        13.3
% ption |   

%**** In most cases, the lower_fence is equal to the minimum. 
%**** In all cases, the difference between the lower_fence and the minimum is less than 0.5.
%%%%%%%%%%%%%%%%
\begin{subfigure}{\textwidth}
    \centering
    \includegraphics[width=0.95\textwidth]{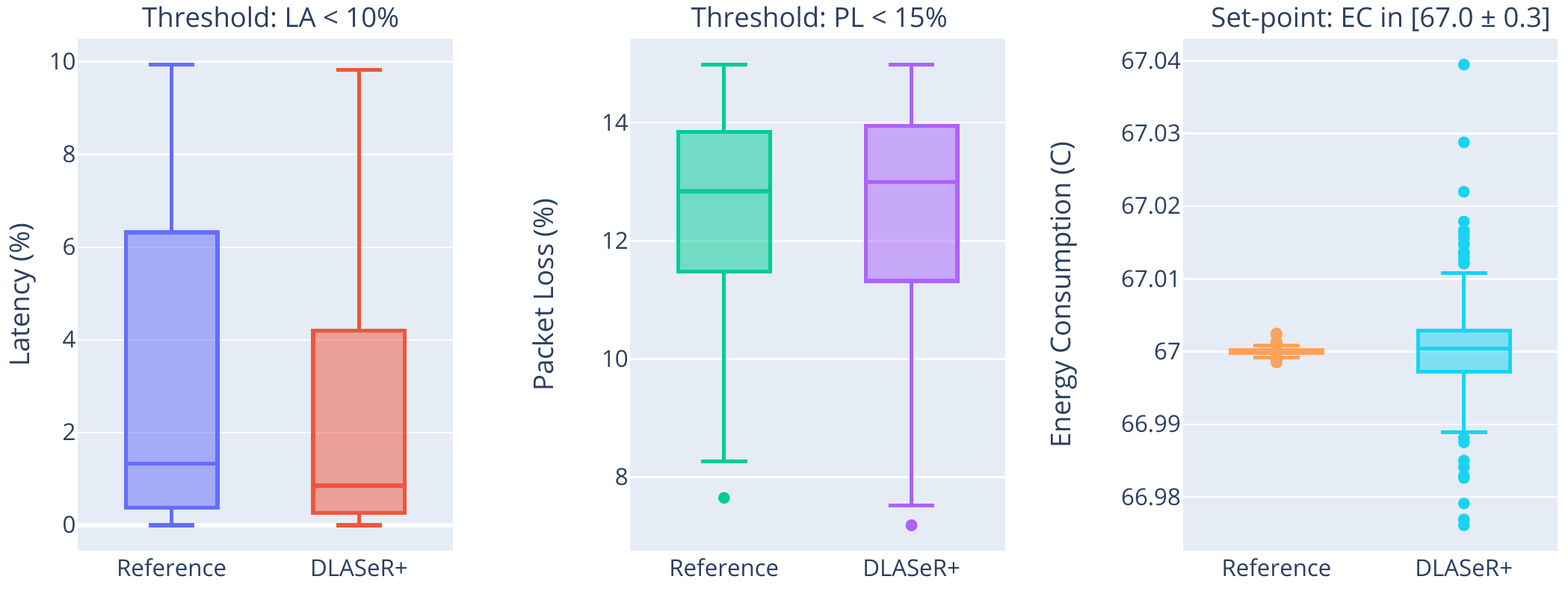}
    \caption{DeltaIoTv2}
\end{subfigure}
%%%%% DeltaIoTv2 TTS %%%%%
%                         min   q1     median      q3    upper_fence      max    
% Latency|  Reference     0.0   0.4     1.3       6.3         9.9         9.9
%           DLASeR+       0.0   0.3     0.9       4.2         9.8         9.8

%                         min   q1     median      q3    upper_fence      max    
% Packet|   Reference     7.6   11.5    12.8      13.8        15.0        15.0
% Loss  |   DLASeR+       7.2   11.3    13.0      13.9        15.0        15.0

%                         min   q1     median      q3    upper_fence      max    
% Energy|   Reference     67.0  67.0    67.0      67.0        67.0        67.0
% Consum|   DLASeR+       67.0  67.0    67.0      67.0        67.0        67.0
% ption |   

%**** In most cases, the lower_fence is equal to the minimum. 
%**** In all cases, the difference between the lower_fence and the minimum is less than 0.2.
%%%%%%%%%%%%%%%%
\caption{Effect on the realization of the adaptation goals for the TTS setting.}
\label{fig:dlaser+-TTS}
\end{figure}

\begin{figure}[htbp]
\centering
\begin{subfigure}{\textwidth}
    \centering
    \includegraphics[width=0.95\textwidth]{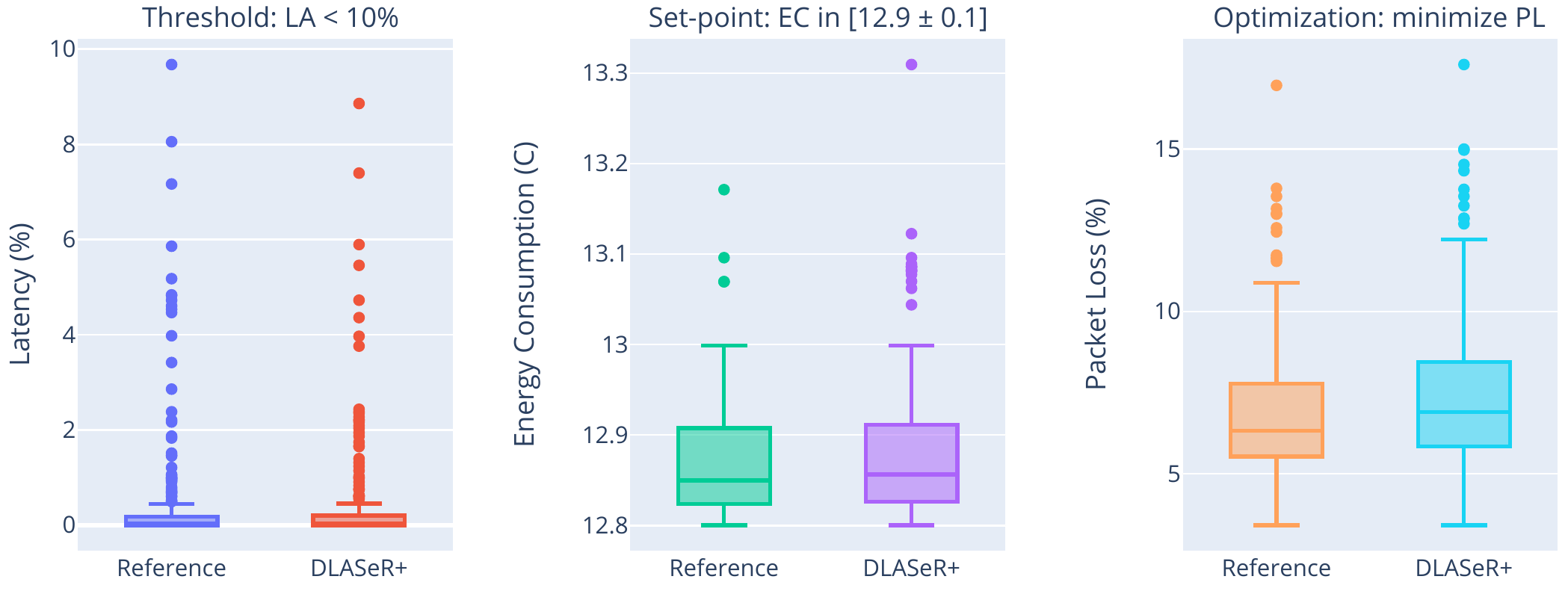}
    \caption{DeltaIoTv1}
\end{subfigure}
%%%%% DeltaIoTv1 TSO %%%%%
%                         min   q1     median      q3    upper_fence      max    
% Latency|  Reference     0.0   0.0     0.0       0.2         0.4         9.7
%           DLASeR+       0.0   0.0     0.0       0.2         0.5         8.6

%                         min   q1     median      q3    upper_fence      max    
% Energy|   Reference     12.8  12.8     12.8      12.9       13.0        13.2
% Consum|   DLASeR+       12.8  12.8     12.9      12.9       13.0        13.3
%ption  |

%                         min   q1     median      q3    upper_fence      max    
% Packet|   Reference     3.4   5.5     6.3        7.8        10.9        17.0
% loss  |   DLASeR+       3.4   5.8     6.9        8.4        12.2        17.6

%**** In most cases, the lower_fence is equal to the minimum. 
%**** In all cases, the difference between the lower_fence and the minimum is less than 0.2.
%%%%%%%%%%%%%%%%
\begin{subfigure}{\textwidth}
    \centering
    \includegraphics[width=0.95\textwidth]{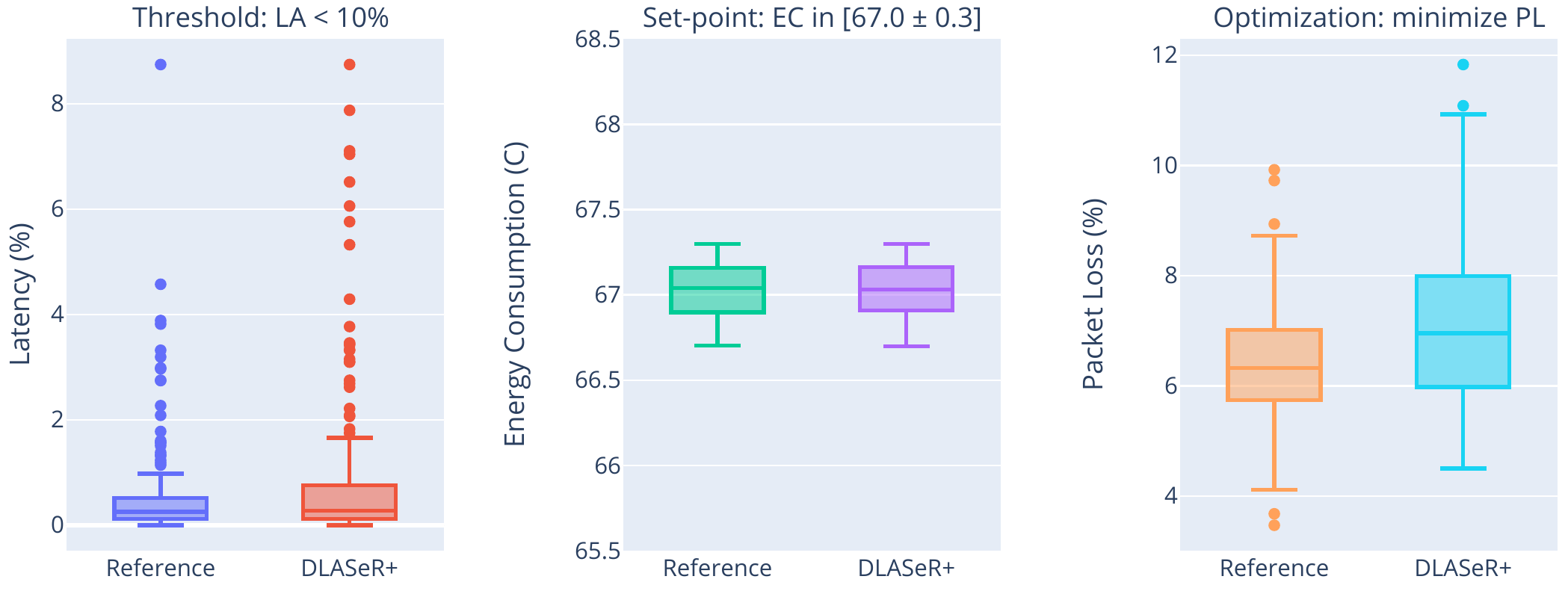}
    \caption{DeltaIoTv2}
\end{subfigure}
%%%%% DeltaIoTv2 TSO %%%%%
%                         min   q1     median      q3    upper_fence      max    
% Latency|  Reference     0.0   0.1     0.3       0.5         1.0         8.7
%           DLASeR+       0.0   0.1     0.3       0.8         1.7         8.7

%                         min   q1     median      q3    upper_fence      max    
% Energy|   Reference     66.7  66.9     67.0      67.2       67.3        67.3
% Consum|   DLASeR+       66.7  66.9     67.0      67.2       67.3        67.3
%ption  |

%                         min   q1     median      q3    upper_fence      max    
% Packet|   Reference     3.5   5.7     6.3        7.0        8.7         9.9
% loss  |   DLASeR+       4.5   6.0     7.0        8.0        11.0        11.8

%**** In most cases, the lower_fence is equal to the minimum. 
%**** In all cases, the difference between the lower_fence and the minimum is less than 0.2.
%%%%%%%%%%%%%%%%
\caption{Effect on the realization of the adaptation goals for the TSO setting.}
\label{fig:dlaser+-TSO}
\end{figure}

\added{
\subsubsection{Effectiveness - Effect on realization of adaptation goals}
To evaluate the effectiveness, we compare the median values of the quality properties that correspond to the adaptation goals over 230 learning cycles (i.e., representing about three days of operation of the IoT networks, see evaluation setup). \newadded{Note that a threshold goal is satisfied if the median of the values over 230 cycles satisfies the goal. This does not necessarily mean that the system satisfies the goal in all cycles.}
%This implies that the system may not respect a goal in some cycles, it will satisfy that goal if the median of the values over 230 cycles satisfies the goal.

The boxplots of Figures~\ref{fig:dlaser+-TTO},~\ref{fig:dlaser+-TTS},~and~\ref{fig:dlaser+-TSO} show the results for the quality properties of the adaptation goals with DLASeR\fplus{} and other approaches. It is important to note that the reference approach exhaustively analyzes the whole adaptation space. This is the ideal case, but practically not always feasible (as in DeltaIoTv2) due to time constraints on the time available to perform adaptation. 

For the setting with TTO goals, see~\autoref{fig:dlaser+-TTO}, the results for packet loss and latency (threshold goals) are similar for all approaches. We observe that DLASeR\fplus{} always satisfies the goals (i.e., all median values are below the  thresholds). For some adaptation cycles, no  configurations are available that satisfy the threshold goals (as shown by the reference approach that exhaustively searches through the complete adaptation space). For energy consumption (optimization goal), the results for DLASeR\fplus{} are slightly higher compared to the reference approach, i.e., an increase of 0.03~C (0.24\%) for DeltaIoTv1 (median of 12.69~C for the reference approach; 12.72~C for DLASeR\fplus{}), and 0.03~C (0.05\%) for DeltaIoTv2 (66.15~C for the reference approach; 66.18~C for DLASeR\fplus{}). 

For the setting with TTS goals, see~\autoref{fig:dlaser+-TTS}, we observe similar results for DLASeR\fplus{} and the reference approach. The threshold goals for packet loss and latency are again always satisfied by DLASeR\fplus{}. The set-point goal of energy consumption is also satisfied for both instances of DeltaIoT \newadded{(DLASeR\fplus{} shows  slightly more variability around the set-point)}.

Finally, for the setting with TSO goals, see~\autoref{fig:dlaser+-TSO}, the results show that the threshold goal for latency is always satisfied with DLASeR\fplus{} and the same applies for the set-point goal for energy consumption. As for the optimization goal, the packet loss of DLASeR\fplus{} is slightly higher compared to the reference approach, respectively 0.57\% for DeltaIoTv1 (median of 6.33\% for the reference approach versus 6.90\% for DLASeR\fplus{}), and 0.62\% for DeltaIoTv2 (6.33\% for the reference approach versus 6.95\% for DLASeR\fplus{}). 

In summary, the results for the quality properties using DLASeR\fplus{} and the reference approach are similar. For settings with two threshold goals and one optimization goal, the results are similar as for two state-of-the-art approaches, the initial DLASeR~\cite{vanapplying} and ML4EAS~\cite{quin2019efficient}. However, contrary to these state-of-the-art approaches, DLASeR\fplus{} realizes adaptation space reduction for combinations of threshold, set-point goals, and  optimization goals with a small to negligible effect on the qualities compared to the reference approach. 
}

\added{
\subsubsection{Efficiency - Learning time}
\autoref{tab:learning_time} presents the results for the analysis time, learning time, and overall time reduction of DLASeR\fplus{} compared to the reference approach. As we can see, the major part of the time is spent on analysis, i.e., the time used for the verification of adaptation options. This is in contrast with the learning time;\footnote{\newadded{Recall that the learning time is the sum of the time used for online prediction and online training, cf. Table~\ref{tab:metrics-research-question}.}} on average 16.45\% of the time is spent on learning for DeltaIoTv1 (total time for verification of the three settings is 4.45s versus 0.894s for learning) and 1.56\% for DeltaIoTv2 (102.95s in total for verification of the three settings versus 1.63s for learning). 
We observe high numbers for time reduction, on average 92.80\% for the three settings of DeltaIoTv1 and 94.84\% for the settings of DeltaIoTv2. For TTO settings, DLASeR\fplus{} realizes better time reductions than the other approaches. Note that the optimal time reduction is 95\%, since we use an exploration rate of 5\%, i.e., in each adaptation cycle, 5\% of the adaptation space is explored.

In sum, the learning time of DLASeR\fplus{} is only a fraction of the total time used for analysis. DLASeR\fplus{} realizes an impressive reduction of the time required for analysis and decision-making compared to the reference approach.  
}

\begin{table}[htbp]
\centering
\begin{tabular}{l|l|l|lll}
\toprule
\textbf{Problem} & \textbf{Setting} & \textbf{Method} & \textbf{Verification Time} & \textbf{Learning Time} & \textbf{Time Reduction} \\ \midrule
\multirow{5}{*}{DeltaIoTv1} &  & DLASeR\fplus{} & 1.95 s  & 0.004 s & 90.72\% \\
                            & TTO & DLASeR         & 2.13 s  & 0.003 s & 89.87\% \\
                            &  & ML4EAS         & 6.99 s  & 0.002 s & 66.80\% \\\cmidrule(r){2-6}
                            & TTS & DLASeR\fplus{} & 1.11 s  & 0.45 s & 94.73\% \\\cmidrule(r){2-6}
                            & TSO & DLASeR\fplus{} & 1.48 s  & 0.44 s & 92.96\% \\\midrule

\multirow{5}{*}{DeltaIoTv2} &  & DLASeR\fplus{} & 36.44 s & 0.45 s & 94.52\% \\
                            & TTO & DLASeR         & 61.21 s & 1.62 s & 90.79\% \\
                            &  & ML4EAS         & 68.44 s & 0.04 s & 89.71\% \\\cmidrule(r){2-6}
                            & TTS & DLASeR\fplus{} & 33.27 s & 0.63 s & 95.00\% \\\cmidrule(r){2-6}
                            & TSO & DLASeR\fplus{} & 33.24 s & 0.55 s & 95.00\% \\\midrule
\bottomrule
\end{tabular}\vspace{5pt}
\caption{Verification and learning time for DLASeR\fplus{}. Time reduction compares the total time used by DLASeR\fplus{} compared to the time used by the reference approach that verifies the complete adaptation space.}
\label{tab:learning_time}
\end{table}

\added{

\subsubsection{Efficiency - Scalability}
To evaluate the scalability of DLASeR\fplus{}, we discuss the difference for the metrics when scaling up the evaluation setting from DeltaIoTv1 with 216 adaptation options to DeltaIoTv2 with 4096 adaptation options; an increase of adaptation space with a factor around 20.
 
For the \textit{coverage of relevant adaptation options} (see~\autoref{tab:effectiveness-metrics}), we observe a decrease in F1-score for the threshold goals of 23.76\%  for the TTO setting (86.92\% for DeltaIoTv1 versus 63.16\% for DeltaIoTv2), and a decrease of 11.41\% for the TSO setting (75.60\% versus 64.18\%). For the TTS settings we notice a small increase of the F1-score for the threshold goals of 3.36\% (57.12\% versus 60.48\%). For set-point goals, we observe a small increase in F1-score of 1.37\% for the setting TTS (33.77\% versus 35.14\%), and a small decrease of 8.55\% for the setting TSO (43.75\% versus 35.20\%). The results show that DeltaIoTv2 is more challenging, but DLASeR\fplus{} scales well for threshold goals. For the optimization goal, we notice a large decrease for Spearman's rho with 61.9\% for the setting with TTO goals (from 83.43\% for DeltaIoTv1 to 21.53\% for DeltaIoTv2), and a small decrease of 0.5\% for the setting with TSO goals (from 96.49\% to 95.99\%). These results suggest that the optimization goal in TTO (energy consumption) might be significantly harder to predict. 

\newadded{For the \textit{average adaptation space reduction} (see Table~\ref{tab:dlaser+-AASR}), we measured a  substantial increase for two configurations and a small decrease for one configuration when scaling up from DeltaIoTv1 to DeltaIoTv2. On average, we measured an increase of 14.63\% over the three configurations (from 56.77\% to 89.03\% for TTO %%
and from 38.27\% to 51.00\% for TSO; from
%38.53\% to 42.26\% for TTS, 
42.88\% to 41.79\% for TTS).} 
%+32,26 +12,73 -1,09
%%
The \textit{average analysis effort reduction} also improved for DeltaIoTv2, with an average of 4.20\% (from 90.11\% to 95.57\% for TTO,
\newadded{from 98.54\% to 99.90\% for TTS, and 
from 94.18\% to 99.95\% for TSO.
%5,46 1,36 5,77
This resulted in an average increase of 2.72\% for the \textit{total reduction} (from 95.72\% to 99.52\% for TTO, from }
\omidadded{99.16\%} to \omidadded{99.94\%} for TTS, and from 96.41\% to 99.98\% for TSO).  
%3, 8 0,78 3,57

For the \textit{effect on realizing the adaptation goals} (see Figure~\ref{fig:dlaser+-TTO},~\ref{fig:dlaser+-TTS},~and~\ref{fig:dlaser+-TSO}), we observe that DLASeR\fplus{} realizes the threshold and set-point goals for all configurations of both DeltaIoTv1 and DeltaIoTv2. For the optimization goal compared to the reference approach, we observe a slight increase of energy consumption for the TTO setting with DLASeR\fplus{} from 0.24\% extra with DeltaIoTv1 (medians 12.69~C versus 12.72~C) to 0.05\% extra with DeltaIoTv2 (medians 66.15~C versus 66.18~C). Similarly, we observe a small increase of packet loss for TSO with DLASeR\fplus{} from 0.57\% extra with DeltaIoTv1 (medians 6.33\% versus 6.90\%) to 0.62\% extra with  DeltaIoTv2 (medians 6.33\% versus 6.95\%).

Finally, the results for the \textit{learning time} (see~\autoref{tab:learning_time}) show that relative part of the time required for learning compared to the time required for analysis decreases from 16.45\% for DeltaIoTv1 to 1.56\% for DeltaIoTv2, while the \textit{total time reduction} improves from 92.80\% to 94.84\% (the numbers are averages over the three settings TTO, TTS, and TSO). The results show that with increasing scale, the total time reduction improves substantially and gets close to optimum for DeltaIoTv2. 

In summary, while some of the indicators for the coverage of relevant adaptation options are slightly inferior for configurations with larger adaptation spaces, the other metrics show no negative effects on the effectiveness and efficiency of DLASeR\fplus{} when the size of the adaptation space is increased (in the evaluating setting with a factor of about 20). 
}

\subsection{Threats to validity} 

The evaluation results show that DLASeR\fplus{} is an effective and efficient approach to reduce large adaptation spaces. 
%The approach realizes the threshold and set-point goals for all the configurations we tested on the DeltaIoT artifact. Compared to the ideal approach, we observe only a a small tradeoff for the quality property of the optimization goal. Yet this is a small cost for the dramatic improvement of adaptation time. For DeltaIoTv2, DLASeR requires on average 50.53 seconds for verification plus 0.92 seconds for online learning training, compared to 16.73 minutes with the reference approach (which is in practice infeasible since it exceeds the cycle time of 9.5 minutes). 
However, the evaluation of DLASeR\fplus{} is subject to a number of validity threats. 

\vspace{5pt}\noindent
\textit{External validity}. Since we evaluated the approach in only one domain \added{with a particular adaptation spaces and adaptation goals, we cannot generalize the conclusions, including the configuration of DLASeR\fplus{}, and the effectiveness and efficiency of adaptation space reduction.} We mitigated this threat to some extent by applying DLASeR\fplus{} to two IoT networks that differ in their topology and size of adaptation space. Nevertheless, more extensive evaluation is required in different domains to strengthen the validity of the results. 
Furthermore, the reduction stages of the unified architecture of DLASeR\fplus{} presented in this paper target only a single optimization goal. Hence, the approach is not directly applicable to systems with multi-objective optimization goals. Additional research will be required to extend DLASeR\fplus{} for systems with such types of goals. We also tested DLASeR\fplus{} for scenarios with up to 4000 adaptation options in one domain. Further research is required to study and evaluate other types of systems with much larger adaptation spaces.  

\vspace{5pt}\noindent
\textit{Internal validity}. We evaluated the effectiveness and efficiency of DLASeR\fplus{} using different metrics. It might be possible that the specifics of the evaluation setting of the applications that we used, in particular the topology of the network, the uncertainties, and the choices for the specific goals that we considered may have an effect on the complexity of the problem of adaptation space reduction. To mitigate this threat to some extent we applied DLASeR\fplus{} to simulated settings of real IoT deployments that were developed in close collaboration with an industry partner in IoT.

\vspace{5pt}\noindent
\textit{Reliability}. For practical reasons, we performed the evaluation of DLASeR\fplus{} in simulation. The uncertainties used in this simulation are based on stochastic models. This may cause a threat that the results may be different if the study would be repeated. We minimized this threat in three ways: (i) the profiles are designed based on field tests, (ii) we evaluated DLASeR\fplus{} over a period of several days, and (iii) the complete package we used for evaluation is available to replicate the study.\footnote{DLASeR website: \url{https://people.cs.kuleuven.be/danny.weyns/software/DLASeR/index.html}}

\section{Related work}\label{sec:relatedwork}

Over the past years, we can observe an increasing interest in using machine learning and related techniques to support self-adaptation~\cite{gheibi2021}. Three examples of initial work are Richert W. et al.~\cite{1370039} that apply reinforcement learning to robots that learn to select tasks that balance their own benefit with the needs of other robots, Sykes et al.~\cite{sykes2013learning} that use a probabilistic rule learning approach to update the environment models of a self-adaptive system at runtime relying on execution traces, and Bencomo et al.~\cite{6615198} that use 
dynamic decision networks to model and support decision-making in self-adaptive systems explicitly taking into account uncertainty. 
%Bijdrage: Maintain updated environment models.
%Verschil: Instead of guiding or adaptation decisions based on environment models, we focus on effective and efficient analysis at runtime. They do not consider scalability of their approach.

%Belangrijk werk dat ML toepast in SAS related to decision-making

%Link: https://people.cs.kuleuven.be/~danny.weyns/papers/2019SEAMSb.pdf

In this section, we focus on the use of  machine learning and other related techniques that are used for adaptation space reduction and efficient decision-making of systems with complex adaptation spaces. We have structured the related work in four  groups based on their main focus: (1) learning used for the reduction of adaptation spaces, (2) learning used for efficient decision-making, (3) search-based techniques to efficiently explore large adaptation spaces, and finally (4) approaches for efficient verification. For each group we present a representative selection of related work. 

\subsection{Learning used for adaptation space reduction}

%Link: https://dl.acm.org/doi/pdf/10.1145/1882291.1882296
Elkhodary et al.~\cite{elkhodary2010fusion} propose the FUSION framework that learns the impact of adaptation decisions on the system's goals. In particular, FUSION utilizes M5 decision trees to learn the utility functions that are associated with the qualities of the system. The results show a significant improvement in analysis. Whereas DLASeR\fplus{} targets the adaptation space, FUSION targets the feature selection space, focusing on proactive latency-aware adaptations relying on a separate model for each utility. 

%Link: https://ieeexplore.ieee.org/stamp/stamp.jsp?arnumber=7572219
Chen et al.~\cite{chen2016self} study feature selection and show that different learning algorithms perform significantly different depending on the types of quality of service attributes considered the way they fluctuate. The work is centered on an adaptive multi-learners technique that  dynamically selects the best learning algorithms at runtime. Similar to our work, the authors focus on efficiency and effectiveness, but the scope of that work is on the features instead of adaptation options.  

Quin et al.~\cite{quin2019efficient} apply classical machine learning techniques, in particular classification and regression, to reduce large adaptation spaces. These techniques require domain expertise to perform feature engineering, which is not required in DLASeR\fplus{} (that only requires model selection). That work also only considers threshold goals based on linear learning models. In contract, our work considers threshold, optimization, and set-point goals based on non-linear deep learning models. 

%Link: http://acme.able.cs.cmu.edu/pubs/uploads/pdf/SEAMS2019-ML.pdf
Jamshidi et al.~\cite{jamshidi2019machine} presents an approach that learns a set of Pareto optimal configurations offline that are then used during operation to generate adaptation plans. The approach reduces adaptation spaces, while the system can still  apply model checking with PRISM~\cite{3-540-46029-2_13} at runtime to quantitatively reason about adaptation decisions. Compared to that work, DLASeR\fplus{} is more versatile by reducing the adaptation space at runtime in a dynamic and learnable fashion.

%Link: https://arxiv.org/pdf/1907.09158.pdf
Metzger et al.~\cite{metzger2019feature} apply 
online learning to explore the adaptation space of self-adaptive systems using feature models.
The authors demonstrate a speedup in convergence of the online learning process. 
The approach is centered on the adaptation of 
rules, whereas DLASeR\fplus{} is centered on  model-based self-adaptation. Furthermore, the work also looks at the evolution of the adaptation space, while DLASeR\fplus{} only considers dynamics in the adaptation space (not its evolution).

%Link: https://bit.ly/36L3zC7
Camara et al.~\cite{camara2020quantitative}
use reinforcement learning to select an adaptation pattern relying on two long-short-term memory deep learning models. Similar to our work, these authors demonstrate the benefits of integrating machine learning with runtime verification. However, the focus differs in the type of goals considered (they only consider threshold goals) and the type of verification used (they use runtime quantitative verification); that work also does not consider scalability.

%Link: http://cse.unl.edu/~hbagheri/publications/2020ICSE.pdf
%Specifiek werk rond adaptation space reduction
Stevens et al.~\cite{stevens2020reducing} present Thallium that exploits a combination of automated formal modeling techniques to significantly reduce the number of states that need to be considered with each adaptation decision. Thallium addresses the adaptation state explosion by applying utility bounds analysis. The (current) solution operates on a Markov decision process that represents the structure of the system itself, independent of the stochastic environment. \added{The authors suggest future work in combining learning-based approaches employed on the reduced adaptation space from Thallium.} 

%DLASeR\fplus{} on the other hand focuses on effective and efficient reduction of adaptation spaces that rely on hybrid quality models taking into account the system and uncertainties of its environment. 

%\subsection{Learning used for the reduction of features}

%mRMR methods for feature selection in performance modelling 

\subsection{Learning used for efficient decision-making}

%Link: https://darkrsw.net/papers/SEAMS2009.pdf
Kim et al.~\cite{kim2009reinforcement} present 
a reinforcement learning-based approach that enables a software system to improve its behavior by learning the results of its behavior and dynamically changing its plans under environmental changes. Compared to DLASeR\fplus{}, the focus is on effective adaptation decision-making, without considering guarantees or the scalability of the proposed approach. 

%Link: https://dl.acm.org/doi/pdf/10.1145/2593929.2593941
Anaya et al.~\cite{anaya2014prediction} present a framework for proactive adaptation that uses predictive models and historical information to model the environment. These predictions are then fed to a reasoning engine to improve the decision-making process. The authors show that their approach outperforms a typical reactive system by evaluating different prediction models (classifiers). Whereas they focus on leveraging proactive techniques to make decision-making more effective, our work  focuses on both the efficiency and effectiveness of adaptation space reduction to improve the decision-making.  

%Specifiek werk dat RL toepast in self-adaptive systems

%Link: https://link.springer.com/article/10.1007/s00766-015-0227-1
Qian et al.~\cite{qian2015rationalism} study goal-driven self-adaptation centered on case-based reasoning for storing and retrieving adaptation rules. Depending on requirements violations and context changes, similar cases are used and if they are not available, goal reasoning is applied. This way, the approach realizes more precise adaptation decisions. The evaluation is done only for threshold goals and the authors only provide some hints to scale up their solutions to large-sized systems. In our work, we explicitly evaluate the effect of scaling up the adaptation space.  

%Link: https://dl.acm.org/doi/pdf/10.1145/2701126.2701191
Nguyen Ho et al.~\cite{ho2015model} rely on model-based reinforcement learning to improve system performance. By utilizing engineering knowledge, the system maintains a model of interaction with its environment and predicts the consequence of its action. DLASeR\fplus{} relies on a different learning techniques. Furthermore, we study the effect of scalability of the adaptation space, which is not done in that paper. 
\subsection{Search-based techniques to explore large adaptation spaces}

%Search based techniques

%https://ieeexplore.ieee.org/stamp/stamp.jsp?tp=&arnumber=6604427

\added{Cheng et al.~\cite{6604427} argue for three overarching techniques that are essential to
address uncertainty in software systems:  model-based development, assurance, and dynamic adaptation. In relation to the work presented in this paper, the authors argue for the application of search-based software engineering
techniques to model-based development, in particular, the use of evolutionary algorithms to support an adaptive system to self-reconfigure safely. In~\cite{ramirez10}, the authors propose Hermes, a genetic algorithmic approach, that adapts the system efficiently in time. In~\cite{ramirez09}, the authors propose Plato, an approach that maps data monitored from the system or the environment into genetic schemes and evolves the system by leveraging genetic algorithms. The main aim of these approaches is ensuring safety under uncertainty in an efficient manner. In contrast, DLASeR\fplus{} is  conceptually different, relying on deep learning to explicitly reduce large adaptation spaces, providing explicit support for different types of adaptation goals.

Le Goues et al.~\cite{6035728} propose GenProg, an automated method for repairing defects in legacy programs. GenProg relies on genetic programming to evolve a program variant that retains required functionality but is not susceptible to a given defect, using existing test suites to encode both the defect and required functionality. The focus of this work is on efficiently producing evolved programs that repair a defect, without introducing substantial degradation in functionality. The focus of DLASeR\fplus{} on the other hand is on reducing large adaptation spaces at architectural level aiming to enhance the efficiency of the decision-making of self-adaptive systems that need to deal with different types of quality properties. 
}

%Link: https://www.infosun.fim.uni-passau.de/publications/docs/NYM+18tse.pdf
Nair et al.~\cite{8469102} present FLASH that aims at efficiently finding good configurations of a software system. FLASH  sequentially explores the configuration space by reflecting on the configurations evaluated so far to determine the next best configuration to explore. FLASH can solve both single-objective and multi-objective optimization problems. Whereas FLASH assumes that the system is stationary, 
%. This could be addressed by building a new model at a specified time interval. Whereas 
DLASeR\fplus{} uses incremental learning to stay up to date during operation; i.e., DLASeR\fplus{} deals with dynamics in the environment at runtime. 

%Link: https://dl.acm.org/doi/pdf/10.1145/3194133.3194145
Kinneer et al.~\cite{kinneer2018managing} propose a planner based on genetic programming that reuses existing plans. Their approach uses stochastic search to deal with unexpected adaptation strategies, specifically by reusing or building upon prior knowledge. Their genetic programming planner is able to handle very large search spaces.
Similar to DLASeR\fplus{}, the evaluation of this work considers efficiency and
effectiveness. However, the technique used is different focusing on planning and that work put particular emphasis on reuse. 

%Link: https://dl.acm.org/doi/pdf/10.1145/3204459
Chen et al.~\cite{chen2018femosaa} present FEMOSAA, a framework that leverages a feature model and a multi-objective evolutionary algorithm to optimize the decision-making of adaptation at runtime. The authors show that FEMOSAA produces statistically better and more balanced results for tradeoff with reasonable overhead compared to other search-based techniques. Compared to DLASeR\fplus{}, the authors use a different analysis technique and rely on feature models. The latter implies that the approach relies on domain engineers to construct a feature model for the self-adaptive system.

\added{
In~\cite{Coker15}, Coker et al. use genetic programming planning and combine this with probabilistic model checking to determine the fitness of plans for a set of quality properties. The proposed search-based approach provides an integrated solution for guiding the decision-making of a self-adaptive system. This approach requires a well-defined objective function. In contrast, 
DLASeR\fplus{} focuses on the reduction of the adaptation space for different types of adaptation goals. With DLASeR\fplus{}, different types of decision-making mechanisms can be combined. 

Pascual et al.~\cite{Pascual13} apply a genetic algorithm to generate automatically at runtime configurations for adapting a system together with
reconfiguration plans. The generated configurations are optimal in terms of  functionality taking into account the available resources (e.g. battery). Concretely, the configurations are defined as variations of the application's software architecture based on a so called feature model. In contrast, DLASeR\fplus{} targets the reduction of large adaptation spaces targeting quality properties of the system that are formulated as adaptation goals. 

}

\subsection{Approaches for efficient verification}

%Link: https://dl.acm.org/doi/10.1145/1985793.1985840
Filieri et al.~\cite{filieri2011}  present a mathematical framework for efficient run-time probabilistic model checking. Before deployment, a set of symbolic expressions that represent satisfaction of the requirements is pre-computed. At runtime, the verification step simply evaluates the formulae by replacing the variables with the real values gathered by monitoring the system. By shifting the cost of model analysis partially to design time, the approach enables more efficient verification at runtime. \added{In later work~\cite{filieri16}, the authors elaborate on this and explain how the mathematical framework supports reasoning about the effects of changes and can drive effective adaptation strategies. 
} Whereas DLASeR\fplus{} focuses on reducing the set of adaptation options during operation, their work focuses on efficient runtime verification by offloading work before system deployment. 

%Link: https://dl.acm.org/doi/10.1145/2593929.2593932
Gerasimou et al.~\cite{gerasimou2014} propose three techniques to speed up runtime quantitative verification, namely caching, lookahead and nearly-optimal reconfiguration. The authors evaluate several combinations of the techniques on various scenarios of self-adaptive systems. The focus of this work is different from DLASeR\fplus{}, but the proposed techniques are complementary and can perfectly be integrated in our work.   

%Specifiek werk rond snellere verificatie

%Link: https://dl.acm.org/doi/pdf/10.1145/3149180
Moreno et al.~\cite{moreno2018flexible}  present an approach for proactive latency-aware adaptation that relies on stochastic dynamic programming to enable more efficient decision-making. Experimental results show that this approach is close to an order of magnitude faster than runtime probabilistic model checking to make adaptation decisions, while preserving the same effectiveness. Whereas our approach focus on reducing the set of adaptation options to improve analysis, their work focuses on fast verification; here too, a system may benefit from a combination of both approaches. 

\added{

Goldsby et al.~\cite{Goldsby2008} and Zhang et al.~\cite{zhang2009modular} present AMOEBA-RT, a run-time approach that provides assurance that  dynamically adaptive systems satisfy their requirements. In AMOEBA-RT, an adaptive program is instrumented with aspects that non-invasively collect state of the system that can then be checked against a set of adaptation properties specified in A-LTL, an extended linear temporal logic. At run-time, the instrumented code sends the collected state information to a runtime model checking server that determines whether the state of the adaptive program satisfies the adaptation properties. The focus of this work is on assuring properties using runtime model checking. In contrast, DLASeR\fplus{} focuses on adaptation space reduction. AMOEBA-RT can be used in tandem with DLASeR\fplus{} to enhance the efficiency of the decision-making process. 

Junges et al.~\cite{Junges21} present a runtime monitoring approach for partially observable systems with non-deterministic and probabilistic dynamics. The approach is based on traces of observations on models that combine non-determinism and probabilities. The authors propose a technique called forward filtering to estimate the possible system states in
partially observable settings along with a pruning strategy to enhance its efficiency. Based on empirical results, the authors propose 
a tractable algorithm based on model checking conditional reachability probabilities as a more tractable alternative. In contrast, DLASeR\fplus{} focuses on the reduction of large adaptation spaces of self-adaptive systems that are subject to uncertainties that can be expressed as parameters of runtime models. Yet, DLASeR\fplus{} can be combined with the proposed approach to enhance the performance of decision-making in self-adaptive systems. 
}

%-------------------------------------------------------

%\section{Why not an auto-encoder}
%- The system context has a sequential behavior
%- Embedding the context data makes not a lot of sense
%Problems:
%-> Auto-encoder does not capture sequential behavior (online learning / recurrent approaches can capture this)
%-> A lot of data that is (highly) correlated is necessary
%   => hard to generate

\section{Conclusions}\label{sec:conclusions}

In this paper, we studied the research question: ``How to reduce large adaptation spaces and rank adaptation options effectively and efficiently for self-adaptive systems with threshold, optimization, and set-point goals?''
To answer this question, we presented DLASeR\fplus{}. DLASeR\fplus{} relies on an integrated deep neural network architecture that shares knowledge of different adaptation goals. The approach is flexible as the core layers can be easily extended with goal-specific heads. The evaluation shows that DLASeR\fplus{} is an effective and efficient approach to reduce large adaptation spaces, including for settings with large adaptation spaces. The approach realizes the threshold and set-point goals for all the configurations we tested on the DeltaIoT artifact. Compared to the theoretical optimal, we observe only a small tradeoff for the quality property of the optimization goal. Yet this is a small cost for the dramatic improvement of adaptation time. 
%To answer this question, we presented DLASeR\fplus{}, a novel approach for the reduction of large adaptation spaces in self-adaptive systems. DLASeR\fplus{} supports the reduction of adaptation spaces for three common types of adaptation goals that are based on thresholds, set-points and an optimization criterion. DLASeR\fplus{} relies on an integrated deep neural network architecture that shares knowledge of different adaptation goals. The evaluation show that DLASeR\fplus{} is an effective and efficient approach to reduce large adaptation spaces. The approach realizes the threshold and set-point goals for all the configurations we tested on the DeltaIoT artifact. Compared to the ideal approach, we observe only a a small tradeoff for the quality property of the optimization goal. Yet this is a small cost for the dramatic improvement of adaptation time. 

We are currently applying DLASeR\fplus{} to service-based systems, which will provide us insights in the effectiveness of the approach beyond the domain of IoT. \added{For these systems, we are studying the reduction of adaptation spaces with sizes far beyond the adaptation spaces used in the evaluation of this paper, posing more challenging learning problems to DLASeR\fplus{}.} We also plan to extend DLASeR\fplus{} for multi-objective optimization goals. \added{Beyond DLASeR\fplus{} and learning-based approaches, we also plan to compare the approach with conceptually different approaches for improving the analysis of large adaptation spaces (as discussed in related work) and perform a tradeoff analysis.} In the mid term, we plan to look into support for dynamically adding and removing adaptation goals. We also plan to explore the use of machine learning in support of self-adaptation in decentralized settings~\cite{Quin21}. In the long term, we aim at investigating how we can define bounds on the guarantees that can be achieved when combining formal analysis techniques, in particular runtime statistical model checking, with machine learning, a starting point is~\cite{GheibiWQ21}. 

%% The bibliography
\bibliographystyle{ACM-Reference-Format}
\bibliography{references}

%% The appendix
%\appendix
%\input{text/appendix.tex}

\end{document}